\documentclass[journal]{IEEEtran}
\usepackage[T1]{fontenc}
\usepackage[latin9]{inputenc}
\usepackage{bm}
\usepackage{algorithm2e}
\usepackage{amsmath}
\usepackage{amsthm}
\usepackage{amssymb}
\usepackage{graphicx}
\usepackage{esint}
\usepackage[sort,compress]{cite}

\makeatletter

\providecommand{\tabularnewline}{\\}

\theoremstyle{plain}
\newtheorem{thm}{\theoremname}
\theoremstyle{definition}
\newtheorem{defn}{\definitionname}
\theoremstyle{remark}
\newtheorem{rem}{\remarkname}
\theoremstyle{plain}
\newtheorem{lem}{\lemmaname}
\theoremstyle{plain}
\newtheorem{prop}{\propositionname}
\theoremstyle{definition}
\newtheorem{example}{\examplename}

\usepackage{algpseudocode}
\usepackage{epstopdf}

\providecommand{\definitionname}{Definition}
\providecommand{\examplename}{Example}
\providecommand{\lemmaname}{Lemma}
\providecommand{\propositionname}{Proposition}
\providecommand{\remarkname}{Remark}
\providecommand{\theoremname}{Theorem}

\usepackage{pbox}

\usepackage{cuted}
\usepackage{flushend}

\@ifundefined{showcaptionsetup}{}{%
 \PassOptionsToPackage{caption=false}{subfig}}
\usepackage{subfig}
\makeatother

\providecommand{\definitionname}{Definition}
\providecommand{\examplename}{Example}
\providecommand{\lemmaname}{Lemma}
\providecommand{\propositionname}{Proposition}
\providecommand{\remarkname}{Remark}
\providecommand{\theoremname}{Theorem}

\begin{document}
\title{Ring Compute-and-Forward over Block-Fading Channels}
\author{Shanxiang Lyu, Antonio Campello, and Cong Ling,~\IEEEmembership{Member,~IEEE}
\thanks{This work was presented in part at the International Symposium on
Information Theory 2017, Aachen, Germany. The work of S. Lyu was supported
by the China Scholarship Council.} \thanks{S. Lyu is with the College of Information Science and Technology,  and
the College of Cyber Security, Jinan University, Guangzhou 510632,
China (e-mail: s.lyu14@imperial.ac.uk).} \thanks{A. Campello and C. Ling are with the Department of Electrical and
Electronic Engineering, Imperial College London, London SW7 2AZ, United
Kingdom (e-mail: accampellojr@gmail.com, cling@ieee.org).}}
\maketitle
\begin{abstract}
The Compute-and-Forward protocol in quasi-static channels normally
employs lattice codes based on the rational integers $\mathbb{Z}$,
Gaussian integers $\mathbb{Z}\left[i\right]$ or Eisenstein integers
$\mathbb{Z}\left[\omega\right]$, while its extension to more general channels often assumes channel state information at transmitters (CSIT). In this paper, we propose a novel
scheme for Compute-and-Forward in block-fading channels without CSIT, which is
referred to as Ring Compute-and-Forward because the fading coefficients
are quantized to the canonical embedding of a ring of algebraic integers.
Thanks to the multiplicative closure of the algebraic lattices employed,
a relay is able to decode an algebraic-integer linear combination
of lattice codewords. We analyze its achievable computation rates
and show it outperforms conventional Compute-and-Forward based on
$\mathbb{Z}$-lattices. By investigating the effect of Diophantine
approximation by algebraic conjugates, we prove that the degrees-of-freedom
(DoF) of the optimized computation rate is ${n}/{L}$, where $n$ is the number of blocks and $L$ is the
number of users. 
\end{abstract}


\section{Introduction}

\IEEEPARstart{E}{fficient} information transmission over wireless
relay networks has been extensively pursued in the past decades, in
which the main issues to address include signal interference and fading.
A number of relaying strategies have been proposed. The decode-and-forward
protocol \cite{Cover1979,Wang2007} decodes at least some parts of
the transmitted messages and removes the additive noise. Its main
drawback is that the decoding performance deteriorates when the number
of transmitters increases. The amplify-and-forward \cite{Borade2007,Patel2007}
and compress-and-forward \cite{Lim2011,Song2013} protocols maintain
signal interference where the relay either transmits a scaled version
of the received signal, or quantizes the received signal before passing
it to the destination. The additive noise can however be amplified
as signals traverse the network. The compute-and-forward (C\&F) \cite{Nazer2011}
protocol harnesses signal interference introduced by the channel and
removes the additive noise. It usually adopts lattice codes at source
nodes so that the relay can decode a linear function of the messages.
The C\&F paradigm has become a popular cooperative communication technique.
In most cases, the underlying channel is assumed to be quasi-static,
which means that the (random) fading coefficients stay constant over
the duration of each codeword.

There have been some works in the literature on C\&F dealing with
more general channel models \cite{Zhan2009,ElBakoury2015,Wang2016}.
In this paper, we investigate C\&F for block-fading channels so as
to achieve higher network throughput. Suppose that source nodes can
transmit information with $n$ different resources (e.g., multiple
carriers using orthogonal frequency-division multiplexing (OFDM)),
and that channel coefficients also remain constant over the duration
of each codeword. \textit{Our model of block-fading channels is essentially
that of parallel independent fading channels defined in \cite[Section 5.4.4]{Tse2012},
which assumes channel state information (CSI) at the receiver only.} While the block length (or coherence time) $T$ in block-fading is dictated by properties of the physical world, and is 
a  design parameter in parallel independent fading, the two models are equivalent if $T$ is large enough (see also \cite{Kositwattanarerk2015,CampelloLingBelfiore2016,Campello2016} for using term ``block-fading''). 
The crux here is that multiple resources offer diversity, which a coding scheme may
utilize to improve performance. 

Closely related to our work are \cite{ElBakoury2015,Wang2016}
where time-varying fading channels were investigated using lattice
codes over the rational integers $\mathbb{Z}$. Yet, the channel model
in \cite{ElBakoury2015,Wang2016} is slightly different in that it
consists of several blocks successive in time, which is better interpreted
as time diversity. Also assuming multiple receive antennas at the
relay, \cite{ElBakoury2015} derived the achievable rates of two integer-forcing
decoders, namely, the arithmetic-mean (AM) decoder and geometric-mean
(GM) decoder, for lattice codes over $\mathbb{Z}$. A practical C\&F
scheme based on root-LDA lattices was proposed in \cite{Wang2016},
where full diversity was observed for two-way relay channels and multiple-hop
line networks. In a multi-input multi-output (MIMO) multiple-access
channel (MAC), \cite{Zhan2009} showed the multiplexing gain in MIMO
C\&F is better than that provided by random coding if CSI is available
at transmitters. Without CSI to perform precoding, however, the multiplexing
gain in \cite{Zhan2009} is no better than that of a single antenna
setting. For this reason, a coding technique with more algebraic structures
is needed for C\&F over such channels. In this paper, we take a modest
step by proposing algebraic lattice codes for C\&F over block-fading
channels (which may be viewed as degenerated MIMO channels where channel
matrices are diagonal), while leaving algebraic lattice codes for
MIMO C\&F as future work.

In quasi-static fading channels, the structure of C\&F codes has been
extended to rings and modules, initiated in \cite{FSK13}. This extension
enlarges the space of code design, which brings several advantages
to C\&F. For example, using more compact rings can result in higher
computation rates, because the rational integers $\mathbb{Z}$ or
Gaussian integers $\mathbb{Z}[i]$ may not be the most suitable ring
to quantize channel coefficients. It has been shown that using the
Eisenstein integers $\mathbb{Z}[\omega]$ \cite{Sun2013,Tunali2015}
or rings from general quadratic number fields \cite{Huang2015b} can
have better computation rates for complex channels. Since the lattice
codes in these extensions are all $\mathcal{O}_{\mathbb{K}}$-modules
($\mathcal{O}_{\mathbb{K}}$ refers to the ring of integers in number
field $\mathbb{K}$), the message space can also be defined over $\mathcal{O}_{\mathbb{K}}$
due to the first isomorphism theorem of modules.

Our goal in this paper is to explore the fundamental limits of C\&F
over block-fading channels by using algebraic lattices built from
number fields of degree $n$ ($n\geq2$). In quasi-static channels,
the C\&F protocol essentially builds on capacity-achieving lattice
codes for the additive white Gaussian-noise (AWGN) channel \cite{Erez2004}.
To perform C\&F in block-fading channels, we employ universal lattice
codes proposed in \cite{CampelloLingBelfiore2016,Campello2016} for
compound block-fading channels. The celebrated Construction A has
been extended to number fields in recent years \cite{Oggier2013,Kositwattanarerk2015,Huang2015b,CampelloLingBelfiore2016,Campello2016b}.
In \cite{Kositwattanarerk2015}, the authors proposed algebraic lattice
codes based on Construction A over $\mathcal{O}_{\mathbb{K}}$ so
that the codes enjoy full diversity; subsequently it was proved in
\cite{CampelloLingBelfiore2016,Campello2016} that such generalized
Construction A can achieve the compound capacity of block-fading channels.
It was also briefly suggested in \cite{Huang2016a} that number-field
constructions as in \cite{Huang2015b,CampelloLingBelfiore2016,Campello2016}
could be advantageous for C\&F in a block-fading scenario.

In this work, we propose a scheme termed Ring C\&F based on such algebraic
lattices. As an extension of \cite{Lyu2017a}, we elaborate the construction
of algebraic lattices for Ring C\&F, and provide a detailed analysis
using the geometry of numbers and Diophantine approximation. The main
contributions of this work are the following:

1) We propose Ring C\&F over block-fading channels based on lattice
$\Lambda^{\mathcal{O}_{\mathbb{K}}}\left(\mathcal{C}\right)$ from
generalized Construction A, which satisfies relation $\mathcal{O}_{\mathbb{K}}^{T}/\Lambda^{\mathcal{O}_{\mathbb{K}}}\left(\mathcal{C}\right)/{\mathcal{I}_{\mathbb{K}}}^{T}$,
where $T$ is the number of channel uses, $\mathcal{O}_{\mathbb{K}}^{T}$,
$\Lambda^{\mathcal{O}_{\mathbb{K}}}\left(\mathcal{C}\right)$ and
${\mathcal{I}_{\mathbb{K}}}^{T}$ denote lattices built from ring
$\mathcal{O}_{\mathbb{K}}$ itself, code $\mathcal{C}$ and ideal
$\mathcal{I}_{\mathbb{K}}$, respectively. Such algebraic lattices
are shown to be $\mathcal{O}_{\mathbb{K}}$-submodules so that they
are multiplicatively closed. The relay aims to decode an algebraic-integer
linear combination of lattice codewords, which means that the channel
coefficient vectors are quantized to a lattice which is the canonical
embedding of the ring of integers $\mathcal{O}_{\mathbb{K}}$. As
a comparison, the lattice partition in a real quasi-static channel
is $\mathbb{Z}^{T}/\Lambda^{\mathbb{Z}}\left(\mathcal{C}\right)/\left(p\mathbb{Z}\right)^{T}$,
in which $p$ is a prime number. Also note the difference from techniques
in \cite{Sun2013,Tunali2015} where channel coefficients are quantized
to complex quadratic ring $\mathcal{O}_{\mathbb{K}}$ itself. Since
the channel coefficients in different fading blocks are unequal with
high probability, it is advantageous to employ the canonical embedding
of $\mathcal{O}_{\mathbb{K}}$ so as to enjoy better quantization
performance.

2) We analyze the computation rates in Ring C\&F based on the universal
coding goodness and quantization goodness of algebraic lattices. The
quantization goodness of algebraic lattices constructed from quadratic
number fields \cite{Huang2015b} is extended to general number fields.
The semi norm-ergodic metric in \cite{Ordentlich2012} is adopted
to handle the effective noise. Regarding the equivalent block-fading
channel, the universal lattice codes in \cite{Campello2016} play
an important role. In order to determine optimal algebraic-integer
coefficients, we resort to solving lattice problems over $\mathbb{Z}$-lattices
and provide a means to assure linear independency of multiple equations
over $\mathcal{O}_{\mathbb{K}}$.

3) We analyze the degrees-of-freedom (DoF) of our proposed coding
scheme. The DoF of C\&F over quasi-static fading channels has been
analyzed using the theory of Diophantine approximation in \cite{Niesen2012,Ordentlich2014,Nazer2016}.
Our analysis of DoF for Ring C\&F requires a new result of Diophantine
approximation by conjugates of an algebraic integer. The original
contribution of our work is the proof of a Khintchin-type result for
Diophantine approximation by conjugate algebraic integers (Lemma \ref{lem: DA_conjugate}).
It is well known that the standard Khintchine and Dirichlet theorems
\cite{Cassels1957} only deal with the approximation of real numbers
by rationals, which are algebraic numbers of degree one. Although
some results on approximating a real number by an algebraic number
are available in literature \cite{Roy04,Roy05}, these results come
with various restrictions which unfortunately do not lend themselves
to our problem at hand. For instance, \cite{Roy04} only addresses
simultaneous approximation of one number by algebraic conjugates or
multiple numbers by non-conjugates of a bounded degree, while \cite{Roy05}
requires the real numbers to be approximated lie in a field of transcendence
degree one.

The rest of this paper is organized as follows. In Section II, we
review some backgrounds on algebraic number theory and C\&F. In Sections
III and IV, we present our Ring C\&F scheme and analyze its computation
rates, respectively. In Section V, we analyze the achievable DoF without
CSI at transmitters. Subsequently Section VI provides some simulation
results. The last section concludes this paper.

Notation: The sets of all rationals, integers, real and complex numbers
are denoted by $\mathbb{Q}$, $\mathbb{Z}$, $\mathbb{R}$ and $\mathbb{C}$,
respectively. $\log$ denotes logarithm with base $2$, and $\log^{+}(x)=\max(\log(x),0)$.
Matrices and column vectors are denoted by uppercase and lowercase
boldface letters, respectively. $\mathrm{dg}(\mathbf{x})$ represents
a matrix filling vector $\mathbf{x}$ in the diagonal entries and
zeros in the others. The operation of stacking the columns of matrix
$\mathbf{X}$ one below the other is denoted by $\mathrm{vec}\left(\mathbf{X}\right)$.
$\|\mathbf{x}\|$ denotes the Euclidean norm of vector $\mathbf{x}$,
while $\|\mathbf{X}\|$ denotes the Frobenius norm of matrix $\mathbf{X}$.
$\otimes$ denotes the Kronecker tensor product, and $\oplus$ denotes
the finite field summation. $\mathcal{Q}_{\Lambda}(\cdot)$ is the
nearest neighbor quantizer to a lattice $\Lambda$. $\mathcal{V}\left(\Lambda\right)\triangleq\left\{ \mathbf{x}\in\mathbb{R}^{T}\mathrel{\Big|}\mathcal{Q}_{\Lambda}(\mathbf{x})=\mathbf{0}\right\} $
denotes the fundamental Voronoi region of lattice $\Lambda$. $\left[\mathbf{X}\right]\mod\thinspace\text{\ensuremath{\Lambda}}$
denotes $\left[\mathrm{vec}\left(\mathbf{X}\right)\right]\mod\thinspace\text{\ensuremath{\Lambda}}$.

\section{Preliminaries}

We first introduce necessary backgrounds on number fields and lattices
(readers are referred to texts \cite{BK:Mollin-ANT,ViterboOggier,BK:Zamir}
for an introduction to these subjects), then review the protocol of
C\&F over quasi-static channels.

\subsection{Number Fields and Lattices}
\begin{defn}[Number field]
Let $\theta$ be a complex number with minimum polynomial $\mathfrak{m}_{\theta}$
of degree $n$. A number field is a field extension $\mathbb{K}\triangleq\mathbb{F}(\theta)$
that defines the minimum field containing the base field $\mathbb{F}$
and the primitive element $\theta$. 
\end{defn}
A number $c$ is called an algebraic integer if its minimal polynomial
$\mathfrak{m}_{c}$ has integer coefficients. The maximal order of
an algebraic number field is its ring of integers. Let $\mathbb{S}$
be the set of algebraic integers, then the ring of integers is $\mathcal{O}_{\mathbb{K}}=\mathbb{K}\cap\mathbb{S}$.
The set $\left\{ \theta_{1},\theta_{2},\thinspace...,\thinspace\theta_{n}\right\} \in\mathcal{O}_{\mathbb{K}}^{n}$
is called an integral basis of $\mathcal{O}_{\mathbb{K}}$ if $\forall c\in\mathcal{O}_{\mathbb{K}}$,
$c=c_{1}\theta_{1}+c_{2}\theta_{2}+\ldots+c_{n}\theta_{n}$ with $c_{i}\in\mathbb{Z}$.
An element $u\in\mathcal{O}_{\mathbb{K}}$ is called a unit if it
is invertible under multiplication. All the units of $\mathcal{O}_{\mathbb{K}}$
form a multiplicative group $\mathcal{U}$, referred to as the unit
group.

An embedding of $\mathbb{K}$ into $\mathbb{C}$ is a homomorphism
into $\mathbb{C}$ that fixes elements in $\mathbb{Q}$. For a number
field of degree $n$, there are in total $n$ embeddings of $\mathbb{K}$
into $\mathbb{C}$: $\sigma_{i}:\thinspace\mathbb{K}\rightarrow\mathbb{C}$,
{$i=1,\ldots,n$,} referred to as canonical embedding. Canonical
embedding establishes a correspondence between an element of an algebraic
number field of degree $n$ and an $n$-dimensional vector in the
Euclidean space. The embeddings of $\theta$, denoted by {$\left\{ \sigma_{i}\left(\theta\right)\right\} _{i=1}^{n}$},
are determined by the roots of $\mathfrak{m}_{\theta}$. We denote
by $r_{1}$ the number of embeddings with image in $\mathbb{R}$ and
by $2r_{2}$ the number of embeddings with image in $\mathbb{C}$.
The pair $\left(r_{1},r_{2}\right)$ is called the signature of $\mathbb{K}$.
In a totally real number field, $\left(r_{1},r_{2}\right)=\left(n,0\right)$.

The following two quantities of an algebraic number are of particular
interest: 
\begin{enumerate}
\item The trace of $\theta$: $\mathrm{Tr}(\theta)\triangleq\sum_{i=1}^{n}\sigma_{i}(\theta)\in\mathbb{F}$; 
\item The norm of $\theta$: $\mathrm{Nr}(\theta)\triangleq\prod_{i=1}^{n}\sigma_{i}(\theta)\in\mathbb{F}.$ 
\end{enumerate}
In this work, we are only concerned with the scenario of real channels
and hence totally real number fields, so we use $\mathbb{F}=\mathbb{Q}$
as the base field. For an extension to complex channels, one can choose
$\mathbb{F}=\mathbb{Q}(i)$ as the base field. 
\begin{defn}[Ideals and prime ideals]
Let $R$ be a commutative ring with identity $1_{R}\neq0$. An ideal
$\mathfrak{I}$ of $R$ is a nonempty subset of $R$ that has the
following two properties: 
\begin{enumerate}
\item $c_{1}+c_{2}\in\mathfrak{I}$ if $c_{1},\thinspace c_{2}\in\mathfrak{I}$; 
\item $c_{1}c_{2}\in\mathfrak{I}$ if $c_{1}\in\mathfrak{I},\thinspace c_{2}\in R$. 
\end{enumerate}
An ideal $\mathfrak{p}$ of $R$ is prime if it has the following
two properties: 
\begin{enumerate}
\item If $c_{1}$ and $c_{2}$ are two elements of $R$ such that their
product $c_{1}c_{2}$ is an element of $\mathfrak{p}$, then either
$c_{1}\in\mathfrak{p}$ or $c_{2}\in\mathfrak{p}$; 
\item $\mathfrak{p}$ is not equal to $R$ itself. 
\end{enumerate}
\end{defn}
Every ideal of $R$ can be decomposed into a product of prime ideals.
In particular, if $p$ is a rational prime, we have $pR=\prod_{i=1}^{g}\mathfrak{p}_{i}^{e_{i}}$
in which $e_{i}$ is the ramification index of prime ideal $\mathfrak{p}_{i}$.
The inertial degree of $\mathfrak{p}_{i}$ is defined as $f_{i}=[R/\mathfrak{p}_{i}\thinspace:\thinspace\mathbb{Z}/p\mathbb{Z}]$,
and it satisfies $\sum_{i=1}^{g}e_{i}f_{i}=n$. Each prime ideal $\mathfrak{p}_{i}$
is said to lie above $p$. 
\begin{defn}[Modules]
A $R$-module is a set $M$ together with a binary operation under
which $M$ forms an Abelian group, and an action of $R$ on $M$ which
satisfies the same axioms as those for vector spaces. 
\end{defn}
Let $D$ be a subset of $R$-module $M$. $D$ forms an $R$-module
basis of $M$ if every element in $M$ can be written as a finite
linear combination of the elements of $D$. The order of the basis
is called the rank of the module. A finite subset $\left\{ d_{1},\ldots\thinspace,d_{m}\right\} $
of distinct elements of $M$ is said to be linearly independent over
$R$ if whenever $\sum_{i=1}^{m}c_{i}d_{i}=\mathbf{0}$ for some $c_{1},\ldots\thinspace,c_{m}\in R$,
then $c_{1}=\cdots=c_{m}=0$.

%We refer to a $R$-module that is discrete ($\mathbf{0}$ is not a limit point)  as a $R$-lattice.
%In this paper, we will be concerned with $\mathbb{Z}$-lattices from $\mathcal{O}_{\mathbb{K}}$-modules.
A real $\mathbb{Z}$-lattice is a discrete $\mathbb{Z}$-submodule
of $\mathbb{R}^{m}$. Such a lattice $\Lambda'$ generated by a basis
$\mathbf{D}=[\mathbf{d}_{1},\ldots\thinspace,\mathbf{d}_{m}]\in\mathbb{R}^{m\times m}$
can be written as a direct sum: 
\[
\Lambda'(\mathbf{D})=\mathbb{Z}\mathbf{d}_{1}+\mathbb{Z}\mathbf{d}_{2}+\cdots+\mathbb{Z}\mathbf{d}_{m}.
\]
With canonical embedding $\sigma$, an $\mathcal{O}_{\mathbb{K}}$-module
$\Lambda$ of rank $m$ can be transformed into a $\mathbb{Z}$-lattice
$\Lambda'$, and we write $\Lambda'=\sigma\left(\Lambda\right)$.
If $\mathbb{K}$ is a totally real number field of degree $n$, then
we have an embedded basis $\mathbf{D}\in\mathbb{R}^{mn\times mn}$,
and we define its discriminant as $\mathrm{disc}_{\mathbb{K}}=|\det(\mathbf{D})|^{2}$.
The successive minima $\lambda_{i}(\Lambda')$ of the $\mathbb{Z}$-lattice
$\Lambda'$ are defined in the usual manner. Analogously, we may define
successive minima of $\Lambda$ over $\mathcal{O}_{\mathbb{K}}$.
\begin{defn}[Successive minima of modules \cite{Rogers-SD}]
\label{def:ALsmp}The $i$th successive minimum of an $\mathcal{O}_{\mathbb{K}}$-module
$\Lambda$ is the smallest real number $r$ such that the ball $\mathcal{B}(\mathbf{0},r)$
contains the canonical embedding of $i$ linearly independent vectors
of $\sigma\left(\Lambda\right)$ over $\mathbb{K}$: 
\[
\lambda_{i}(\Lambda)=\inf\left\{ r\mathrel{\Big|}\dim\left(\mathrm{span}_{\mathbb{K}}\left(\sigma^{-1}\left(\sigma\left(\Lambda\right)\cap\mathcal{B}(\mathbf{0},r)\right)\right)\right)\geq i\right\} .
\]
\end{defn}
Notice that $\lambda_{1}(\Lambda)=\lambda_{1}(\Lambda')$, and in
general $\lambda_{i}(\Lambda)\geq\lambda_{i}(\Lambda')$ for $i>1$.
Also, if $\mathbf{x}_{1},\ldots,\mathbf{x}_{m}$ are linearly independent
over $\mathbb{K}$ and achieve the successive minima of $\Lambda$,
then the embeddings $\sigma(\mathbf{x}_{1}),\ldots,\sigma(\mathbf{x}_{m})$
are linearly independent and primitive in the Euclidean lattice $\Lambda'$.

For any real $\mathbb{Z}$-lattice $\Lambda'(\mathbf{D})$ with $\mathbf{D}\in\mathbb{R}^{m\times m}$,
Minkowski's first theorem states that \cite{Lekkerkerker1987} 
\begin{equation}
\lambda_{1}^{2}(\Lambda')\leq\kappa_{m}|\det(\mathbf{D})|^{\frac{2}{m}},\label{eq:min1}
\end{equation}
and Minkowski's second theorem states that 
\begin{equation}
\prod_{i=1}^{m}\lambda_{i}^{2}(\Lambda')\leq\kappa_{m}^{m}|\det(\mathbf{D})|^{2},\label{eq:min2}
\end{equation}
where $\kappa_{m}\triangleq\sup_{\Lambda'(\mathbf{D})}\lambda_{1}(\Lambda')^{2}/|\mathrm{det}(\mathbf{D})|^{2/m}$
is called Hermite's constant.

Analogous bounds exist for the successive minima of $\mathcal{O}_{\mathbb{K}}$-module
$\Lambda$. Obviously, 
\begin{equation}
\lambda_{1}^{2}(\Lambda)\leq\kappa_{mn}|\det(\mathbf{D})|^{\frac{2}{mn}},
\end{equation}
since the first minimum is identical. Applying Minkowski's second
theorem to \cite[Theorem 2]{Fieker10} yields 
\begin{equation}
\prod_{i=1}^{m}\lambda_{i}^{2n}\left(\Lambda\right)\leq\kappa_{mn}^{mn}|\det(\mathbf{D})|^{2}.\label{eq:mst-rogers}
\end{equation}

\subsection{C\&F over Quasi-Static Fading Channels}

Consider an AWGN network with $L$ source nodes which cannot collaborate with each other and are noiselessly connected to a final destination. We assume that all
source nodes are operating with the same message space $W$ (over
finite fields \cite{Nazer2011} or rings \cite{FSK13}), and the same message
rate $R_{\mathrm{mes}}=\frac{1}{T}\log(|W|)$. Let $\left(\Lambda_{c}^{\mathbb{Z}},\Lambda_{f}^{\mathbb{Z}}\right)$
be a pair of nested lattices in the partition chain $\mathbb{Z}^{T}/\Lambda_{f}^{\mathbb{Z}}/\Lambda_{c}^{\mathbb{Z}}/\left(p\mathbb{Z}\right)^{T}$,
in which $p$ is a prime number growing with the lattice dimension.
A message $\mathbf{w}_{l}\in W$ is mapped bijectively into a lattice
code via $\mathbf{x}_{l}=\mathcal{E}(\mathbf{w}_{l})\in\gamma\Lambda_{f}^{\mathbb{Z}}$,
satisfying a power constraint of $\left\Vert \mathbf{x}_{l}\right\Vert ^{2}\leq TP$.
$\gamma$ denotes a parameter to control the transmission power, and
$P$ denotes the signal power, hence the signal-to-noise ratio (SNR)
if the noise variance is normalized.

The noisy observation at a relay is 
\begin{equation}
\mathbf{y}=\sum_{l=1}^{L}h_{l}\mathbf{x}_{l}+\mathbf{z},\label{eq:fund_model}
\end{equation}
where the channel coefficients $\mathbf{h}=[h_{1},\ldots\thinspace,h_{L}]^{\top}\in\mathbb{R}^{L}$,
and the additive noise $\mathbf{z}\sim\mathcal{N}(\mathbf{0},\mathbf{I}_{T})$.
The relay aims to compute a finite field equation 
\begin{equation}
\mathbf{u}=\bigoplus_{l=1}^{L}a_{l}\mathbf{w}_{l}\label{eq:INDmes}
\end{equation}
with coefficient vector $\mathbf{a}=[a_{1},\ldots\thinspace,a_{L}]^{\top}\in\mathbb{Z}^{L}$
and forward $\mathbf{u},\mathbf{a}$ to the destination. Each $\mathbf{u}$
corresponds to a lattice equation $\left[\sum_{l=1}^{L}a_{l}\mathbf{x}_{l}\right]\mod\thinspace\gamma\Lambda_{c}^{\mathbb{Z}}$
as they are isomorphic. By first estimating the lattice equation and
then map it to a finite field, the forwarded message from the relay
is written as $\hat{\mathbf{u}}=\mathcal{D}\left(\mathbf{y}\mid\mathbf{h},\mathbf{a}\right)$.
We say equation $\mathbf{u}=\bigoplus_{l=1}^{L}a_{l}\mathbf{w}_{l}$
is decoded with probability of error $\delta$ if $\mathrm{Pr}\left(\mathbf{u}\neq\hat{\mathbf{u}}\right)<\delta.$
\begin{defn}[Achievable Computation Rate for a Chosen $\mathbf{a}$ at a Relay] \label{static-achieve-rate}
For a given channel coefficient vector $\mathbf{h}$ and {a chosen}
coefficient vector $\mathbf{a}$, the computation rate $R_{\mathrm{comp}}\left(\mathbf{h},\mathbf{a}\right)$
is achievable at a relay if for any $\delta>0$ and $T$ large enough,
there exist encoders $\mathcal{E}_{1},\ldots\thinspace\mathcal{E}_{L}$
and decoders $\mathcal{D}$ such that the relay can recover its desired
equation with error probability bound $\delta$ if the underlying
message rate $R_{\mathrm{mes}}$ satisfies: 
\[
R_{\mathrm{mes}}<R_{\mathrm{comp}}\left(\mathbf{h},\mathbf{a}\right).
\]
\end{defn}

\begin{thm}[\cite{Nazer2011}]
	\label{thm:nazerThm}There is a sequence of nested lattice codebooks
	$\left\{ \Lambda_{f}^{\mathbb{Z}},\Lambda_{c}^{\mathbb{Z}}\right\} $
	of length $T$, such that by setting $T\rightarrow\infty$, the following
	computation rate is achievable: 
	\begin{equation}
	R_{\mathrm{comp}}\left(\mathbf{h},\mathbf{a}\right)=\frac{1}{2}\max_{\alpha\in\mathbb{R}}\log^{+}\left(\frac{P}{|\alpha|^{2}+P\left\Vert \alpha\mathbf{h}-\mathbf{a}\right\Vert ^{2}}\right).\label{eq:staticrate}
	\end{equation}
\end{thm}

Upon receiving $L$ linearly independent equations in the form of
(\ref{eq:INDmes}), the destination estimates the messages by inverting
the equations. %\begin{equation}
%[\hat{\mathbf{w}}_{1}^{\top},\ldots\thinspace\hat{\mathbf{w}}_{L}^{\top}]=[\mathbf{a}_{1}^{\top},\ldots\thinspace\mathbf{a}_{L}^{\top}]^{-1}[\mathbf{u}_{1}^{\top},\ldots\thinspace\mathbf{u}_{L}^{\top}].\label{eq:modulo equation}
%\end{equation}
The maximum information rate that the destination can receive through
the AWGN network is dictated by the computation rates at the relays.
\begin{defn}[Achievable Computation Rate of the AWGN Network] \label{static-achieve-Network}
	Given $\left\{ \mathbf{h}_{l}\right\} _{l=1}^{L}$, and $\left\{ \mathbf{a}_{l}\right\} _{l=1}^{L}$ from $L$ relays such that the morphism of $\left\{ \mathbf{a}_{l}\right\} _{l=1}^{L}$ is invertible in the message space, the achievable computation rate of the AWGN network is $\min_{l} R_{\mathrm{comp}}\left(\mathbf{h}_{l},\mathbf{a}_{l}\right)$.
\end{defn}
 
 To characterize the the growth of computation rate w.r.t. SNR, 
define the DoF as
\begin{equation}
d_{\mathrm{comp}}=\lim_{P\rightarrow\infty}\frac{\max_{\mathbf{a}}R_{\mathrm{comp}}\left(\mathbf{h},\mathbf{a}\right)}{\frac{1}{2}\log\left(1+P\right)}.\label{eq:dof definition}
\end{equation}
Using the theory of Diophantine approximation, Niesen and Whiting
\cite{Niesen2012} showed that 
\[
d_{\mathrm{comp}}\leq\left\{ \begin{array}{ll}
\frac{1}{2}, & L=2;\\
\frac{2}{L+1}, & L>2.
\end{array}\right.
\]
This has subsequently been improved by Ordenlitch, Erez and Nazer
\cite{Ordentlich2014} to 
\[
d_{\mathrm{comp}}=\frac{1}{L}.
\]

\section{Ring C\&F}

\begin{figure}[t]
\center

\includegraphics[width=0.45\textwidth]{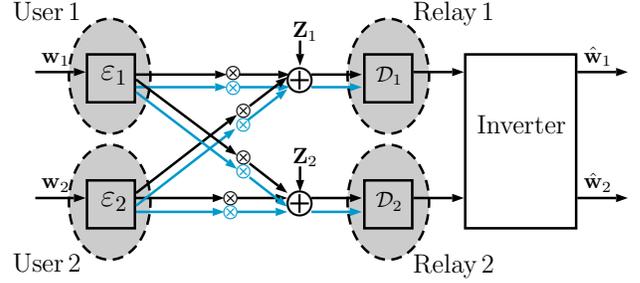}

\caption{Compute-and-Forward over block-fading channels with $2$ users and
$2$ relays.}
\label{fig8 block MAC} 
\end{figure}
In this work, we consider a block-fading scenario where diversity
is supplied in $n$ blocks and fading coefficients remain constant
in each frame of coherence time $T$. That is, the fading process
experienced by a codeword $\mathbf{x}_{l}$ of user $l$ consists
of $n$ blocks $\{\underset{T}{\underbrace{h_{1,l},h_{1,l},\ldots,h_{1,l}}}\}$,
$\{\underset{T}{\underbrace{h_{2,l},h_{2,l},\ldots,h_{2,l}}}\}$,
$\cdots$, $\{\underset{T}{\underbrace{h_{n,l},h_{n,l},\ldots,h_{n,l}}}\}$
in parallel. Thus the received signal at a relay can be written in
matrix form as 
\begin{equation}
\mathbf{Y}=\sum_{l=1}^{L}\mathbf{H}_{l}\mathbf{X}_{l}+\mathbf{Z},\label{eq: fad3}
\end{equation}
where $\mathbf{Y}\in\mathbb{R}^{n\times T}$, $\mathbf{H}_{l}=\mathrm{dg}\left(h_{1,l},\ldots\thinspace,h_{n,l}\right)$
denotes the channel coefficients from user $l$ to the relay, $\mathbf{X}_{l}\in\mathbb{R}^{n\times T}$
denotes a transmitted codeword to be designed in the sequel, and $\mathbf{Z}\in\mathbb{R}^{n\times T}$
is the additive noise with entries drawn from $\mathcal{N}\left(0,1\right)$. The index of the relay is dropped in the equation for simplicity of notation.
The C\&F diagram for this model with two users (source nodes) and
two relays is shown in Fig. \ref{fig8 block MAC}. In the figure,
the encoded messages $\mathcal{E}\left(\mathbf{w}_{1}\right)$ and
$\mathcal{E}\left(\mathbf{w}_{2}\right)$ are both transmitted by
using two sub-channels in parallel, which are respectively denoted
by black and blue arrows. Relays $1$ and $2$ forward two linearly
independent equations to the destination which subsequently recovers
message $\hat{\mathbf{w}}_{1}$, $\hat{\mathbf{w}}_{2}$ by inverting
the equations.

Next, we present our Ring C\&F scheme, which contains message encoding
based on algebraic lattices (such that
the degree of the number field equals to the number of blocks in the block-fading model), and decoding algebraic-integer
linear combinations of lattice codewords. The ``goodness''
properties of algebraic lattices are shown in the last subsection.

\subsection{Encoding}

We follow \cite{Kositwattanarerk2015,Huang2015b,CampelloLingBelfiore2016}
to build lattices from Construction A over number fields. Choose a
prime ideal $\mathfrak{p}$ lying above rational prime $p$ with inertial
degree $f$ so that we have an isomorphism $\mathcal{O}_{\mathbb{K}}/\mathfrak{p}\cong\mathbb{F}_{p^{f}}$.
Let $\mathcal{C}$ be a $(T,\thinspace k)$ linear code over $\mathbb{F}_{p^{f}}$
where $k<T$. Let $\rho:\thinspace\mathcal{O}_{\mathbb{K}}\rightarrow\mathbb{F}_{p^{f}}$
be a component-wise ring homomorphism defined by reduction modulo
the ideal $\mathfrak{p}$. Generalized Construction A from code $\mathcal{C}$
is defined as 
\begin{equation}
\Lambda^{\mathcal{O}_{\mathbb{K}}}(\mathcal{C})=\rho^{-1}\left(\mathcal{C}\right),\label{eq:firstconstruction}
\end{equation}
which is a free $\mathbb{Z}$-module\footnote{A free module is a module that has a basis.}
of rank $nT$. The coding lattice $\Lambda^{\mathbb{Z}}(\mathcal{C})$
is the canonical embedding of $\mathcal{O}_{\mathbb{K}}$ module $\Lambda^{\mathcal{O}_{\mathbb{K}}}(\mathcal{C})$
into the Euclidean space.

We first build a pair of nested lattices $\left(\Lambda_{f}^{\mathbb{Z}},\Lambda_{c}^{\mathbb{Z}}\right)$
based on a pair of nested linear codes $(\mathcal{C}_{f},\thinspace\mbox{\ensuremath{\mathcal{C}}}_{c})$.
Let $k_{c}<k_{f}<T$. Define $\mathcal{C}_{f}=\left\{ \mathbf{G}_{f}\mathbf{w}_{f}\mathrel{\Big|}\mathbf{w}_{f}\in\mathbb{F}_{p^{f}}^{k_{f}}\right\} $
and $\mathcal{C}_{c}=\left\{ \mathbf{G}_{c}\mathbf{w}_{c}\mathrel{\Big|}\mathbf{w}_{c}\in\mathbb{F}_{p^{f}}^{k_{c}}\right\} ,$
where $\mathbf{G}_{f}=\left[\mathbf{G}_{c},\mathbf{G}'\right]\in\mathbb{F}_{p^{f}}^{T\times k_{f}}$,
and $\mathbf{G}_{c}\in\mathbb{F}_{p^{f}}^{T\times k_{c}}$. These
codes are then lifted from $\mathbb{F}_{p^{f}}^{T}$ to $\mathcal{O}_{\mathbb{K}}^{T}$:
\[
\Lambda_{f}^{\mathcal{O}_{\mathbb{K}}}=\mathcal{\rho}^{-1}(\mathcal{C}_{f}),\ \ \Lambda_{c}^{\mathcal{O}_{\mathbb{K}}}=\mathcal{\rho}^{-1}(\mathcal{C}_{c}),
\]
which produce $\mathbb{Z}$-lattices $\Lambda_{f}^{\mathbb{Z}}$ and
$\Lambda_{c}^{\mathbb{Z}}$ with canonical embeddings. The volumes
of the Voronoi regions of $\Lambda_{f}^{\mathbb{Z}}$ and $\Lambda_{c}^{\mathbb{Z}}$
are $\mathrm{Vol}\left(\Lambda_{f}^{\mathbb{Z}}\right)=p^{(T-k_{f})f}\mathrm{disc}_{\mathbb{K}}^{T/2}$
and $\mathrm{Vol}\left(\Lambda_{c}^{\mathbb{Z}}\right)=p^{(T-k_{c})f}\mathrm{disc}_{\mathbb{K}}^{T/2}$,
respectively. Let $\left\{ \phi_{1},\thinspace...,\thinspace\phi_{n}\right\} $
be an integral basis of $\mathcal{O}_{\mathbb{K}}$. Since every ideal
of $\mathcal{O}_{\mathbb{K}}$ is a free $\mathbb{Z}$-module of rank
$n$, a basis of ideal $\mathfrak{p}$ can be represented by $\left\{ \mu_{1},\ldots,\mu_{n}\right\} $
where $\mu_{i}=\sum_{j=1}^{n}\mu_{ij}\phi_{j},\thinspace\mu_{ij}\in\mathbb{Z}$.
Thus the generator matrices of $\mathcal{O}_{\mathbb{K}}$ and $\mathfrak{p}$
are respectively given by 
\begin{align*}
 & \Phi=\left[\begin{array}{ccc}
\sigma_{1}(\phi_{1}) & \cdots & \sigma_{1}(\phi_{n})\\
\sigma_{2}(\phi_{1}) & \cdots & \sigma_{2}(\phi_{n})\\
\vdots & \vdots & \vdots\\
\sigma_{n}(\phi_{1}) & \cdots & \sigma_{n}(\phi_{n})
\end{array}\right],\\
 & \Phi_{\mathfrak{p}}=\left[\begin{array}{ccc}
\sum_{j=1}^{n}\mu_{1j}\sigma_{1}(\phi_{j}) & \cdots & \sum_{j=1}^{n}\mu_{nj}\sigma_{1}(\phi_{j})\\
\sum_{j=1}^{n}\mu_{1j}\sigma_{2}(\phi_{j}) & \cdots & \sum_{j=1}^{n}\mu_{nj}\sigma_{2}(\phi_{j})\\
\vdots & \vdots & \vdots\\
\sum_{j=1}^{n}\mu_{1j}\sigma_{n}(\phi_{j}) & \cdots & \sum_{j=1}^{n}\mu_{nj}\sigma_{n}(\phi_{j})
\end{array}\right].
\end{align*}
Let the canonical representations of $\mathcal{C}_{f}$ , $\mathcal{C}_{c}$
be $\mathbf{G}_{f}=[\mathbf{I}_{k_{f}},\mathbf{A}_{f}^{\top}]^{\top}$,
$\mathbf{G}_{c}=[\mathbf{I}_{k_{c}},\mathbf{A}_{c}^{\top}]^{\top}$,
it was shown in \cite{Kositwattanarerk2015} that the generator matrices
of $\Lambda_{f}^{\mathbb{Z}}$ and $\Lambda_{c}^{\mathbb{Z}}$ are
respectively given by 
\begin{align*}
 & \mathbf{M}_{f}=\left[\begin{array}{cc}
\mathbf{I}_{k_{f}}\otimes\Phi & \mathbf{0}_{nk_{f},n(T-k_{f})}\\
\mathbf{A}_{f}\otimes\Phi & \mathbf{I}_{T-k_{f}}\otimes\Phi_{\mathfrak{p}}
\end{array}\right],\\
 & \mathbf{M}_{c}=\left[\begin{array}{cc}
\mathbf{I}_{k_{c}}\otimes\Phi & \mathbf{0}_{nk_{c},n(T-k_{c})}\\
\mathbf{A}_{c}\otimes\Phi & \mathbf{I}_{T-k_{c}}\otimes\Phi_{\mathfrak{p}}
\end{array}\right].
\end{align*}

For each user, a message $\mathbf{w}\in\mathbb{F}_{p^{f}}^{k_{f}-k_{c}}$
is encoded into $\mathbf{\tilde{x}}\in\Lambda_{f}^{\mathcal{O}_{\mathbb{K}}}$
as
\begin{equation}
\mathbf{\tilde{x}}=\mathcal{E}\left(\mathbf{w}\right)\triangleq\gamma\left[\rho^{-1}\left(\mathbf{G}'\mathbf{w}\right)\right]\mod\thinspace\Lambda_{c}^{\mathcal{O}_{\mathbb{K}}},\label{eq:encodedpoints}
\end{equation}
with a transmission rate $R_{\mathrm{mes}}=\frac{(k_{f}-k_{c})f}{T}\log(p)$.
The actually transmitted codeword is obtained by apply component-wise
canonical embedding to $\mathbf{\tilde{x}}$, which yields its matrix
form 
\begin{equation}
\mathbf{X}=\gamma\left[\begin{array}{c}
\sigma_{1}\left({\mathbf{\tilde{x}}}^{\top}\right)\\
\sigma_{2}\left({\mathbf{\tilde{x}}}^{\top}\right)\\
\vdots\\
\sigma_{n}\left({\mathbf{\tilde{x}}}^{\top}\right)
\end{array}\right]\in\mathbb{R}^{n\times T}.\label{eq:alge_semble}
\end{equation}
Again, $\gamma$ denotes a power scaling factor as before. In the
construction, it is possible to map messages to lattice points and
back while preserving linearity. 
\begin{prop}
\label{prop:mappingLemma}The encoding function $\mathcal{E}\left(\mathbf{w}\right)$
defines a bijection between messages $\mathbf{w}\in\mathbb{F}_{p^{f}}^{k_{f}-k_{c}}$
and lattice points inside $\Lambda_{f}^{\mathbb{Z}}\cap\mathcal{V}(\Lambda_{c}^{\mathbb{Z}})$. 
\end{prop}
\begin{IEEEproof}
As $\rho^{-1}\left(\mathcal{C}\right)$ defines a lattice, there is
a unique correspondence between a codeword $\mathbf{G}'\mathbf{w}_{i}$
and a lattice coset $\Lambda_{c}^{\mathbb{Z}}+\mathbf{x}_{i}^{*}$,
where the set of representatives $\left\{ \mathbf{x}_{i}^{*}\right\} $
satisfy $\left|\left\{ \mathbf{x}_{i}^{*}\right\} \right|=p^{(k_{f}-k_{c})f}$,
and $\mathbf{x}_{i}^{*}\notin\Lambda_{c}^{\mathbb{Z}}$ if $\mathbf{x}_{i}^{*}\neq\mathbf{0}$.
We only need to show points in different cosets would not collide
after modulo $\Lambda_{c}^{\mathbb{Z}}$, in which 
\[
\left[\rho^{-1}\left(\mathbf{G}'\mathbf{w}_{i}\right)\right]\mod\thinspace\Lambda_{c}^{\mathbb{Z}}=\mathbf{x}_{i}^{*}+\arg\min_{\hat{\mathbf{x}}_{i}\in\Lambda_{c}^{\mathbb{Z}}}\left\Vert \mathbf{x}_{i}^{*}+\hat{\mathbf{x}}_{i}\right\Vert ^{2}.
\]
Since $\mathbf{x}_{i}^{*}-\mathbf{x}_{j}^{*}\notin\Lambda_{c}^{\mathbb{Z}}$
{for $i\neq j$}, there is no $\hat{\mathbf{x}}_{i}\in\Lambda_{c}^{\mathbb{Z}}$
such that $\mathbf{x}_{i}^{*}-\mathbf{x}_{j}^{*}+\hat{\mathbf{x}}_{i}\in\Lambda_{c}^{\mathbb{Z}}$,
and the proposition is proved. 
\end{IEEEproof}
As usual, we apply dithering from the set $\left\{ \mathrm{vec}\left(\mathbf{D}_{l}\right)\right\} _{l=1}^{L}$
where each $\mathrm{vec}\left(\mathbf{D}_{l}\right)$ is uniformly
distributed over $\mathcal{V}\left(\Lambda_{c}^{\mathbb{Z}}\right)$.
To simplify the presentation, however, we defer their presence until
Section \ref{sec:Rate-analysis}.

\subsection{Decoding}

The following lemma is the crux of our decoding algorithm, which {says
codewords} $\mathbf{X}_{l}$'s are not only closed in $\gamma\Lambda_{f}^{\mathbb{Z}}$
under $\mathbb{Z}$-linear combinations, but more generally under
$\mathcal{O}_{\mathbb{K}}$-linear combinations. 
\begin{lem}
\label{prop:ScalarAbsorb}Let $a_{l}\in\mathcal{O}_{\mathbb{K}}$,
and $\mathbf{A}_{l}=\mathrm{dg}(\sigma_{1}(a_{l}),\thinspace...,\thinspace\sigma_{n}(a_{l}))$
for $1\leq l\leq L$. The physical layer codewords are closed under
the action of ring elements, i.e., $\sum_{l=1}^{L}\Big(\mathbf{A}_{l}\mathbf{X}_{l}\Big)\in\gamma\Lambda_{f}^{\mathbb{Z}}.$ 
\end{lem}
\begin{IEEEproof}
We let $\gamma=1$ for clarity. We first show that $\Lambda^{\mathcal{O}_{\mathbb{K}}}(\mathcal{C})$
constructed from (\ref{eq:firstconstruction}) is an $\mathcal{O}_{\mathbb{K}}$-submodule.
The definitions of rings and ideals show that $\mathcal{O}_{\mathbb{K}}$,
$\mathfrak{p}$ are both $\mathcal{O}_{\mathbb{K}}$-modules of rank
$1$. It then follows from \cite[p. 338]{Dummit2003} that the Cartesian
product $\mathcal{O}_{\mathbb{K}}^{T}$ is a free $\mathcal{O}_{\mathbb{K}}$-module
of rank $T$, based on component-wise addition and multiplication
by elements of $\mathcal{O}_{\mathbb{K}}$. Since $\mathcal{O}_{\mathbb{K}}/\mathfrak{p}\cong\mathbb{F}_{p^{f}}$
and $\mathcal{C}$ is a subgroup of $\mathbb{F}_{p^{f}}^{T}$, $\rho^{-1}\left(\mathcal{C}\right)$
becomes an $\mathcal{O}_{\mathbb{K}}$-submodule \cite[p. 342]{Dummit2003}
of $\mathcal{O}_{\mathbb{K}}^{T}$ {which satisfies} $a_{l}\rho^{-1}\left(\mathcal{C}\right)\subset\rho^{-1}\left(\mathcal{C}\right)$,
$\forall\thinspace a_{l}\in\mathcal{O}_{\mathbb{K}}$. It follows
from a component-wise ring homomorphism $\sigma(\cdot):\thinspace\mathbb{K}\rightarrow\mathbb{R}^{n}$
that $\mathbf{A}_{l}\mathbf{X}_{l}\in\Lambda_{f}^{\mathbb{Z}}$. Lastly,
the additive closure of lattice points clearly holds. 
\end{IEEEproof}
Based on Lemma \ref{prop:ScalarAbsorb}, the decoder aims to extract
an algebraic combination of lattice codewords from the scaled signal
\begin{equation}
\mathbf{B}\mathbf{Y}=\underset{\mathrm{lattice\thinspace codeword}}{\underbrace{\sum_{l=1}^{L}\mathbf{A}_{l}\mathbf{X}_{l}}}+\underset{\mathrm{effective\thinspace noise}}{\underbrace{\mathbf{B}\sum_{l=1}^{L}\mathbf{H}_{l}\mathbf{X}_{l}-\sum_{l=1}^{L}\mathbf{A}_{l}\mathbf{X}_{l}+\mathbf{B}\mathbf{Z}}},\label{eq: deeffect-1}
\end{equation}
where $\mathbf{B}=\mathrm{dg}(b_{1},\thinspace...,\thinspace b_{n}),$
$b_{i}\in\mathbb{R}$ is an minimum mean square error (MMSE) matrix.
We refer to

\begin{equation}
\left[\sum_{l=1}^{L}\mathbf{A}_{l}\mathbf{X}_{l}\right]\mod\thinspace\gamma\Lambda_{c}^{\mathbb{Z}}\label{eq:desired equation}
\end{equation}
as an algebraic lattice equation. With some {decoding procedures}
to be specified in the next section, we proceed by assuming (\ref{eq:desired equation})
is available. Then each relay can extract a finite field equation
\begin{align}
\mathbf{u} & =\left[\left(\mathbf{G}'^{\top}\mathbf{G}'\right)^{-1}\mathbf{G}'^{\top}\rho\left(\left[\gamma^{-1}\sum_{l=1}^{L}\mathbf{A}_{l}\mathbf{X}_{l}\right]\mod\thinspace\Lambda_{c}^{\mathbb{Z}}\right)\right]\mod\thinspace\mathbb{F}_{p^{f}}\nonumber \\
 & =\sum_{l=1}^{L}\left[\left(\mathbf{G}'^{\top}\mathbf{G}'\right)^{-1}\mathbf{G}'^{\top}\rho\left(a_{l}\right)\rho\left(\gamma^{-1}\mathbf{X}_{l}\right)\right]\mod\thinspace\mathbb{F}_{p^{f}}\nonumber \\
 & =\bigoplus_{l=1}^{L}\rho\left(a_{l}\right)\mathbf{w}_{l},\label{eq:desired equation-1}
\end{align}
where the second equality is from the property of ring homomorphism
$\rho\left(\cdot\right)$, and the third equality is due to Proposition~\ref{prop:mappingLemma}
such that we have a bijection $\rho\left(\gamma^{-1}\mathbf{X}_{l}\right)=\mathbf{G}'\mathbf{w}_{l}$.

In practice,
all relays forward their decoded messages $\hat{\mathbf{u}}$'s and
coefficients $\left\{ \mathbf{A}_{l}\right\} _{l=1}^{L}$'s to the
destination, where $\hat{\mathbf{u}}=\mathcal{D}\left(\mathbf{Y}\mid\left\{ \mathbf{H}_{l}\right\} _{l=1}^{L},\left\{ \mathbf{A}_{l}\right\} _{l=1}^{L}\right)$ denotes an estimated
message. Upon collecting $L$ linearly independent equations from
those relays, the destination can estimate messages $\mathbf{w}_{1},\ldots\thinspace,\mathbf{w}_{L}$.

To explain the rationale, we give two examples below. Example \ref{exa:absorb}
demonstrates how multiplications are closed. Example \ref{exa:workout}
shows {the} information flow from users to a destination. 
\begin{example}
\label{exa:absorb}Let $\mathrm{vec}\left(\mathbf{X}_{l}\right)=\mathbf{M}_{f}\mathbf{z}_{l}$,
$\mathbf{z}_{l}\in\mathbb{Z}^{nT}$. The closure of $\mathbf{A}_{l}\times\Lambda^{\mathbb{Z}}\subset\Lambda^{\mathbb{Z}}$
implies that $\mathrm{vec}\left(\mathbf{A}_{l}\mathbf{X}_{l}\right)=\mathbf{M}_{f}\mathbf{z}_{l}',$
$\mathbf{z}_{l}'\in\mathbb{Z}^{nT}$, where $\mathbf{z}_{l}=\mathbf{z}_{l}'$
if and only if $a_{l}=1$. For instance, in a quadratic field $\mathbb{K}=\mathbb{Q}\left(\sqrt{3}\right)$,
let the lattice basis be $\mathbf{M}_{f}=\left[\begin{array}{cc}
1 & \sqrt{3}\\
1 & -\sqrt{3}
\end{array}\right]$ and the multiplication coefficient be $a_{l}=1+\sqrt{3}$. Then for
any $\mathbf{z}_{l}\in\mathbb{Z}^{2}$, one has 
\[
\left[\begin{array}{cc}
1+\sqrt{3} & 0\\
0 & 1-\sqrt{3}
\end{array}\right]\mathbf{M}_{f}\mathbf{z}_{l}=\mathbf{M}_{f}\mathbf{z}_{l}'
\]
with $\mathbf{z}_{l}'=\left[\begin{array}{cc}
1 & 3\\
1 & 1
\end{array}\right]\mathbf{z}_{l}\in\mathbb{Z}^{2}$. 
\end{example}
\begin{example}
\label{exa:workout}Consider quadratic field $\mathbb{K}=\mathbb{Q}\left(\sqrt{5}\right)$.
Choose $p=5$, so the ideal factorization becomes $p\mathcal{O}_{\mathbb{K}}=\mathfrak{p}^{2}$,
where $\mathfrak{p}=\frac{5-\sqrt{5}}{2}\mathbb{Z}+\frac{-5+3\sqrt{5}}{2}\mathbb{Z}$.
For the isomorphism $\mathbb{F}_{p}\cong\mathcal{O}_{\mathbb{K}}/\mathfrak{p}$,
the five coset representatives in $\mathbb{R}^{2}$ corresponding
to $\mathbb{F}_{5}$ are 
\begin{align*}
 & [0,0]^{\top},[1,1]^{\top},[\frac{-1-\sqrt{5}}{2},\frac{-1+\sqrt{5}}{2}]^{\top},\\
 & [\frac{1-\sqrt{5}}{2},\frac{1+\sqrt{5}}{2}]^{\top},[-1,-1]^{\top}.
\end{align*}
Let the two uncoded messages be $w_{1}=2$ for User $1$ and $w_{2}=3$
for User $2$. {For} $\gamma=1$, the transmitted lattice points
are 
\begin{eqnarray*}
\mathbf{X}_{1} & = & \mathcal{E}\left(w_{1}\right)=[\frac{-1-\sqrt{5}}{2},\frac{-1+\sqrt{5}}{2}]^{\top},\\
\mathbf{X}_{2} & = & \mathcal{E}\left(w_{2}\right)=[\frac{1-\sqrt{5}}{2},\frac{1+\sqrt{5}}{2}]^{\top}.
\end{eqnarray*}
For convenience, suppose the channel coefficients {are exactly taken
from} $\mathcal{O}_{\mathbb{K}}$. In Relay $1$, we receive $\mathbf{V}_{1}=\sum_{l=1}^{2}\mathbf{A}_{l}^{(1)}\mathbf{X}_{l}$
with 
\begin{align*}
 & \mathbf{A}_{1}^{(1)}=\mathrm{dg}\left(2+17\sqrt{5},2-17\sqrt{5}\right),\\
 & \mathbf{A}_{2}^{(1)}=\mathrm{dg}\left(13+\sqrt{5},13-\sqrt{5}\right).
\end{align*}
Its decoded message is $\hat{u}_{1}=\mathcal{D}\left(\mathbf{V}_{1}\right)=3$.
Similarly in Relay $2$, we receive $\mathbf{V}_{2}=\sum_{l=1}^{2}\mathbf{A}_{l}^{(2)}\mathbf{X}_{l}$
with

\begin{align*}
 & \mathbf{A}_{1}^{(2)}=\mathrm{dg}\left(\frac{15+9\sqrt{5}}{2},\frac{15-9\sqrt{5}}{2}\right),\\
 & \mathbf{A}_{2}^{(2)}=\mathrm{dg}\left(2+17\sqrt{5},2-17\sqrt{5}\right).
\end{align*}
Its decoded message is $\hat{u}_{2}=\mathcal{D}\left(\mathbf{V}_{2}\right)=1$.
Then Relays 1 and 2 forward messages $\hat{u}_{1}$, $\hat{u}_{2}$
along with coefficients $\rho\left(a_{1}^{(1)}\right)$, $\rho\left(a_{2}^{(1)}\right)$,
$\rho\left(a_{1}^{(2)}\right)$ and $\rho\left(a_{2}^{(2)}\right)$.
Namely, the destination also receives a finite field matrix 
\[
\mathbf{A}_{p}\triangleq\rho\left(\mathbf{A}\right)=\left[\begin{array}{cc}
\rho\left(a_{1}^{(1)}\right) & \rho\left(a_{2}^{(1)}\right)\\
\rho\left(a_{1}^{(2)}\right) & \rho\left(a_{2}^{(2)}\right)
\end{array}\right]=\left[\begin{array}{cc}
2 & 3\\
0 & 2
\end{array}\right],
\]
and accordingly obtains a solution 
\[
[\hat{w}_{1},\hat{w}_{2}]^{\top}=\mathbf{A}_{p}^{-1}[\hat{u}_{1},\hat{u}_{2}]^{\top}=[2,3]^{\top}.
\]
\end{example}
\begin{rem}
As in \cite[Theorem. 11]{Nazer2011}, we may choose large $p$ in
Ring C\&F such that if $\mathbf{A}$ has full rank over $\mathcal{O}_{\mathbb{K}}$
(i.e., linear independence over a number field), then $\mathbf{A}_{p}$
also has full rank over $\mathbb{F}_{p}$ (i.e., linear independence
over a finite field) with high probability. The sufficient and necessary
condition for ensuring $\mathbf{A}$ has full rank over $\mathcal{O}_{\mathbb{K}}$
in Example \ref{exa:workout} is 
\begin{equation}
\det(\mathbf{A})=\det\left[\begin{array}{cc}
a_{1}^{(1)} & a_{2}^{(1)}\\
a_{1}^{(2)} & a_{2}^{(2)}
\end{array}\right]\neq0.\label{eq:indp_condition}
\end{equation}
Obviously, this condition can be extended to cases $L>2$. 
\end{rem}

\subsection{Goodness of Algebraic Lattices}
\begin{defn}[Moments]
The second moment of a lattice $\Lambda^{\mathbb{Z}}\subseteq\mathbb{R}^{nT}$
is $\tilde{\sigma}^{2}\left(\Lambda^{\mathbb{Z}}\right)\triangleq\frac{\int_{\mathcal{V}\left(\Lambda^{\mathbb{Z}}\right)}\left\Vert \mathbf{x}\right\Vert ^{2}\mathrm{d}\mathbf{x}}{nT|\mathcal{V}\left(\Lambda^{\mathbb{Z}}\right)|}$,
and the normalized second moment of $\Lambda^{\mathbb{Z}}$ is $G\left(\Lambda^{\mathbb{Z}}\right)\triangleq\frac{\tilde{\sigma}^{2}\left(\Lambda^{\mathbb{Z}}\right)}{|\mathcal{V}\left(\Lambda^{\mathbb{Z}}\right)|^{2/(nT)}}$. 
\end{defn}
\begin{defn}[Quantization goodness]
\label{def:msegood} A sequence of lattices $\Lambda^{\mathbb{Z}}\subseteq\mathbb{R}^{nT}$
is called good for MSE quantization if 
\[
\lim_{T\rightarrow\infty}G\left(\Lambda^{\mathbb{Z}}\right)=\frac{1}{2\pi e}.
\]
\end{defn}
The existence of such lattices has been shown in \cite{Zamir1996}.
For lattices built from Construction A over quadratic fields, the
quantization goodness has been proved in \cite{Huang2015b} following
\cite{Ordentlich2012}. In the following theorem, we extend the quantization
goodness to lattices constructed from general number fields, whose
proof is given in Appendix \ref{sec:Proofquan}. 
\begin{thm}[]
\label{thm:quangood}There exist a sequence of lattices in the ensemble
(\ref{eq:alge_semble}) which are good for MSE quantization. 
\end{thm}
\begin{defn}[Universal coding goodness]
\label{def:polty} For a block-fading channel in the form of $\mathbf{y}=\mathbf{H}\mathbf{x}+\mathbf{z}$,
with channel $\mathbf{H}\in\mathrm{dg}\left(\mathbb{R}^{n}\right)\otimes\mathbf{I}_{T}$,
codeword $\mathbf{x}\in\Lambda^{\mathbb{Z}}$, and noise $\mathbf{z}\in\mathbb{R}^{nT}$
admitting $\mathcal{N}(\mathbf{0},\sigma_{\mathbf{z}}^{2}\mathbf{I}_{nT})$,
define the generalized volume-to-noise ratio (VNR) as 
\[
\mu\left(\mathbf{H}\Lambda^{\mathbb{Z}}\right)\triangleq\frac{\left(\det\left(\mathbf{H}\right)|\mathcal{V}\left(\Lambda^{\mathbb{Z}}\right)|\right)^{\frac{2}{nT}}}{\sigma_{\mathbf{z}}^{2}}.
\]
A sequence of lattices $\Lambda^{\mathbb{Z}}\subseteq\mathbb{R}^{nT}$
is called universally good for coding if for any $\mu\left(\mathbf{H}\Lambda^{\mathbb{Z}}\right)>2\pi e$,
the error probability of estimating $\mathbf{x}$ given $\mathbf{H}$
satisfies $P_{e}(\Lambda^{\mathbb{Z}},\mathbf{H})\rightarrow0$ for
all $\mathbf{H}$ . 
\end{defn}
%The existence of a sequence of lattices which are good for a specific
%$\mathbf{H}$ has been implicitly shown in \cite{Gamal2004a} following
%\cite{Loeliger1997}. Regarding the universal coding goodness, we
%need the following theorem.

\begin{thm}[\cite{CampelloLingBelfiore2016,Campello2016}]
\label{thm:codinggood}There exist a sequence of lattices in the
ensemble (\ref{eq:alge_semble}) which are universally good for coding
in block-fading channels. 
\end{thm}
Coding over algebraic lattices and coding over $\mathbb{Z}$-lattices
have some differences, which we highlight in the following. 
\begin{enumerate}
\item Relation to coding using a rank-$nT$ $\mathbb{Z}$-lattice. The algebraic
lattice $\Lambda^{\mathbb{Z}}$ is a special case of rank-$nT$ $\mathbb{Z}$-lattices.
Its extraordinary feature is that $\mathrm{dg}(\sigma\left(a_{l}\right))\times\Lambda^{\mathbb{Z}}\subset\Lambda^{\mathbb{Z}}$.
It also has a constant lower bound on $d_{\mathrm{min}}(\Lambda^{\mathbb{Z}})\triangleq\min_{\mathbf{x}\in\Lambda^{\mathbb{Z}}\backslash\mathbf{0}}\prod_{j=1}^{n}\left(\sum_{t=(j-1)T+1}^{jT}x_{t}^{2}\right)$,
so the lattice enjoys full diversity in block fading channels \cite{Kositwattanarerk2015}.
On the contrary, for an arbitrary lattice constructed from a random
Construction A over $\mathbb{Z}$, e.g., $\Lambda'$, it may have
$d_{\mathrm{min}}(\Lambda')=0$.
\item Relation to coding using $n$ rank-$T$ $\mathbb{Z}$-lattices. If
we just transmit $n$ short lattice codewords of length $T$, {then}
we will lose diversity and coding gain. %For instance, if the transmitters
%aim to control outage probability $\leq1\%$,  each of these $n$ blocks must require the feedback
%of their $1\%$ outage rates,  while   we only need to return
%the total $1\%$ outage rate information to each user.
\end{enumerate}

\section{\label{sec:Rate-analysis}Achievable Computation Rate}

The main results in this section are Theorems \ref{thm:rate} and
\ref{prop:dof_upper}, whose proofs will be given in the subsections. 
We reemphasize here that our results only require channel knowledge at the receivers, not at the transmitters. 

We begin by defining $\mathbf{a}\triangleq\left[a_{1},\ldots\thinspace,a_{L}\right]^{\top}\in\mathcal{O}_{\mathbb{K}}^{L}$
, $\mathbf{h}_{j}\triangleq[h_{j,1},\ldots\thinspace,h_{j,L}]^{\top}\in\mathbb{R}^{L}$,
and $\left\{ \mathbf{H}_{l}\right\} $ as the shorthand notation of
$\left\{ \mathbf{H}_{l}\right\} _{l=1}^{L}$. The definitions of the achievable computation rates in one relay and the whole block-fading network are the same as those in Definitions \ref{static-achieve-rate} and \ref{static-achieve-Network}, except that the channel coefficient here is $\left\{ \mathbf{H}_{l}\right\} $, and the coefficient vector $\mathbf{a}$ is algebraic. 
\begin{thm}
\label{thm:rate}  With our coding scheme
in block-fading channels, the following computation rate for a chosen $\mathbf{a}$ at a relay is achievable
as $T\rightarrow\infty$: 
\begin{align}
 & R_{\mathrm{comp}}\left(\left\{ \mathbf{H}_{l}\right\} ,\mathbf{a}\right)=\nonumber \\
 & \frac{n}{2}\max_{\mathbf{b}}\log^{+}\left(\frac{nP}{\sum_{j=1}^{n}\left(|b_{j}|^{2}+P\left\Vert b_{j}\mathbf{h}_{j}-\sigma_{j}(\mathbf{a})\right\Vert ^{2}\right)}\right);\label{eq:am}
\end{align}
and by optimizing $\mathbf{b}$ in (\ref{eq:am}), we have: 
\begin{equation}
R_{\mathrm{comp}}\left(\left\{ \mathbf{H}_{l}\right\} ,\mathbf{a}\right)=\frac{n}{2}\log^{+}\left(\frac{n}{\sum_{j=1}^{n}\sigma_{j}(\mathbf{a})^{\top}\mathbf{M}_{j}\sigma_{j}(\mathbf{a})}\right),\label{eq:am-1}
\end{equation}
where $\mathbf{M}_{j}=\mathbf{I}-\frac{P}{P\left\Vert \mathbf{h}_{j}\right\Vert ^{2}+1}\mathbf{h}_{j}\mathbf{h}_{j}^{\top}$. 
\end{thm}

\begin{rem}
	If we confine $\mathbf{a}\in\mathbb{Z}^{L}$ in the above theorem,
	then obviously $R_{\mathrm{comp}}\left(\left\{ \mathbf{H}_{l}\right\} ,\mathbf{a}\mathrel{\Big|}\mathbf{a}\in\mathbb{Z}^{L}\right)\leq R_{\mathrm{comp}}\left(\left\{ \mathbf{H}_{l}\right\} ,\mathbf{a}\mathrel{\Big|}\mathbf{a}\in\mathcal{O}_{\mathbb{K}}^{L}\right)$,
	namely, the rate achieved by $\mathbb{Z}$-lattice codes of length
	$nT$ can only be lower.
\end{rem}

The above theorem leads to the computation rate of the block-fading network, which is simply the minimum computation rate among $L$ relays while making the set of combination coefficients invertible. In the following, we focus on understanding the computation rate at one relay, as well as its extension to the multiple access scenario.

Evaluating the $\mathcal{O}_{\mathbb{K}}$ coefficient vector $\mathbf{a}$
is {crucial in understanding} the performance limit of the computation
rate. Our goal is to find one coefficient vector or multiple coefficient
vectors minimizing the so-called additive Humbert form \cite{Leibak2005}
\begin{equation}
F(\mathbf{a})=\sum_{j=1}^{n}\sigma_{j}(\mathbf{a})^{\top}\mathbf{M}_{j}\sigma_{j}(\mathbf{a}).\label{eq:deno_rate}
\end{equation}
With Cholesky decomposition of the $L\times L$ matrix $\mathbf{M}_{j}=\bar{\mathbf{M}}_{j}^{\top}\bar{\mathbf{M}}_{j}$,
we may write $F(\mathbf{a})=\sum_{j=1}^{n}\left\Vert \bar{\mathbf{M}}_{j}\sigma_{j}(\mathbf{a})\right\Vert ^{2}$.
This induces a squared distance over an $\mathcal{O}_{\mathbb{K}}$-module
$\Lambda^{\mathcal{O}_{\mathbb{K}}}\left(\left\{ \bar{\mathbf{M}}_{j}\right\} \right)$,
whose generator matrix is given by the tuple $\left\{ \bar{\mathbf{M}}_{j}\right\} $,
and multiplication in the module is defined over the embedded space.

Let $\mathbf{a}_{1},\ldots\thinspace,\mathbf{a}{}_{L}$ be the coefficient
vectors of the $L$ $\mathcal{O}_{\mathbb{K}}$-successive minima
of $\Lambda^{\mathcal{O}_{\mathbb{K}}}\left(\left\{ \bar{\mathbf{M}}_{j}\right\} \right)$.
Define the equation rate w.r.t. the $i$th coefficient vector $\mathbf{a}_{i}$
as 
\begin{equation}
R_{\mathrm{achv,}i}\left(\left\{ \mathbf{H}_{l}\right\} \right)=\frac{n}{2}\log^{+}\left(\frac{n}{F(\mathbf{a}_{i})}\right).\label{eq:achirate}
\end{equation}
We refer to $R_{\mathrm{achv,}1}\left(\left\{ \mathbf{H}_{l}\right\} \right)$
as the optimized (in the sense of optimizing the coefficient vectors)
computation rate, and $\sum_{i=1}^{L}R_{\mathrm{achv,}i}\left(\left\{ \mathbf{H}_{l}\right\} \right)$
as the optimized computation sum-rate. 

\begin{thm}
\label{prop:dof_upper} The optimized computation rate satisfies 
\begin{align}
 & R_{\mathrm{achv,}1}\left(\left\{ \mathbf{H}_{l}\right\} \right)\geq\nonumber \\
 & \frac{1}{2L}\sum_{j=1}^{n}\log^{+}\left(1+P\left\Vert \mathbf{h}_{j}\right\Vert ^{2}\right)-\frac{n}{2}\log^{+}\left(\frac{\kappa_{nL}}{n}\left(\mathrm{disc}_{\mathbb{K}}\right)^{1/n}\right);\label{eq:ach1}
\end{align}
and the optimized computation sum-rate  satisfies: 
\begin{align}
 & \sum_{i=1}^{L}R_{\mathrm{achv,}i}\left(\left\{ \mathbf{H}_{l}\right\} \right)\geq\nonumber \\
 & \frac{1}{2}\sum_{j=1}^{n}\log^{+}\left(1+P\left\Vert \mathbf{h}_{j}\right\Vert ^{2}\right)-\frac{nL}{2}\log^{+}\left(\frac{\kappa_{nL}}{n}\left(\mathrm{disc}_{\mathbb{K}}\right)^{1/n}\right).\label{eq:achi}
\end{align}
\end{thm}

\begin{rem}
	While Eq. (\ref{eq:achi}) serves as a characterization of the performance of the $L$ best linearly independent combinations, 
	our coding technique  should be further generalized (for this equation) to allow for $L$ fine lattices (one per user) as well as a form of successive interference cancellation at the receiver in order to create effective channels that only involve the subset of lattices that can tolerate the increased varying noise faced when decoding each linear combination. For quasi-static channels, such a scheme is developed by Ordentlich et al. in 
	\cite{Ordentlich2014}. Our generalization follows in the same manner.
\end{rem}

\begin{rem}
	Theorem \ref{prop:dof_upper}  resembles its quasi-static counterpart in \cite[Theorem 3]{Ordentlich2014},
	\cite[Theorem 6]{Nazer2016}. The sum-rate  
	is understood in the context of block-fading MAC, whose sum capacity
	is 
	\[\frac{1}{2}\sum_{j=1}^{n}\log^{+}\left(1+P\left\Vert \mathbf{h}_{j}\right\Vert ^{2}\right).\]
    The theorem shows that, for any SNR, the computation rate 
	and sum-rate
	are never much smaller than the symmetric capacity and sum-capacity
	of block-fading MAC. Since the gaps are determined by $n$, $L$ and $\mathrm{disc}_{\mathbb{K}}$, one should choose a number field with the smallest possible discriminant.
\end{rem}

\subsection{Proof of Theorem \ref{thm:rate}}

With dithering, the transmitted codeword is given by $\tilde{\mathbf{X}}_{l}=\left[\mathbf{X}_{l}+\gamma\mathbf{D}_{l}\right]\mod\thinspace\gamma\Lambda_{c}^{\mathbb{Z}}$.
The signal $\mathrm{vec}\left(\tilde{\mathbf{X}}_{l}\right)$ is then
uniformly distributed over $\gamma\mathcal{V}\left(\Lambda_{c}^{\mathbb{Z}}\right)$
and is statistically independent of $\mathrm{vec}\left(\mathbf{X}_{l}\right)$
according to the Crypto lemma \cite[Lemma 1]{Erez2004}. After MMSE
scaling as well as removing the dithers, we have 
\begin{align}
 & \mathbf{B}\mathbf{Y}-\gamma\sum_{l=1}^{L}\mathbf{A}_{l}\mathbf{D}_{l}\nonumber \\
 & =\sum_{l=1}^{L}\mathbf{B}\mathbf{H}_{l}\tilde{\mathbf{X}}_{l}+\mathbf{B}\mathbf{Z}-\gamma\sum_{l=1}^{L}\mathbf{A}_{l}\mathbf{D}_{l}\nonumber \\
 & =\sum_{l=1}^{L}\mathbf{A}_{l}\mathbf{X}_{l}+\sum_{l=1}^{L}\mathbf{B}\mathbf{H}_{l}\tilde{\mathbf{X}}_{l}+\mathbf{B}\mathbf{Z}-\sum_{l=1}^{L}\mathbf{A}_{l}\left(\mathbf{X}_{l}+\gamma\mathbf{D}_{l}\right).\label{eq:rate1f}
\end{align}
To proceed, we need the following lemma.
\begin{lem}
\label{lem:CloseLemma} If $\mathbf{A}=\mathrm{dg}\left(\sigma_{1}(a),\thinspace...,\thinspace\sigma_{n}(a)\right)$
with $a\in\mathcal{O}_{\mathbb{K}}$ and $\mathbf{S}\in\mathbb{R}^{n\times T}$,
then 
\begin{equation}
\left[\mathbf{A}\mathbf{S}\right]\mod\thinspace\gamma\Lambda_{c}^{\mathbb{Z}}=\left[\mathbf{A}\left[\mathbf{S}\right]\mod\thinspace\gamma\Lambda_{c}^{\mathbb{Z}}\right]\mod\thinspace\gamma\Lambda_{c}^{\mathbb{Z}}.\label{eq:simpleLemma}
\end{equation}
\end{lem}
\begin{IEEEproof}
Write $\mathbf{S}=\mathbf{X}+\mathbf{S}'$, where $\mathbf{X}$ is
the closest lattice vector of $\mathbf{S}$ in $\gamma\Lambda_{c}^{\mathbb{Z}}$.
Then clearly both sides of Eq. (\ref{eq:simpleLemma}) equal $\left[\mathbf{A}\mathbf{S}'\right]\mod\thinspace\gamma\Lambda_{c}^{\mathbb{Z}}$,
because $\Lambda_{c}^{\mathbb{Z}}$ is also multiplicatively closed,
similarly to Lemma \ref{prop:ScalarAbsorb}. 
\end{IEEEproof}
Thus, the last term of Eq. (\ref{eq:rate1f}) satisfies 
\[
\sum_{l=1}^{L}\mathbf{A}_{l}\left(\mathbf{X}_{l}+\gamma\mathbf{D}_{l}\right)\mod\thinspace\gamma\Lambda_{c}^{\mathbb{Z}}=\sum_{l=1}^{L}\mathbf{A}_{l}\tilde{\mathbf{X}}_{l}\mod\thinspace\gamma\Lambda_{c}^{\mathbb{Z}},
\]
so we obtain 
\begin{align}
 & \mathbf{Y}_{\mathrm{eff}}\triangleq\mathbf{B}\mathbf{Y}-\gamma\sum_{l=1}^{L}\mathbf{A}_{l}\mathbf{D}_{l}\mod\thinspace\gamma\Lambda_{c}^{\mathbb{Z}},\nonumber \\
 & =\underset{\mathrm{lattice\thinspace codeword}}{\underbrace{\sum_{l=1}^{L}\mathbf{A}_{l}\mathbf{X}_{l}}}+\nonumber \\
 & \mathbf{E}_{\mathbf{a}}\cdot\underset{\mathrm{effective\thinspace noise\ \ \mathbf{Z}_{\mathrm{eff}}}}{\underbrace{\mathbf{E}_{\mathbf{a}}^{-1}\left(\sum_{l=1}^{L}\left(\mathbf{B}\mathbf{H}_{l}-\mathbf{A}_{l}\right)\tilde{\mathbf{X}}_{l}+\mathbf{B}\mathbf{Z}\right)}}\mod\thinspace\gamma\Lambda_{c}^{\mathbb{Z}},\label{eq:channel model}
\end{align}
in which $\mathbf{E}_{\mathbf{a}}=\mathrm{dg}\left(\left[E_{1},E_{2},\ldots\thinspace,E_{n}\right]\right)$
with
\begin{align*}
E_{n} & =\frac{\sqrt{|b_{n}|^{2}+P\left\Vert b_{n}\mathbf{h}_{n}-\sigma_{n}(\mathbf{a})\right\Vert ^{2}}}{\prod_{j=1}^{n}\left(\sqrt{|b_{j}|^{2}+P\left\Vert b_{j}\mathbf{h}_{j}-\sigma_{j}(\mathbf{a})\right\Vert ^{2}}\right)^{\frac{1}{n}}},
\end{align*}
and $\mathbf{Z}_{\mathrm{eff}}=\mathbf{E}_{\mathbf{a}}^{-1}\left(\sum_{l=1}^{L}\left(\mathbf{B}\mathbf{H}_{l}-\mathbf{A}_{l}\right)\tilde{\mathbf{X}}_{l}+\mathbf{B}\mathbf{Z}\right)$
represents an effective noise.
We then use the semi norm-ergodicity in \cite{Ordentlich2012} to
characterize $\mathbf{Z}_{\mathrm{eff}}$. 
\begin{defn}[Semi norm-ergodicity \cite{Ordentlich2012}]
A random vector $\mathbf{x}$ of length $T$ is called semi norm-ergodic
with effective variance $\frac{1}{T}\mathbb{E}\left\Vert \mathbf{x}\right\Vert ^{2}$
if for any $\epsilon,\delta>0$, and $T$ large enough, 
\[
\mathrm{Pr}\left(\mathbf{x}\notin\mathcal{B}\left(\mathbf{0},\sqrt{\left(1+\delta\right)\mathbb{E}\left(\left\Vert \mathbf{x}\right\Vert ^{2}\right)}\right)\right)\leq\epsilon.
\]
\end{defn}
In Appendix \ref{sec:Proofgau}, we show that:
\begin{lem}
The random vector \label{lem:seminorm}$\mathrm{vec}\left(\mathbf{Z}_{\mathrm{eff}}\right)$
is semi norm-ergodic with effective variance 
\begin{equation}
\sigma_{\mathrm{eff}}^{2}\triangleq\prod_{j=1}^{n}\left(|b_{j}|^{2}+P\left\Vert b_{j}\mathbf{h}_{j}-\sigma_{j}(\mathbf{a})\right\Vert ^{2}\right)^{\frac{1}{n}}.\label{eq:GM}
\end{equation}
\end{lem}

The matrix $\mathbf{E}_{\mathbf{a}}^{-1}$ can be viewed as the channel matrix in Definition \ref{def:polty}.
By inspection of the proof of Theorem \ref{thm:codinggood} in \cite{Campello2016},
it is not difficult to see that Theorem \ref{thm:codinggood} also
holds for semi norm-ergodic noise, similarly to \cite{Ordentlich2012}.
We omit the details. Therefore, there exist a sequence of lattices
in the ensemble (\ref{eq:alge_semble}) such that the decoding error
probability vanishes as $T\rightarrow\infty$ as long as the VNR 
\begin{equation}
\frac{\left(\det\left(\mathbf{E}_{\mathbf{a}}^{-1}\right)\mathrm{Vol}\left(\gamma\Lambda_{f}^{\mathbb{Z}}\right)\right)^{\frac{2}{nT}}}{\sigma_{\mathrm{eff}}^{2}}>2\pi e.\label{eq:vnr}
\end{equation}
On the other hand, the quantization goodness in Theorem \ref{thm:quangood}
implies 
\begin{equation}
\frac{P}{\mathrm{Vol}\left(\gamma\Lambda_{c}^{\mathbb{Z}}\right)^{\frac{2}{nT}}}<\frac{1+\delta}{2\pi e}\label{eq:m2v}
\end{equation}
for any $\delta>0$ if $T$ is large enough. It follows from (\ref{eq:vnr})
and (\ref{eq:m2v}) that any computation rate up to 
\begin{equation}
\frac{1}{T}\log\left(\frac{\mathrm{Vol}(\gamma\Lambda_{c}^{\mathbb{Z}})}{\mathrm{Vol}(\gamma\Lambda_{f}^{\mathbb{Z}})}\right)<\frac{n}{2}\log\left(\frac{P}{\sigma_{\mathrm{eff}}^{2}}\right)\label{eq:rate_std1}
\end{equation}
is achievable.

The effective noise variance $\sigma_{\mathrm{eff}}^{2}$ represents the geometric
mean (GM) of the noise variances in all the blocks. The final rate expression based on this form is given by \footnote{Here, $\prod_{j=1}^{n}\sigma_{j}(\mathbf{a})^{\top}\mathbf{M}_{j}\sigma_{j}(\mathbf{a})$
	is called a multiplicative Humbert form \cite{Baeza1997}.} :
\begin{align}
& R_{\mathrm{comp}}\left(\left\{ \mathbf{H}_{l}\right\} ,\mathbf{a}\right)=\frac{n}{2}\log^{+}\left(\frac{1}{\prod_{j=1}^{n}\left(\sigma_{j}(\mathbf{a})^{\top}\mathbf{M}_{j}\sigma_{j}(\mathbf{a})\right)^{1/n}}\right)\nonumber \\
& =\frac{1}{2}\log^{+}\left(\frac{1}{\prod_{j=1}^{n}\sigma_{j}(\mathbf{a})^{\top}\mathbf{M}_{j}\sigma_{j}(\mathbf{a})}\right).\label{eq:gm-1}
\end{align}
Since the 
the optimization of the algebraic integer vector in a
multiplicative
form is complicated,
we upper bound $\sigma_{\mathrm{eff}}^{2}$ by the arithmetic mean (AM)
\[
\sigma_{\mathrm{AM}}^{2}\triangleq\frac{1}{n}\sum_{j=1}^{n}\left(|b_{j}|^{2}+P\left\Vert b_{j}\mathbf{h}_{j}-\sigma_{j}(\mathbf{a})\right\Vert ^{2}\right)
\]
to reach (\ref{eq:am}), following (\ref{eq:rate_std1}). This enables the applications of a nice algorithmic framework based on successive minima in the next subsection. Lastly, 
the details of
deriving (\ref{eq:am-1}) are given in Appendix \ref{sec:Proofratep}.

\subsection{\label{sub:findinga}Searching the Optimal Coefficients}

In this subsection, we show that $F(\mathbf{a})$ can be written as
the squared distance of a $\mathbb{Z}$-lattice vector, and explain the relation between ${\mathbb{Z}}$-successive minima and $\mathcal{O}_{\mathbb{K}}$-successive minima.
These results enable the application of conventional 
lattice algorithms over $\mathbb{Z}$ to find one or multiple coefficient vectors at a relay.
We refer readers to \cite{Micciancio2002,Sah-TIT,Lyu2017} for these algorithms.

First, each $\Lambda^{\mathcal{O}_{\mathbb{K}}}\left(\left\{ \bar{\mathbf{M}}_{j}\right\} \right)$
has a corresponding $\mathbb{Z}$-lattice $\Lambda^{\mathbb{Z}}\left(\Phi_{\bar{\mathbf{M}}}\right)$
that belongs to a submodule of $\mathbb{R}^{nL}$, whose generator
matrix is 
\[
\Phi_{\bar{\mathbf{M}}}=\bar{\mathbf{M}}(\Phi\otimes\mathbf{I}_{L}),
\]
where 
\[
\bar{\mathbf{M}}=\left[\begin{array}{ccc}
\bar{\mathbf{M}}_{1} & \cdots & \mathbf{0}\\
\mathbf{0} & \cdots & \mathbf{0}\\
\vdots & \vdots & \vdots\\
\mathbf{0} & \cdots & \bar{\mathbf{M}}_{n}
\end{array}\right],
\]
and recall that $\Phi=[\sigma(\phi_{1}),\ldots,\sigma(\phi_{n})]$
and $\{\phi_{1},\ldots,\phi_{n}\}$ is an integral basis of $\mathcal{O}_{\mathbb{K}}$.
To show this more explicitly, note that there exists a bijective mapping
$\Psi:\thinspace\mathbb{Z}^{nL}\rightarrow\mathcal{O}_{\mathbb{K}}^{L}$
defined by 
\begin{align}
 & \mathbf{a}=\Psi\left(\tilde{\mathbf{a}}\right)\nonumber \\
 & =\left[\sum_{k=1}^{n}\phi_{k}\tilde{a}_{(k-1)L+1},\sum_{k=1}^{n}\phi_{k}\tilde{a}_{(k-1)L+2},\ldots,\sum_{k=1}^{n}\phi_{k}\tilde{a}_{kL}\right]^{\top};\label{eq:Psi}
\end{align}
since $\sigma_{j}$ is a ring homomorphism, it follows that 
\[
\sigma_{j}\left(\mathbf{a}\right)=\left[\sum_{k=1}^{n}\sigma_{j}\left(\phi_{k}\right)\tilde{a}_{(k-1)L+1},\ldots,\sum_{k=1}^{n}\sigma_{j}\left(\phi_{k}\right)\tilde{a}_{kL}\right]^{\top}.
\]
Thus, $F(\mathbf{a})=\left\Vert \Phi_{\bar{\mathbf{M}}}\tilde{\mathbf{a}}\right\Vert ^{2}$
($\tilde{\mathbf{a}}\in\mathbb{Z}^{nL}$) represents the squared distance
of a point in $\Lambda^{\mathbb{Z}}\left(\Phi_{\bar{\mathbf{M}}}\right)$.

Second, if multiple message equations are required at one relay, a search algorithm over
$\mathbb{Z}$-lattice $\Lambda^{\mathbb{Z}}\left(\Phi_{\bar{\mathbf{M}}}\right)$
has to ensure their coefficient vectors $\mathbf{a}_{1},\ldots\thinspace,\mathbf{a}{}_{L}$
are linearly independent over $\mathcal{O}_{\mathbb{K}}$. For the
highest rates, it suffices to search for the $\mathcal{O}_{\mathbb{K}}$-successive
minima. This constraint can be incorporated into an enumeration algorithm,
which keeps increasing the search radius until linear independence
is satisfied. {The question that arises here is whether we
can use the first few successive minima of a $\mathbb{Z}$-module
to find those of an $\mathcal{O}_{\mathbb{K}}$-module}.

Let $\tilde{\mathbf{a}}_{i}$ be the vector giving the $i$-th successive
minima $\lambda_{i}\left(\Phi_{\bar{\mathbf{M}}}\right)$ of $\mathbb{Z}$-lattice
$\Lambda^{\mathbb{Z}}\left(\Phi_{\bar{\mathbf{M}}}\right)$. It may
happen that 
\[
\mathrm{dim}\left(\mathrm{span}_{\mathcal{O}_{\mathbb{K}}}\left(\Psi\left(\left[\tilde{\mathbf{a}}{}_{1},\ldots\thinspace,\tilde{\mathbf{a}}{}_{L}\right]\right)\right)\right)<L.
\]
For example, choose $\mathbb{K}=\mathbb{Q}\left(\sqrt{3}\right)$.
Let \textbf{$\tilde{\mathbf{a}}{}_{1}=\left[1,2,1,1\right]^{\top}$},\textbf{
$\tilde{\mathbf{a}}{}_{2}=\left[6,9,4,5\right]^{\top}$}; after mapping
them back to $\mathcal{O}_{\mathbb{K}}^{2}$, we have $\mathbf{a}_{1}=\left[1+\sqrt{3},2+\sqrt{3}\right]^{\top}$,
$\mathbf{a}_{2}=\left[6+4\sqrt{3},9+5\sqrt{3}\right]^{\top}$. Since
$\left(3+\sqrt{3}\right)\mathbf{a}_{1}=\mathbf{a}_{2}$, one concludes
that $\mathbf{a}_{1}$ and $\mathbf{a}_{2}$ are not independent over
$\mathcal{O}_{\mathbb{K}}$.

Nevertheless, we have the following result: 
\begin{prop}
\label{prop:algeIndependence}Let the mapping $\Psi$ be defined as
in \eqref{eq:Psi}. Suppose $\mathbb{Z}$-coefficient vectors $\tilde{\mathbf{a}}_{1},\ldots\thinspace,\tilde{\mathbf{a}}_{nL}$
produce the $nL$ successive minima $\lambda_{1}\left(\Phi_{\bar{\mathbf{M}}}\right),\ldots\thinspace,\lambda_{nL}\left(\Phi_{\bar{\mathbf{M}}}\right)$
of $\mathbb{Z}$-lattice $\Lambda^{\mathbb{Z}}\left(\Phi_{\bar{\mathbf{M}}}\right)$.
Then $\{\Psi(\tilde{\mathbf{a}}_{1}),\ldots\thinspace,\Psi(\tilde{\mathbf{a}}_{nL})\}$
contains the $L$ $\mathcal{O}_{\mathbb{K}}$-successive minima.
\end{prop}
\begin{IEEEproof}
Write the $\mathbb{Z}$-coefficient matrix $\mathbf{T}=\left[\tilde{\mathbf{a}}{}_{1},\ldots\thinspace,\tilde{\mathbf{a}}{}_{nL}\right]$.
From the definition of successive minima, $\mathbf{T}\in\mathbb{Z}^{nL\times nL}$
is a full-rank matrix such that $\Phi_{\bar{\mathbf{M}}}\mathbf{T}=\bar{\mathbf{M}}(\Phi\otimes\mathbf{I}_{L})\mathbf{T}$
yields $\lambda_{1}\left(\Phi_{\bar{\mathbf{M}}}\right),\ldots\thinspace,\lambda_{nL}\left(\Phi_{\bar{\mathbf{M}}}\right)$
of $\mathbb{Z}$-lattice $\Lambda^{\mathbb{Z}}\left(\Phi_{\bar{\mathbf{M}}}\right)$.
Notice that the $L\times nL$ algebraic-integer matrix $\left[\mathbf{a}_{1},\ldots\thinspace,\mathbf{a}{}_{nL}\right]=\Psi\left(\mathbf{T}\right)$
simply consists of the the first $L$ rows of $(\Phi\otimes\mathbf{I}_{L})\mathbf{T}$;
in fact we have 
\begin{equation}
\left[\mathbf{a}_{1},\ldots\thinspace,\mathbf{a}{}_{nL}\right]=\left[\phi_{1}\mathbf{I}_{L},\ldots,\phi_{n}\mathbf{I}_{L}\right]\mathbf{T}.\label{eq:algebraic}
\end{equation}
Since $\{\phi_{1},\ldots,\phi_{n}\}$ is an integral basis of $\mathcal{O}_{\mathbb{K}}$,
the matrix $\left[\phi_{1}\mathbf{I}_{L},\ldots,\phi_{n}\mathbf{I}_{L}\right]$
obviously has rank $L$. Then it follows from the rank identity 
\[
\mathrm{rank}(\mathbf{C}_{1}\mathbf{C}_{2})=\mathrm{rank}(\mathbf{C}_{1})
\]
for full-rank matrix $\mathbf{C}_{2}$ that the matrix $\left[\mathbf{a}_{1},\ldots\thinspace,\mathbf{a}{}_{nL}\right]$
is of rank $L$. Therefore, there exist exactly $L$ vectors in $\{\mathbf{a}_{1},\ldots\thinspace,\mathbf{a}{}_{nL}\}$
which are linearly independent over $\mathcal{O}_{\mathbb{K}}$. Thus,
the $L$ $\mathcal{O}_{\mathbb{K}}$-successive minima must be contained
in the set $\{\Psi(\tilde{\mathbf{a}}_{1}),\ldots\thinspace,\Psi(\tilde{\mathbf{a}}_{nL})\}$. 
\end{IEEEproof}
\begin{figure}[t]
\center

\includegraphics[clip,width=0.45\textwidth]{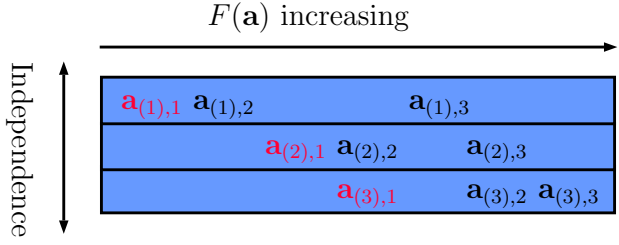}

\caption{Illustration of $\mathcal{O}_{\mathbb{K}}$-successive minima for
$L=3$ and $n=3$. Among the $9$ successive minima of the embedded
real lattice, those marked in red are coefficient vectors of the first
three successive minima over $\mathcal{O}_{\mathbb{K}}$.}
\label{fig_indpendence} 
\end{figure}
The proposition shows searching for $L$ $\mathcal{O}_{\mathbb{K}}$-independent
lattice points inside ball $\mathcal{B}\left(\mathbf{0},\lambda_{nL}\left(\Phi_{\bar{\mathbf{M}}}\right)\right)$
is possible. We further explain Proposition \ref{prop:algeIndependence}
in Fig. \ref{fig_indpendence}. Suppose $L=3$ and $n=3$. There are
$9$ successive minima in the embedded real lattice $\Lambda(\Phi_{\bar{\mathbf{M}}})$,
and their corresponding algebraic coefficient vectors are denoted
by $\mathbf{a}_{(1),1},\ldots\thinspace,\mathbf{a}_{(3),3}$, where
the vectors in the same row are linearly dependent over $\mathcal{O}_{\mathbb{K}}$.
The $\mathbf{a}_{(1),1},\thinspace\mathbf{a}_{(2),1},\thinspace\mathbf{a}_{(3),1}$
marked in red are coefficient vectors of the first three successive
minima over $\mathcal{O}_{\mathbb{K}}$.

%If we only intend to approximately find these $L$ points by using lattice reduction over $\mathcal{O}_{\mathbb{K}}$, we can show that performing lattice reduction on $\Lambda^{\mathcal{O}_{\mathbb{K}}}\left(\left\{ \bar{\mathbf{M}}_{j}\right\} \right)$  is no more than finding   $\mathbf{A}^{*} \in\mathcal{O}_{\mathbb{K}}^{L\times L}$ with $\det{\left(\mathbf{A}^{*}\right)} \in \mathcal{U}$, so that the new lattice basis consists of the tuple $\left\{ \bar{\mathbf{M}}_{1}\sigma_{1}\left(\mathbf{A}^{*}\right),\ldots, \bar{\mathbf{M}}_{n}\sigma_{n}\left(\mathbf{A}^{*}\right)\right\}$.
%In this reduction, the reduced lattice $\overline{\Lambda}^{\mathcal{O}_{\mathbb{K}}}\left(\left\{ \bar{\mathbf{M}}_{j}\sigma_{j}\left(\mathbf{A}^{*}\right)\right\} \right)$ is not a sub-lattice because
%\begin{equation}
%\frac{\mathrm{Vol}\left(\Lambda^{\mathcal{O}_{\mathbb{K}}}\left(\left\{ \bar{\mathbf{M}}_{j}\right\} \right)\right)}{\mathrm{Vol}\left(\overline{\Lambda}^{\mathcal{O}_{\mathbb{K}}}\left(\left\{ \bar{\mathbf{M}}_{j}\sigma_{j}\left(\mathbf{A}^{*}\right)\right\} \right)\right)}=\prod_{j=1}^{n}|\det\left(\sigma_{j}\left(\mathbf{A}^{*}\right)\right)|=|\mathrm{Nr}\left(\det\left(\mathbf{A}^{*}\right)\right)|=1.\label{eq:nested_module_1}
%\end{equation}

\subsection{Proof of Theorem \ref{prop:dof_upper}}

To derive the optimized computation rate and sum-rate, we only need
to apply Minkowski's first and second theorems to $\Lambda^{\mathcal{O}_{\mathbb{K}}}\left(\left\{ \bar{\mathbf{M}}_{j}\right\} \right)$.
First, by applying Sylvester's determinant identity to each $|\det(\bar{\mathbf{M}}_{i})|$,
one has 
\begin{align*}
|\det\left(\bar{\mathbf{M}}\right)| & =\prod_{j=1}^{n}|\det(\bar{\mathbf{M}}_{i})|=\prod_{j=1}^{n}\left(1+P\left\Vert \mathbf{h}_{j}\right\Vert ^{2}\right)^{-1/2}.
\end{align*}
Consequently the volume of $\Lambda^{\mathcal{O}_{\mathbb{K}}}\left(\left\{ \bar{\mathbf{M}}_{j}\right\} \right)$
becomes 
\begin{align*}
 & |\det(\Phi_{\bar{\mathbf{M}}})|=|\det(\bar{\mathbf{M}})||\det(\Phi\otimes\mathbf{I}_{L})|\\
 & =\left(\mathrm{disc}_{\mathbb{K}}\right)^{L/2}\prod_{j=1}^{n}\left(1+P\left\Vert \mathbf{h}_{j}\right\Vert ^{2}\right)^{-1/2}.
\end{align*}
The shortest lattice vector of $\Lambda^{\mathbb{Z}}\left(\Phi_{\bar{\mathbf{M}}}\right)$
is the embedding of the shortest lattice vector from $\Lambda^{\mathcal{O}_{\mathbb{K}}}\left(\left\{ \bar{\mathbf{M}}_{j}\right\} \right)$.
Then it follows from Minkowski's first theorem over $\mathbb{Z}$-lattices
that $\lambda_{1}^{2}\left(\Lambda^{\mathcal{O}_{\mathbb{K}}}\left(\left\{ \bar{\mathbf{M}}_{j}\right\} \right)\right)\leq\kappa_{nL}|\det(\Phi_{\bar{\mathbf{M}}})|^{2/(nL)}$,
which yields 
\begin{align}
\lambda_{1}^{2}\left(\Lambda^{\mathcal{O}_{\mathbb{K}}}\left(\left\{ \bar{\mathbf{M}}_{j}\right\} \right)\right) & \leq\kappa_{nL}\left(\mathrm{disc}_{\mathbb{K}}\right)^{1/n}\prod_{j=1}^{n}\left(1+P\left\Vert \mathbf{h}_{j}\right\Vert ^{2}\right)^{-1/(nL)}.\label{eq:MIN11}
\end{align}
By substituting (\ref{eq:MIN11}) into the rate expression (\ref{eq:am-1}),
we obtain 
\begin{align}
 & R_{\mathrm{achv,1}}\nonumber \\
 & =\frac{n}{2}\log^{+}\left(\frac{n}{\lambda_{1}^{2}\left(\Lambda^{\mathcal{O}_{\mathbb{K}}}\left(\left\{ \bar{\mathbf{M}}_{j}\right\} \right)\right)}\right)\nonumber \\
 & \geq\frac{n}{2}\log^{+}\left(\frac{n\prod_{j=1}^{n}\left(1+P\left\Vert \mathbf{h}_{j}\right\Vert ^{2}\right)^{1/(nL)}}{\kappa_{nL}\left(\mathrm{disc}_{\mathbb{K}}\right)^{1/n}}\right)\nonumber \\
 & =\underset{\frac{1}{L}\times\textrm{MAC capacity}}{\underbrace{\frac{1}{2L}\sum_{j=1}^{n}\log^{+}\left(1+P\left\Vert \mathbf{h}_{j}\right\Vert ^{2}\right)}}-\underset{\textrm{constant}}{\underbrace{\frac{n}{2}\log^{+}\left(\frac{\kappa_{nL}}{n}\left(\mathrm{disc}_{\mathbb{K}}\right)^{1/n}\right)}}.\label{eq:rate lower bound}
\end{align}

Meanwhile, from Minkowski's second theorem \eqref{eq:mst-rogers},
we have 
\begin{equation}
\prod_{j=1}^{L}\lambda_{j}^{2n}\left(\Lambda^{\mathcal{O}_{\mathbb{K}}}\left(\left\{ \bar{\mathbf{M}}_{j}\right\} \right)\right)\leq\kappa_{nL}^{nL}|\det(\Phi_{\bar{\mathbf{M}}})|^{2}.\label{eq:mst_rogers}
\end{equation}
Finally, after substituting (\ref{eq:mst_rogers}) into (\ref{eq:am-1}),
we have: 
\begin{align}
 & \sum_{i=1}^{L}R_{\mathrm{achv,}i}\left(\left\{ \mathbf{H}_{l}\right\} \right)\nonumber \\
 & =\sum_{i=1}^{L}\frac{n}{2}\log^{+}\left(\frac{n}{\lambda_{i}^{2}\left(\Lambda^{\mathcal{O}_{\mathbb{K}}}\left(\left\{ \bar{\mathbf{M}}_{j}\right\} \right)\right)}\right)\nonumber \\
 & \geq\frac{n}{2}\log^{+}\left(\frac{n^{L}\prod_{j=1}^{n}\left(1+P\left\Vert \mathbf{h}_{j}\right\Vert ^{2}\right)^{1/n}}{\kappa_{nL}^{L}\left(\mathrm{disc}_{\mathbb{K}}\right)^{L/n}}\right)\nonumber \\
 & =\underset{\textrm{MAC capacity}}{\underbrace{\frac{1}{2}\sum_{j=1}^{n}\log^{+}\left(1+P\left\Vert \mathbf{h}_{j}\right\Vert ^{2}\right)}}-\underset{\textrm{constant}}{\underbrace{\frac{1}{2}\log^{+}\left(\frac{\kappa_{nL}^{nL}}{n^{nL}}\left(\mathrm{disc}_{\mathbb{K}}\right)^{L}\right)}}.\label{eq:sum_achv}
\end{align}

\section{DoF Analysis}

Define DoF associated with $R_{\mathrm{achv,}i}$ as 
\begin{equation}
d_{\mathrm{achv},i}=\lim_{P\rightarrow\infty}\frac{R_{\mathrm{achv,}i}}{\frac{1}{2}\log\left(1+P\right)}.\label{eq:dof definition}
\end{equation}
The main result of this section is: 
\begin{thm}
\label{thm:maindof}For almost all $\left\{ \mathbf{H}_{l}\right\} $
w.r.t. the Lebesgue measure, the DoF's of the optimized computation
rate and sum-rate are respectively $d_{\mathrm{achv},1}=\frac{n}{L}$
and $\sum_{i=1}^{L}d_{\mathrm{achv},i}=n$. 
\end{thm}
%Although $d_{\mathrm{achv},1}=\frac{n}{L}$ is only optimal for MAC\footnote{??},
%this is unlikely to be improved unless we further assume CSI at transmitters
%so that signal alignment \cite{Niesen2012} is possible.
\begin{IEEEproof}[Proof of Theorem \ref{thm:maindof}]
As a direct consequence of Theorem \ref{prop:dof_upper}, the lower
bounds of DoF's are: 
\[
d_{\mathrm{achv},1}\geq\frac{n}{L},\ \ \sum_{i=1}^{L}d_{\mathrm{achv},i}\geq n.
\]
We will show in Theorem \ref{prop:dof_lower} that $d_{\mathrm{achv},1}\leq\frac{n}{L}$,
which is due to Lemma \ref{lem: DA_conjugate} on Diophantine approximation
of a real vector by algebraic conjugates. The block-fading MAC capacity
can upper bound the sum DoF's, which yields $\sum_{i=1}^{L}d_{\mathrm{achv},i}\leq n$.
Consequently, along with $d_{\mathrm{achv},1}\geq d_{\mathrm{achv},2}\geq d_{\mathrm{achv},L}$,
we have 
\[
d_{\mathrm{achv},1}=\cdots=d_{\mathrm{achv},L}=\frac{n}{L}.
\]
\end{IEEEproof}
\begin{thm}
\label{prop:dof_lower}For almost all $\left\{ \mathbf{H}_{l}\right\} $
w.r.t. the Lebesgue measure, the DoF associated to the first computation
rate satisfies $d_{\mathrm{achv},1}\leq\frac{n}{L}.$ 
\end{thm}
\begin{lem}
\label{lem: DA_conjugate} Let $\psi:\thinspace\mathbb{N}\rightarrow\mathbb{R}^{+}$
be an approximation function. Then for almost all $\left\{ \mathbf{H}_{l}\right\} $
w.r.t. the Lebesgue measure, and for all $q\in\mathcal{O}_{\mathbb{K}}$,
there exists a constant $c_{\left\{ \mathbf{H}_{l}\right\} }'>0$
such that 
\begin{equation}
\max_{l\in\{1,\ldots,L\}}\min_{a\in\mathcal{O}_{\mathbb{K}}}\left\Vert \mathbf{H}_{l}-\mathrm{dg}\left(\sigma(a/q)\right)\right\Vert \geq c_{\left\{ \mathbf{H}_{l}\right\} }'\psi(|\mathrm{Nr}(q)|)\label{eq:lagerthan}
\end{equation}
if $\sum_{k=1}^{\infty}\psi(k)^{nL}k^{L}<\infty.$ 
\end{lem}
Lemma \ref{lem: DA_conjugate} generalizes the classical Khintchine-Groshev
theorem from $\mathbb{Z}$ to $\mathcal{O}_{\mathbb{K}}$. The proof
is given in Appendix \ref{sec:Algebraic-approximation}. Note that
the approximation function in (\ref{eq:lagerthan}) can decay as fast
as $\psi(|\mathrm{Nr}(q)|)=|\mathrm{Nr}(q)|^{-\left(\frac{1+L}{nL}+\delta\right)}$
for any $\delta>0$. Lemma \ref{lem: DA_conjugate} also indicates
that, all points in the set $q\mathcal{U}$ have the same approximation-error
bound $c_{\left\{ \mathbf{H}_{l}\right\} }'\psi(|\mathrm{Nr}(q)|)$.

We proceed to prove Theorem \ref{prop:dof_lower}, where the technique
is to generalize the approach in \cite{Ordentlich2014,Nazer2016}
to vectors of algebraic conjugates.
\begin{IEEEproof}[Proof of Theorem \ref{prop:dof_lower}]
First rewrite the denominator $\sigma_{\mathrm{AM}}^{2}$ in (\ref{eq:am})
explicitly as {a} trade-off between ``range'' and ``accuracy'':
\begin{equation}
\frac{1}{n}\underset{\mathrm{range}}{\underbrace{\left\Vert \mathbf{B}\right\Vert ^{2}}}+\frac{P}{n}\underset{\mathrm{accuracy}}{\underbrace{\left\Vert \mathbf{B}\left[\mathbf{H}_{1},\ldots\thinspace,\mathbf{H}_{L}\right]-\left[\mathrm{dg}\left(\sigma\left(a_{1}\right)\right),\ldots\thinspace,\mathrm{dg}\left(\sigma\left(a_{L}\right)\right)\right]\right\Vert ^{2}}}.\label{eq:da_term}
\end{equation}
Let $\mathcal{V}_{\mathbf{0}}$ stand for the Voronoi region of $\mathbf{0}$
in the embedded lattice $\sigma(\mathcal{O}_{\mathbb{K}})$. In the
shortest vector problem (SVP), one aims to find a shortest nonzero vector, so the coefficients
cannot be $\sigma\left(a_{1}\right)=\cdots=\sigma\left(a_{L}\right)=\mathbf{0}$.
By rearranging the order of $a_{1},\ldots\thinspace,a_{L}$ if necessary,
we can assume that $\sigma\left(a_{1}\right)\neq\mathbf{0}$. Then
the analysis falls into two cases depending on whether $\mathbf{B}\mathbf{H}_{1}\in\mathcal{V}_{\mathbf{0}}$.

i) If $\mathbf{B}\mathbf{H}_{1}\in\mathcal{V}_{\mathbf{0}}$, then
$\left\Vert \mathbf{B}\mathbf{H}_{1}-\mathrm{dg}\left(\sigma\left(a_{1}\right)\right)\right\Vert $
is lower bounded by the packing radius of lattice $\sigma(\mathcal{O}_{\mathbb{K}})$, which is $\frac{\lambda_{1}\left(\mathcal{O}_{\mathbb{K}}\right)}{2}$.
Based on this,
we have 
\begin{align}
 & \sigma_{\mathrm{AM}}^{2}\geq\frac{1}{n}\left\Vert \mathbf{B}\right\Vert ^{2}+\frac{P}{n}\left\Vert \mathbf{B}\mathbf{H}_{1}-\mathrm{dg}\left(\sigma\left(a_{1}\right)\right)\right\Vert ^{2}\nonumber \\
 & >P\frac{\lambda_{1}^{2}\left(\mathcal{O}_{\mathbb{K}}\right)}{4n}>P^{\frac{L-1}{L}}\frac{\lambda_{1}^{2}\left(\mathcal{O}_{\mathbb{K}}\right)}{4n},\label{eq:noise1}
\end{align}
where the first inequality is from 
\begin{align*}
& \left\Vert \mathbf{B}\left[\mathbf{H}_{1},\ldots\thinspace,\mathbf{H}_{L}\right]-\left[\mathrm{dg}\left(\sigma\left(a_{1}\right)\right),\ldots\thinspace,\mathrm{dg}\left(\sigma\left(a_{L}\right)\right)\right]\right\Vert ^{2}\\
& \geq\left\Vert \mathbf{B}\mathbf{H}_{1}-\mathrm{dg}\left(\sigma\left(a_{1}\right)\right)\right\Vert ^{2}.
\end{align*}

ii) If $\mathbf{B}\mathbf{H}_{1}\notin\mathcal{V}_{\mathbf{0}}$,
we have $\mathbf{B}\mathbf{H}_{1}=\mathrm{dg}\left(\sigma(q)+{\bm{\varphi}}\right)$
for $\mathbf{0}\neq\sigma(q)\in\sigma(\mathcal{O}_{\mathbb{K}})$,
${\bm{\varphi}}\in\mathcal{V}_{\mathbf{0}}$. The ``accuracy'' term
for two vectors $\mathbf{H}_{1}$ and $\mathbf{H}_{l}$ satisfies
\begin{align}
 & \left\Vert \mathbf{B}\left[\mathbf{H}_{1},\mathbf{H}_{l}\right]-\left[\mathrm{dg}\left(\sigma\left(a_{1}\right)\right),\mathrm{dg}\left(\sigma\left(a_{l}\right)\right)\right]\right\Vert ^{2}\nonumber \\
 & \geq\left\Vert {\bm{\varphi}}\right\Vert ^{2}+\left\Vert \tilde{\mathbf{H}}_{l}\mathrm{dg}(\sigma(q)+{\bm{\varphi}})-\mathrm{dg}\left(\sigma\left(a_{l}\right)\right)\right\Vert ^{2},\label{eq:trnoise1}
\end{align}
where $\tilde{\mathbf{H}}_{l}=\mathbf{H}_{1}^{-1}\mathbf{H}_{l}$.
The r.h.s. of (\ref{eq:trnoise1}) is a quadratic function of ${\bm{\varphi}}$.
To attain its minimum, we solve the following equation 
\[
\partial\left(\left\Vert {\bm{\varphi}}\right\Vert ^{2}+\left\Vert \tilde{\mathbf{H}}_{l}\mathrm{dg}(\sigma(q)+{\bm{\varphi}})-\mathrm{dg}\left(\sigma\left(a_{l}\right)\right)\right\Vert ^{2}\right)/\partial{\bm{\varphi}}=\mathbf{0}
\]
to get ${\bm{\varphi}}=(\mathbf{I}+\tilde{\mathbf{H}}_{l}^{2})^{-1}\left(\tilde{\mathbf{H}}_{l}\sigma\left(a_{l}\right)-\tilde{\mathbf{H}}_{l}^{2}\sigma(q)\right).$
{Substitute} this back into (\ref{eq:trnoise1}), we have 
\begin{align*}
 & \left\Vert \mathbf{B}\left[\mathbf{H}_{1},\mathbf{H}_{l}\right]-\left[\mathrm{dg}\left(\sigma\left(a_{1}\right)\right),\mathrm{dg}\left(\sigma\left(a_{l}\right)\right)\right]\right\Vert ^{2}\\
 & \geq\left\Vert \left(\tilde{\mathbf{H}}_{l}^{2}+\mathbf{I}_{n}\right)^{-1}\left(\tilde{\mathbf{H}}_{l}\mathrm{dg}\left(\sigma(q)\right)-\mathrm{dg}\left(\sigma\left(a_{l}\right)\right)\right)\right\Vert ^{2}\\
 & \geq
h_l^*
 \left\Vert \tilde{\mathbf{H}}_{l}\mathrm{dg}\left(\sigma(q)\right)-\mathrm{dg}\left(\sigma\left(a_{l}\right)\right)\right\Vert ^{2},
\end{align*}
where $h_l^* \triangleq  \min_{\tilde{h}_{l}\in\tilde{\mathbf{H}}_{l}}\frac{1}{\left(\tilde{h}_{l}^{2}+1\right)^{2}}$.
Thus, for almost all channel realizations it holds that 
\begin{align}
 &   \left\Vert \mathbf{B}\left[\mathbf{H}_{1},\ldots\thinspace,\mathbf{H}_{L}\right]-\left[\mathrm{dg}\left(\sigma\left(a_{1}\right)\right),\ldots\thinspace,\mathrm{dg}\left(\sigma\left(a_{L}\right)\right)\right]\right\Vert ^{2}\nonumber \\
 & \geq  \max_{l\in\left\{ 2,\ldots\thinspace,L\right\} }\left(h_l^* \left\Vert \tilde{\mathbf{H}}_{l}\mathrm{dg}\left(\sigma(q)\right)-\mathrm{dg}\left(\sigma\left(a_{l}\right)\right)\right\Vert ^{2}\right)\nonumber \\
 & \geq  \max_{l\in\left\{ 2,\ldots\thinspace,L\right\} }\left(h_l^* \min_{i}|\sigma_{i}(q)|^{2}\left\Vert \tilde{\mathbf{H}}_{l}-\mathrm{dg}(\sigma(q))^{-1}\mathrm{dg}\left(\sigma\left(a_{l}\right)\right)\right\Vert ^{2}\right)\nonumber \\
 & {\geq}  c_{\left\{ \mathbf{H}_{l}\right\} }''\min_{i}|\sigma_{i}(q)|^{2}|\mathrm{Nr}(q)|^{-\frac{2}{n}-\frac{2}{n(L-1)}}\label{eq:lowerRate}
\end{align}
where the last inequality is due to Lemma \ref{lem: DA_conjugate},
and $c_{\left\{ \mathbf{H}_{l}\right\} }''$ depends on the realizations
of $\left\{ \mathbf{H}_{l}\right\} $.

To analyze the ``range'' term of (\ref{eq:da_term}), we specify
the gap among the embeddings of $q$: $\varrho\triangleq\frac{\min_{i}|\sigma_{i}(q)|^{2}}{\max_{i}|\sigma_{i}(q)|^{2}}$.
Then the analysis follows that of \cite{Ordentlich2014}. Since $\left\Vert \mathbf{B}\mathbf{H}_{1}\right\Vert ^{2}\geq\left\Vert \sigma(q)\right\Vert ^{2}/4$
if $\mathbf{B}\mathbf{H}_{1}\notin\mathcal{V}_{\mathbf{0}}$, the
first term of (\ref{eq:da_term}) satisfies 
\begin{equation}
\frac{1}{n}\left\Vert \mathbf{B}\right\Vert ^{2}\geq\frac{1}{n\max_{i}|h_{1,i}|^{2}}\sum_{i=1}^{n}|b_{i}h_{1,i}|^{2}\geq\frac{\varrho\max_{i}|\sigma_{i}(q)|^{2}}{4\max_{i}|h_{1,i}|^{2}}.\label{eq:first_range_b}
\end{equation}
Hereby we substitute (\ref{eq:lowerRate}) and (\ref{eq:first_range_b})
into (\ref{eq:da_term}): 
\begin{align}
& \sigma_{\mathrm{AM}}^{2}  \geq  \frac{\varrho}{4\max_{i}|h_{1,i}|^{2}}\max_{i}|\sigma_{i}(q)|^{2}+\frac{\varrho c_{\left\{ \mathbf{H}_{l}\right\} }''P}{n}\max_{i}|\sigma_{i}(q)|^{-\frac{2}{L-1}}\nonumber \\
 & \geq  \rho_{\mathrm{min}}^* \left(\max_{i}|\sigma_{i}(q)|^{2}+P\max_{i}|\sigma_{i}(q)|^{-\frac{2}{L-1}}\right)\nonumber \\
 & \geq  \rho_{\mathrm{min}}^* \left(\left(\frac{1}{L-1}\right)^{\frac{L-1}{L}}P^{\frac{L-1}{L}}+\left(\frac{1}{L-1}\right)^{-\frac{1}{L}}P^{\frac{L-1}{L}}\right)\label{eq:noise22}
\end{align}
where $\rho_{\mathrm{min}}^* \triangleq \min\left\{ \frac{\varrho}{4\max_{i}|h_{1,i}|^{2}},\frac{\varrho c_{\left\{ \mathbf{H}_{l}\right\} }''}{n}\right\}$, and the last inequality follows from defining $x\triangleq\max_{i}|\sigma_{i}(q)|^{2}$
and {noticing} that the convex function $f\left(x\right)\triangleq x^{2}+Px^{-\frac{2}{L-1}}$
attains its minimum at root $x=\left(\frac{P}{L-1}\right)^{\frac{L-1}{2L}}$.

Finally, the lower bounds (\ref{eq:noise1}) and (\ref{eq:noise22})
on noise variance $\sigma_{\mathrm{AM}}^{2}$ in both cases admit
the {inequality} of $\sigma_{\mathrm{AM}}^{2}\geq c'''P^{\frac{L-1}{L}}$
for some constant $c'''$. {Substitute} this lower bound on noise
into {the rate expression} (\ref{eq:am}) and {the DoF expression}
(\ref{eq:dof definition}), one can show that $d_{\mathrm{achv},1}\leq\frac{n}{L}$. 
\end{IEEEproof}

\section{Numerical Results}

\begin{figure*}
\centering

\subfloat[Quadratic fields, $n=2$.]{\centering \includegraphics[width=0.45\textwidth]{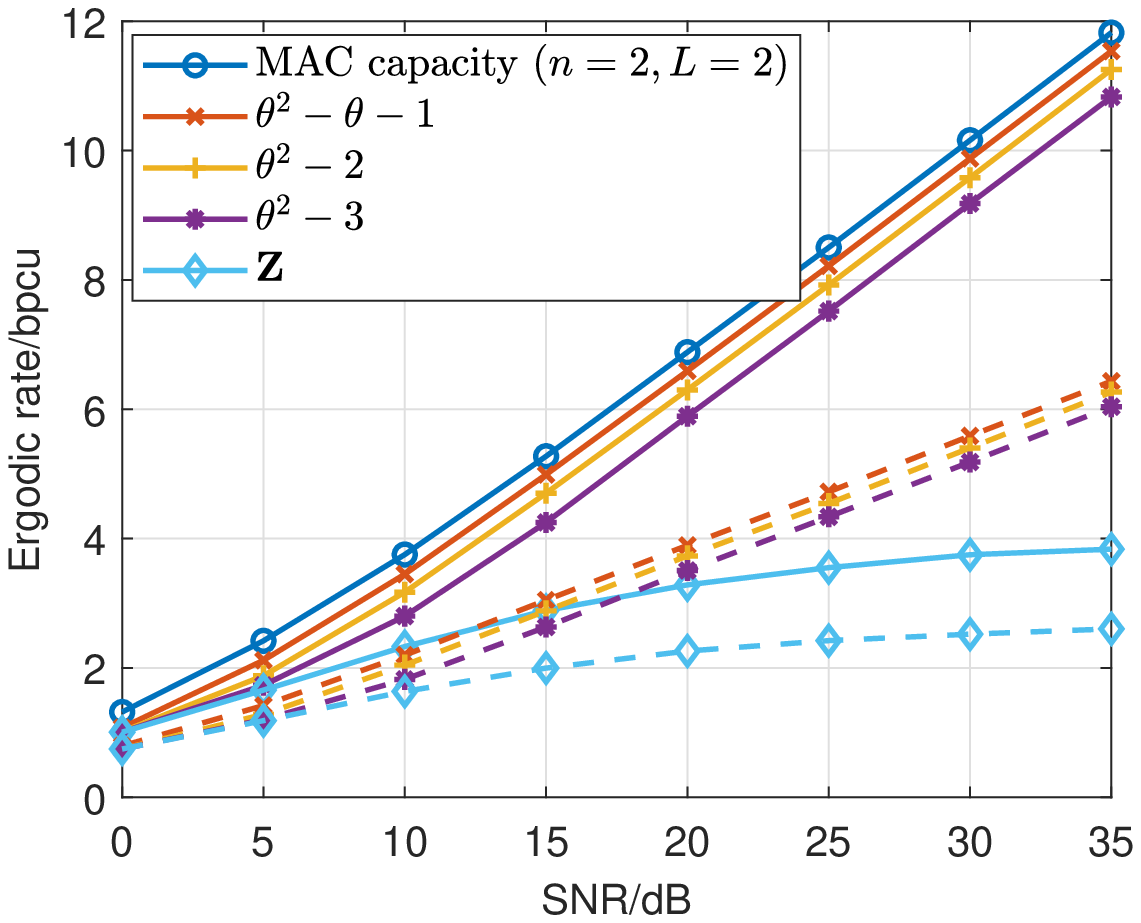}

}\subfloat[Cubic fields, $n=3$.]{\centering \includegraphics[width=0.45\textwidth]{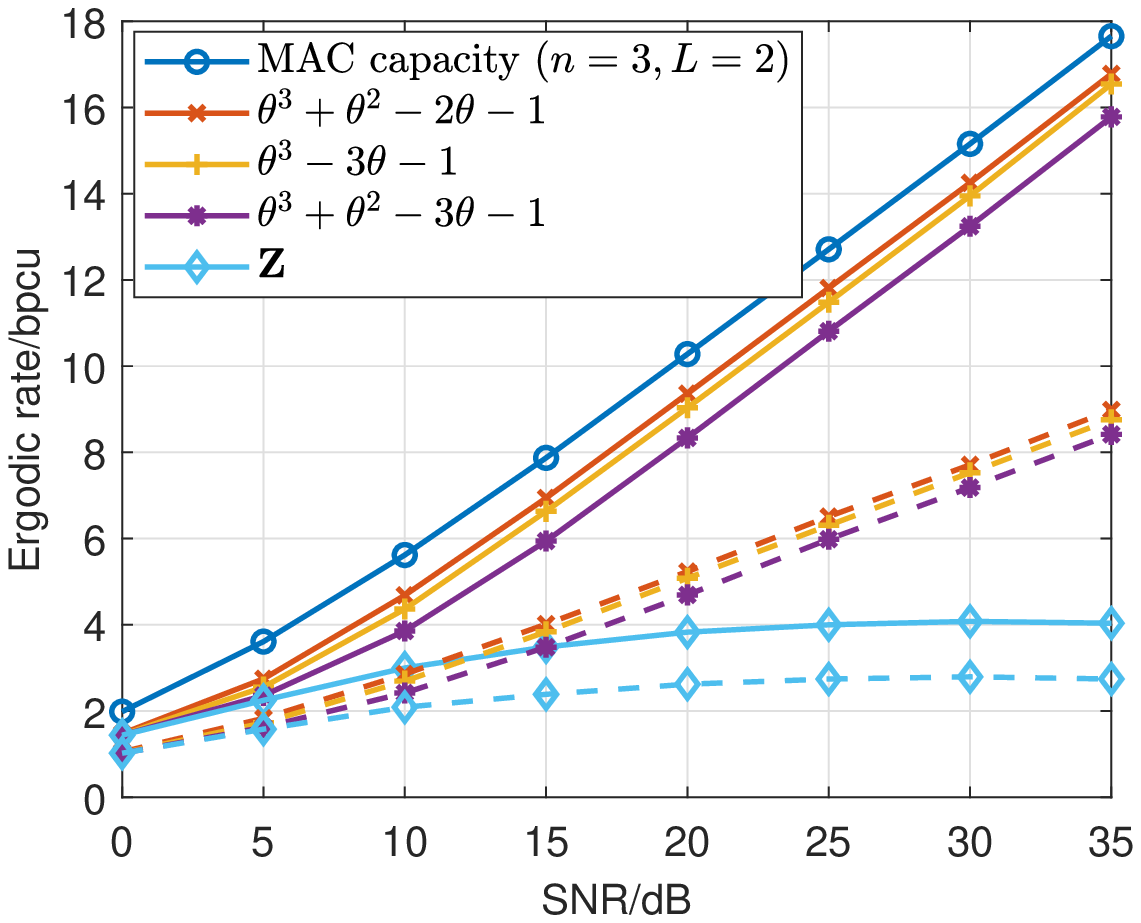}

}

\subfloat[Quartic fields, $n=4$.]{\centering \includegraphics[width=0.45\textwidth]{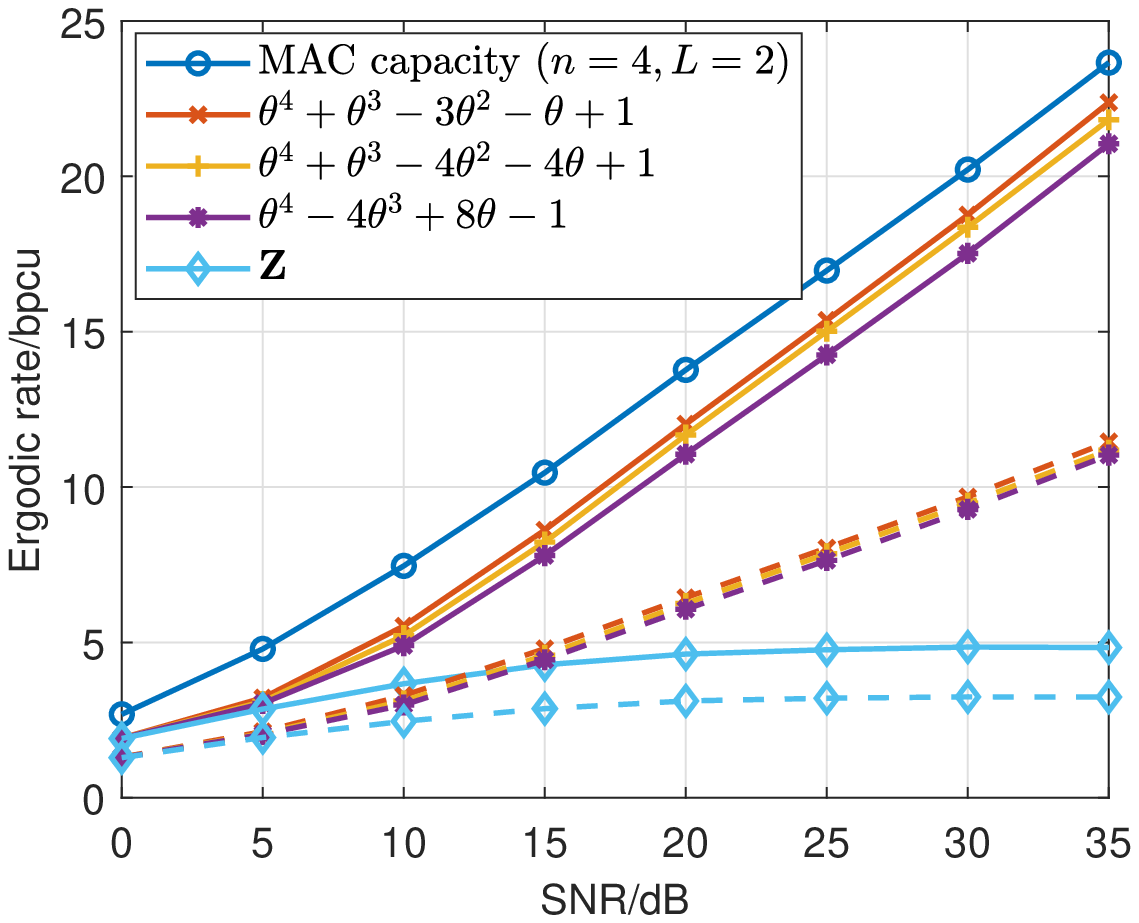}

}\subfloat[Quintic fields, $n=5$.]{\centering \includegraphics[width=0.45\textwidth]{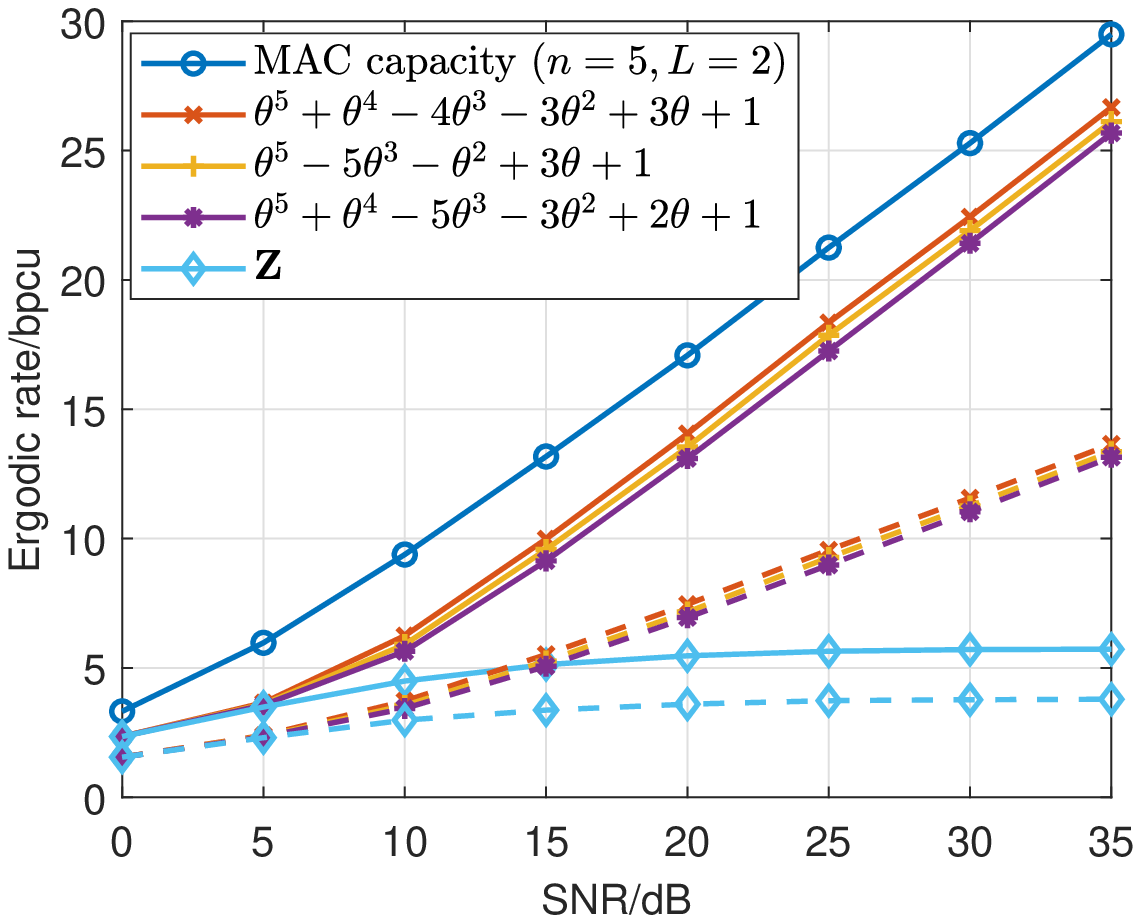}

}

\subfloat[Cyclotomic fields, $n=11$.]{\centering \includegraphics[width=0.45\textwidth]{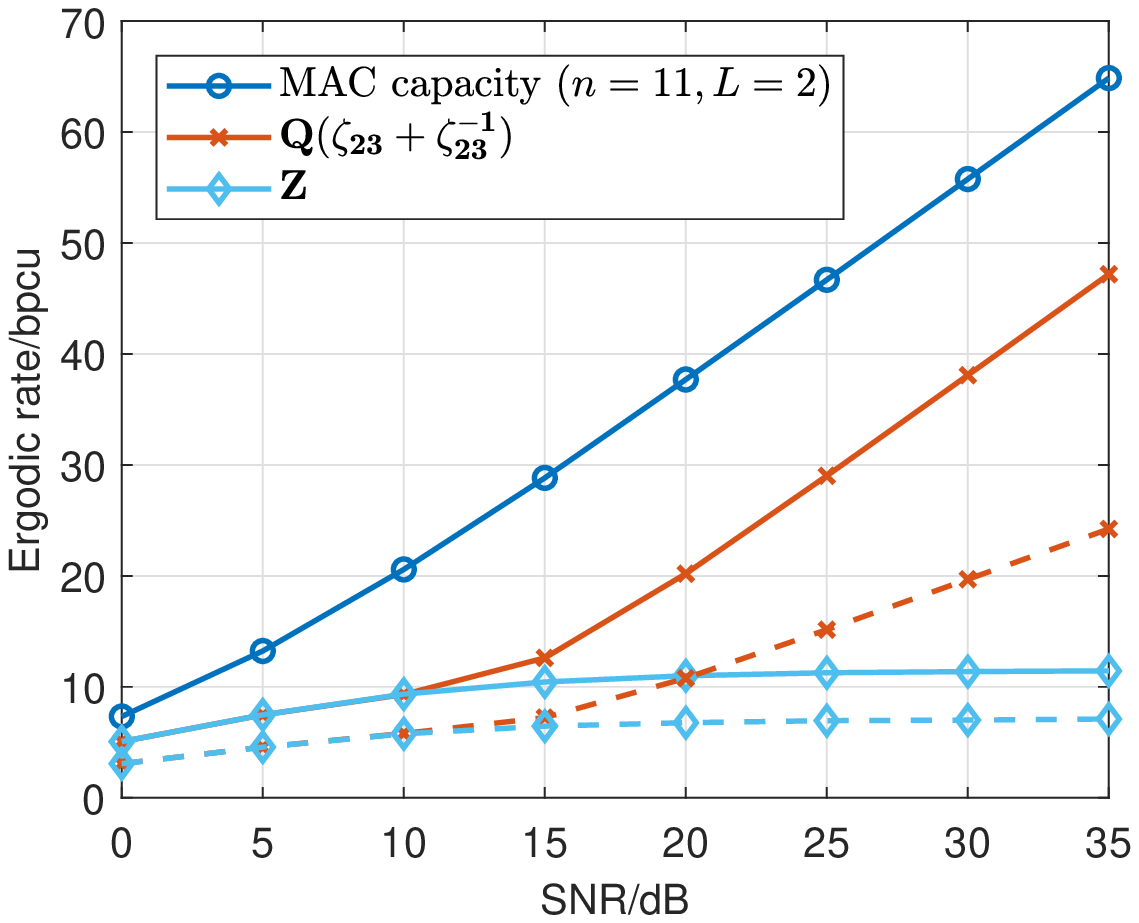}

}\subfloat[Cyclotomic fields, $n=14$.]{\centering \includegraphics[width=0.45\textwidth]{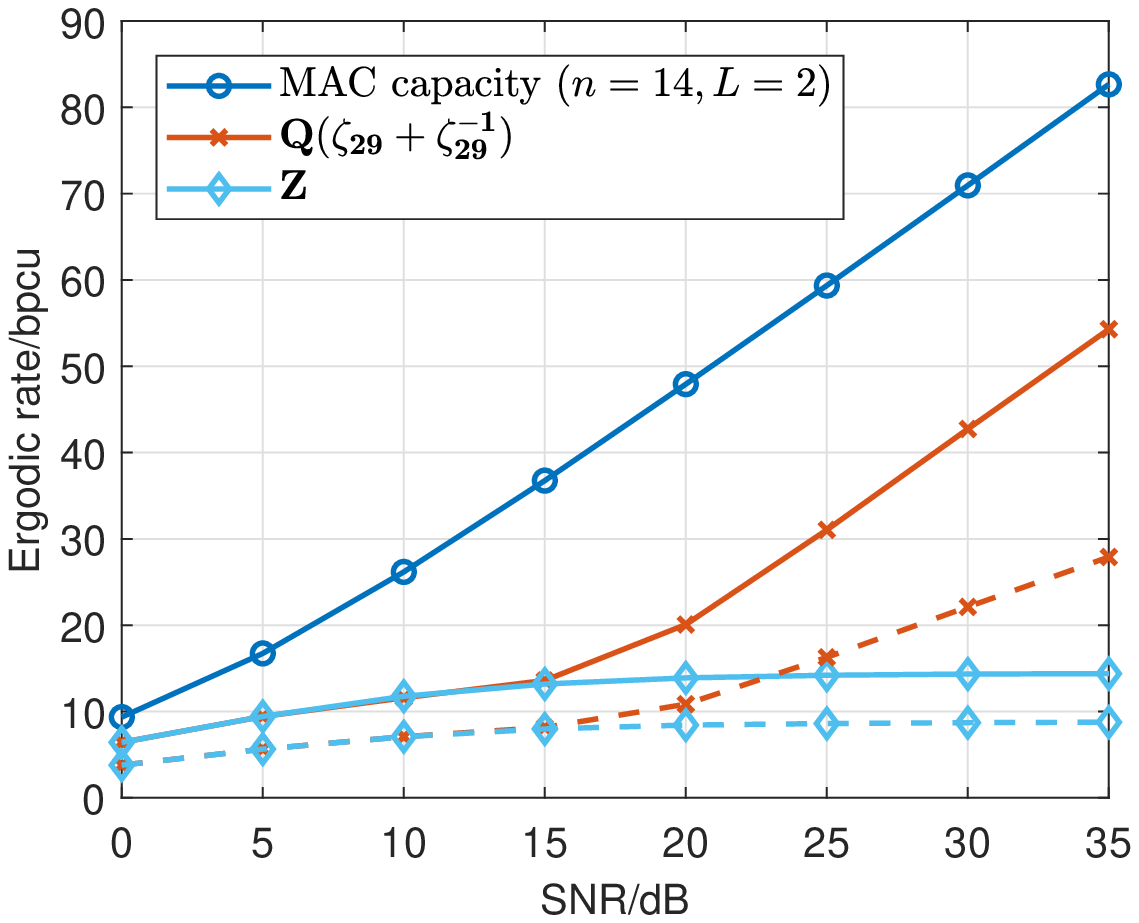}

}

\caption{The ergodic computation rates (dashed lines) and sum-rates (solid
lines) based on number fields of different degrees.}
\label{fig_total} 
\end{figure*}
In this section, we present numerical results to evaluate the performance
of Ring C\&F. Notice that there are many number fields \cite{Pohst1997,Washington1996}
available to construct lattice codes:

i) For relatively small $n$, we can enumerate all totally real number
fields with small discriminants. Tables \ref{tab:title1} to \ref{tab:title2-2}
in Appendix \ref{sec:AlgebraicFields} present this enumeration from
quadratic to quintic number fields. According to the principle of
small discriminants shown in Theorem \ref{prop:dof_upper}, the highest
computation rates should come from quadratic to quintic number fields
with minimal polynomials $\mathfrak{m}_{\theta}=\theta^{2}-\theta-1$,
$\mathfrak{m}_{\theta}=\theta^{3}+\theta^{2}-2\theta-1$, $\mathfrak{m}_{\theta}=\theta^{4}+\theta^{3}-3\theta^{2}-\theta+1$
and $\mathfrak{m}_{\theta}=\theta^{5}+\theta^{4}-4\theta^{3}-3\theta^{2}+3\theta+1$,
respectively.

ii) For relatively large $n$, we can use the maximal real sub-field
of a cyclotomic number field. A cyclotomic field $\mathbb{Q}\left(\zeta_{k}\right)$
is a number field obtained by adjoining $\zeta_{k}$ to $\mathbb{Q}$,
where $\zeta_{k}$ represents a primitive $k$th root of unity. Its
degree {is} $n=\varphi\left(k\right)/2$, where $\varphi\left(\cdot\right)$
is Euler's totient function. Table \ref{tab:cyclo} in Appendix \ref{sec:AlgebraicFields}
shows the properties of maximal real sub-fields $\mathbb{Q}\left(\zeta_{k}+\zeta_{k}^{-1}\right)$
{with degrees} $n=11,\thinspace14$.

In Fig. \ref{fig_total}, we compare the optimized computation rate
and sum-rate of ring C\&F and classic C\&F, in terms of ergodic rate
metrics defined as $\mathbb{E}\left(R_{\mathrm{achv,}1}\left(\left\{ \mathbf{H}_{l}\right\} \right)\right)$
and $\mathbb{E}\left(\sum_{i=1}^{L}R_{\mathrm{achv,}i}\left(\left\{ \mathbf{H}_{l}\right\} \right)\right)$.
The expectation is taken over $2 \times 10^3$ Monte Carlo runs,
with channel coefficients admitting $\mathcal{N}\left(0,1\right)$
entries. The ``$\mathbb{Z}$'' curve in Fig. \ref{fig_total} denotes
the classic C\&F using length-$nT$ $\mathbb{Z}$-lattice codes. The
``$\mathfrak{m}_{\theta}$'' curves, e.g., $\theta^{2}-\theta-1$,
denote ring C\&F based on field $\mathbb{Q}\left(\theta\right)$.
For cyclotomic number fields, we mark them with $\mathbb{Q}\left(\zeta_{k}+\zeta_{k}^{-1}\right)$.
The simulation starts by choosing $L=2$, $n=2$ in Fig. \ref{fig_total}-(a),
then repeats by choosing $n=3,4,5,11,14$ in Fig. \ref{fig_total}-(b)
to Fig. \ref{fig_total}-(f). Simulations can be made for the setting
of larger $L$ in the same manner.

In Fig. \ref{fig_total}-(a), significant performance gains can be
observed for Ring C\&F. The quadratic field with minimal polynomial
$\mathfrak{m}_{\theta}=\theta^{2}-\theta-1$ performs superiorly to
all other quadratic fields, and its sum-rate is within $1\mathrm{dB}$
gap to the MAC capacity. The DoF's of Ring C\&F for computation rates
and sum-rates are respectively $1$ and $2$. The classic C\&F using
$\mathbb{Z}$ gives very poor rates. It falls behind Ring C\&F with
$\mathfrak{m}_{\theta}=\theta^{2}-3$ by more than $25\mathrm{dB}$
and increasing SNR results in little performance gain. Similar observations
can be made from Fig. \ref{fig_total}-(b) to Fig. \ref{fig_total}-(f).
They confirm that fields with minimal polynomials $\mathfrak{m}_{\theta}=\theta^{3}+\theta^{2}-2\theta-1$,
$\mathfrak{m}_{\theta}=\theta^{4}+\theta^{3}-3\theta^{2}-\theta+1$
and $\mathfrak{m}_{\theta}=\theta^{5}+\theta^{4}-4\theta^{3}-3\theta^{2}+3\theta+1$
are indeed the best for $n=3,4,5$. As predicted by the parameters
in Theorem \ref{prop:dof_upper}, the gaps between the computation
sum-rates of $\mathbb{Q}\left(\zeta_{23}+\zeta_{23}^{-1}\right)$,
$\mathbb{Q}\left(\zeta_{29}+\zeta_{29}^{-1}\right)$ and MAC capacities
are much larger than those of quadratic fields, but their optimality
in DoF is preserved. We further explain why the classic C\&F has roughly
$0$ DoF. From the law of large numbers, we have approximation $\sum_{j=1}^{n}\mathbf{a}^{\top}\mathbf{M}_{j}\mathbf{a}\approx\frac{n\left(nP+1-P\right)}{nP+1}\mathbf{a}^{\top}\mathbf{a}$
for relatively large $n$; thus increasing SNR $P$ does not improve
the rate.

The Ring C\&F scheme can be extended to integer-forcing (IF) for time-varying
channels \cite{ElBakoury2015}. Suppose the channel experiences $n$
successive blocks $\hat{\mathbf{H}}_{1},\ldots\thinspace,\hat{\mathbf{H}}_{n}\in\mathbb{R}^{L\times L}$
(i.e., $L$ single-antenna transmitters and one receiver with $L$
antennas) over the duration of a codeword. We can use our algebraic
lattices to show that the following rate is achievable in IF: 
\begin{align*}
 & R_{\mathrm{IF}}\left(\left\{ \hat{\mathbf{H}}_{l}\right\} \right)=\\
 & \max_{\underset{\mathrm{rank}\left[\mathbf{a}_{1},\ldots\thinspace,\mathbf{a}_{L}\right]=L}{\left[\mathbf{a}_{1},\ldots\thinspace,\mathbf{a}_{L}\right]\in\mathcal{O}_{\mathbb{K}}^{L\times L}}}\min_{l\in\{1,\ldots L\}}\frac{1}{2}\log^{+}\left(\frac{nP}{\sum_{j=1}^{n}\sigma_{j}(\mathbf{a}_{l})^{\top}\mathbf{F}_{j}\sigma_{j}(\mathbf{a}_{l})}\right),
\end{align*}
in which $\mathbf{F}_{j}=\left(P^{-1}\mathbf{I}+\hat{\mathbf{H}}_{j}^{\top}\hat{\mathbf{H}}_{j}\right)^{-1/2}$.
The difference from \cite[Theorem 1]{ElBakoury2015} is that $\mathbf{a}_{l}\in\mathcal{O}_{\mathbb{K}}^{L}$
rather than $\mathbb{Z}^{L}$. Again, we compare the ring-based IF
and $\mathbb{Z}$-based IF in terms of ergodic rate $\mathbb{E}\left(R_{\mathrm{IF}}\left(\left\{ \hat{\mathbf{H}}_{l}\right\} \right)\right)$.
The channel capacity which equals the rate of joint maximum likelihood
(ML) decoding is 
\[
\min_{\mathcal{S}\subset\left\{ 1,\ldots,L\right\} }\frac{1}{2n|\mathcal{S}|}\sum_{j=1}^{n}\log\left(\det\left(\mathbf{I}+P\hat{\mathbf{H}}_{j,\mathcal{S}}\hat{\mathbf{H}}_{j,\mathcal{S}}^{\top}\right)\right)
\]
with $\hat{\mathbf{H}}_{j,\mathcal{S}}$ being a submatrix consists
of the $\mathcal{S}$ columns of $\hat{\mathbf{H}}_{j}$.

\begin{figure*}
\centering

\subfloat[Quadratic fields, $n=2$.]{\centering \includegraphics[width=0.45\textwidth]{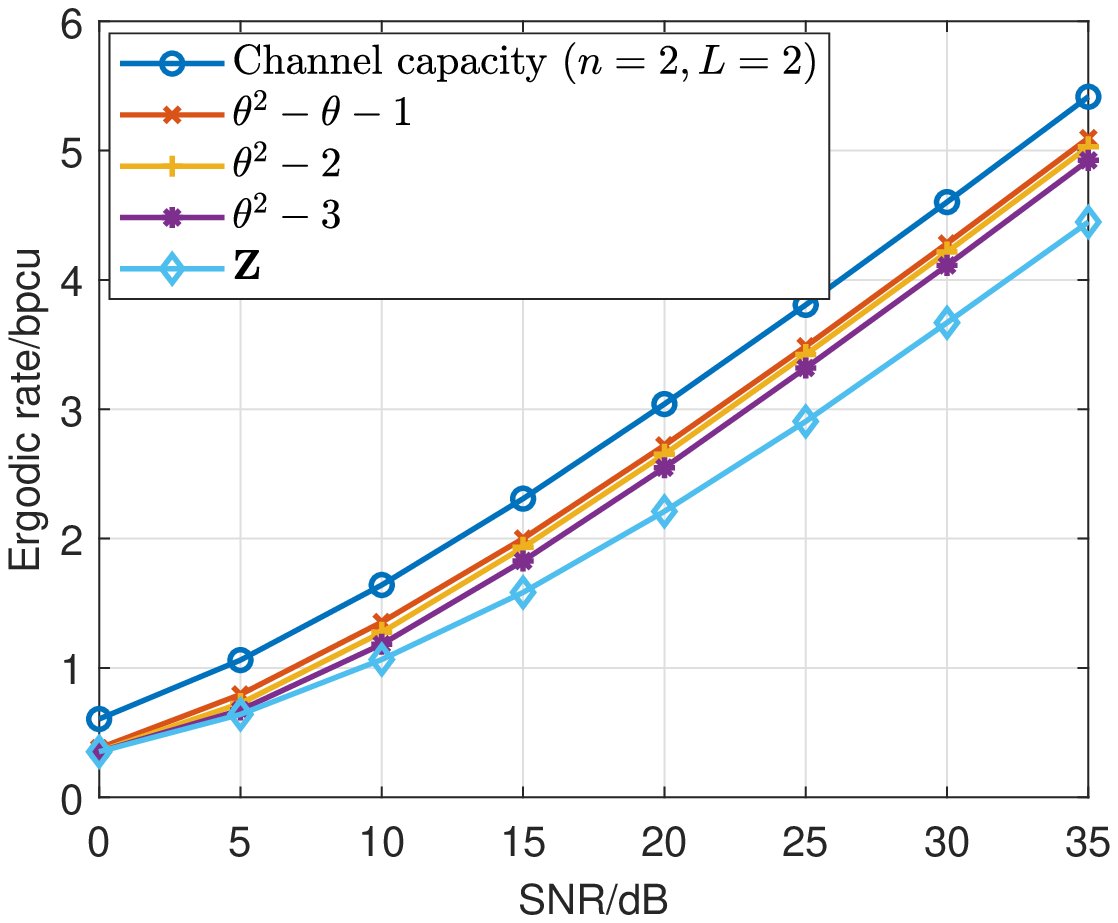} 

}\subfloat[Cyclotomic field, $n=11$.]{\centering \includegraphics[width=0.45\textwidth]{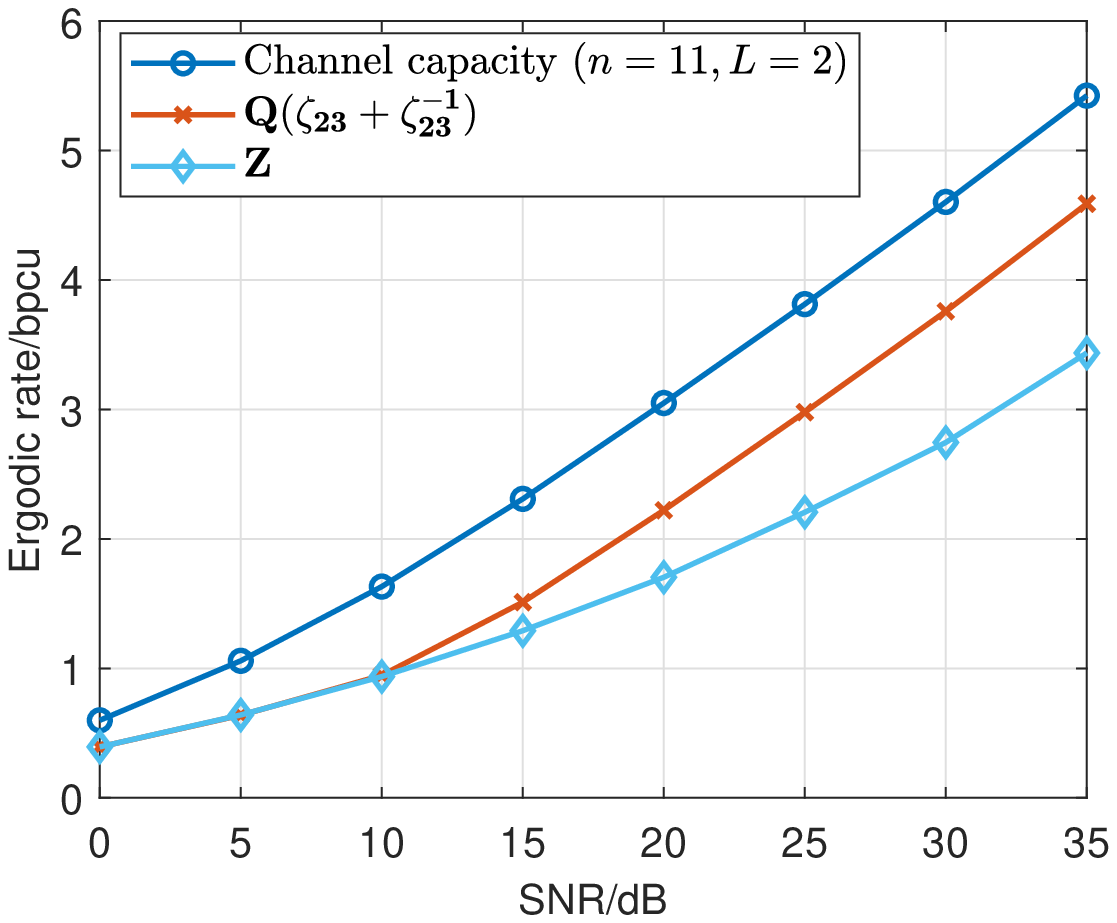}

}\caption{Ergodic rates of IF receivers with channel variation based on number
fields of different degrees.}
\label{fig_IFvar} 
\end{figure*}

As shown in Fig. \ref{fig_IFvar}, unlike the C\&F setting, IF based
on $\mathbb{Z}$ still has full DoF, thanks to the cooperation among
all receive antennas. However, IF using $\mathbb{Q}\left(\sqrt{3}\right)$,
$\mathbb{Q}\left(\sqrt{2}\right)$, and $\mathbb{Q}\left(\sqrt{5}\right)$
in Fig. \ref{fig_IFvar}-(a) provides approximately $4-5\mathrm{dB}$
gain compared to that based on $\mathbb{Z}$. The gain rises to around
$8\mathrm{dB}$ in Fig. \ref{fig_IFvar}-(b) for a large block size
of $n=11$. Thus, similarly to C\&F, the ring structure offers significant
gains in IF.

\section{Conclusions}

The class of algebraic lattices for C\&F proposed in this paper are
built from Construction A over number fields. These lattices enjoy
the advantage of closure under multiplication by algebraic integers.
Since the embeddings of an algebraic integer are different, it helps
to quantize block fading channels in a finer manner. Their achievable
rates outperform those of $\mathbb{Z}$-lattices.

Although we relaxed the GM $\sigma_{\mathrm{eff}}^{2}$ in (\ref{eq:rate_std1})
to the AM so that the problem was reduced to finding the successive
minima of a lattice, an important open question is how to minimize
the GM (a product form) efficiently, and how to analyze its Diophantine
approximation.

Metric Diophantine approximation associated with $\mathcal{O}_{\mathbb{K}}$-modules
studied in this paper is more involved than that associated with $\mathbb{Z}$-lattices.
We only addressed the convergent part of the Khintchine-Groshev theorem,
while the divergent part was not used. We leave the divergent part
of Lemma \ref{lem: DA_conjugate} as another open problem.

\appendices{}

\section{\label{sec:Proofquan}Proof of Quantization goodness}

Our proof follows the steps in \cite{Ordentlich2012} with some adjustments:
i) The prime number $p$ is chosen to grow as $O\left(T^{3n/2}\right)$
rather than $O\left(T^{3/2}\right)$, to compensate for the factor
$p^{T-k}$ in the volume of the coarse lattice, while it is $p^{nT-k}$
in \cite{Ordentlich2012}. ii) We count the number of lattice points
inside a ball for a number field lattice $\sigma\left(\mathcal{O}_{\mathbb{K}}^{T}\right)$
rather than an integer lattice $\mathbb{Z}^{nT}$.

Let $V_{nT}$ be the volume of an $nT$-dimensional unit ball. Set
the inertial degree $f=1$, and the scaling factor $\gamma=p^{-1/n}\mathrm{disc}_{\mathbb{K}}^{-1/\left(2n\right)}\sqrt{nT}$.
Write (\ref{eq:firstconstruction}) explicitly as $\rho^{-1}\left(\mathcal{C}\right)=\mathcal{M}\left(\mathcal{C}\right)+\mathfrak{p}^{T}$,
where $\mathcal{M}\left(\cdot\right)$ maps $\mathbb{F}_{p}$ onto
the coset leaders of each $\mathcal{O}_{\mathbb{K}}/\mathfrak{p}$ based
on component-wise isomorphism. The scaled lattice
is 
\[
\gamma\Lambda_{c}^{\mathcal{O}_{\mathbb{K}}}=\gamma\mathcal{M}\left(\mathbf{G}\mathbf{w}\right)+\gamma\mathfrak{p}^{T},
\]
where the volume of its embedded lattice satisfies $\mathrm{Vol}\left(\gamma\Lambda_{c}^{\mathbb{Z}}\right)\geq\gamma^{nT}{\mathrm{disc}_{\mathbb{K}}}^{T/2}p^{\left(T-k\right)}$
and the equality holds only if the generator matrix $\mathbf{G}\in\mathbb{F}_{p}^{T\times k}$
of $\mathcal{C}$ has full rank.

Since obviously $\lim_{T\rightarrow\infty}G\left(\gamma\Lambda_{c}^{\mathbb{Z}}\right)\geq{1}/\left({2\pi e}\right)$,
we are left with the task of showing that that for any $\delta>0$,
$\epsilon>0$, 
\begin{equation}
\mathrm{Pr}\left(\frac{\tilde{\sigma}^{2}\left(\gamma\Lambda_{c}^{\mathbb{Z}}\right)}{\mathrm{Vol}\left(\gamma\Lambda_{c}^{\mathbb{Z}}\right)^{2/(nT)}}>\frac{1}{2\pi e}+\delta\right)<\epsilon\label{eq:probsmall}
\end{equation}
with large enough $T$. Letting $0<\alpha<\log\left(nT\right)$, and
\begin{equation}
k\triangleq\frac{nT}{2\log\left(p\right)}\log\left(V_{nT}^{-2/\left(nT\right)}2^{\alpha}\right),\label{eq:defpk}
\end{equation}
we have 
\begin{equation}
\mathrm{Vol}\left(\gamma\Lambda_{c}^{\mathbb{Z}}\right)^{2/(nT)}=nTV_{nT}^{2/\left(nT\right)}2^{-\alpha}.\label{eq:eqVolume_APA}
\end{equation}
Denote by $r_{o}$ the covering radius of the embedded lattice $\sigma\left(\mathcal{O}_{\mathbb{K}}^{T}\right)$.
Since $\mathrm{Vol}\left(\sigma\left(\mathcal{O}_{\mathbb{K}}^{T}\right)\right)={\mathrm{disc}_{\mathbb{K}}}^{T/2}$,
the number of points of ${\mathrm{disc}_{\mathbb{K}}}^{-1/\left(2n\right)}\sigma\left(\mathcal{O}_{\mathbb{K}}^{T}\right)$
inside a ball can be measured with volumes. Then we can adapt \cite[Lemma 1]{Ordentlich2012}
from $\mathbb{Z}^{nT}$ to $\sigma\left(\mathcal{O}_{\mathbb{K}}^{T}\right)$
to get the following lemma. 
\begin{lem}
\label{lem_points} For any $\mathbf{x}\in\mathbb{R}^{nT}$ and $r>0$,
the number of points of a scaled lattice ${\mathrm{disc}_{\mathbb{K}}}^{-1/\left(2n\right)}\sigma\left(\mathcal{O}_{\mathbb{K}}^{T}\right)$
inside $\mathcal{B}\left(\mathbf{x},r\right)$ can be bounded as 
\begin{align*}
	&\left|{\mathrm{disc}_{\mathbb{K}}}^{-1/\left(2n\right)}\sigma\left(\mathcal{O}_{\mathbb{K}}^{T}\right)\cap\mathcal{B}\left(\mathbf{x},r\right)\right| \\
	&
	\geq\mathrm{Vol}\left(\mathcal{B}\left(\mathbf{x},r-{\mathrm{disc}_{\mathbb{K}}}^{-1/\left(2n\right)}r_{o}\right)\right).
\end{align*}
 
\end{lem}
 Assume the source $\mathbf{x}$ is uniformly distributed over
a fundamental region of lattice $\sigma\left(\mathfrak{p}^{T}\right)$.
For a target $\mathbf{x}\in\mathbb{R}^{nT}$, its distance to
the closest lattice point equals to that modulo the coarse lattice:
\begin{align*}
d\left(\mathbf{x},\gamma\Lambda_{c}^{\mathbb{Z}}\right) & =\min_{\mathbf{c}\in\mathcal{C}\left(\mathbf{G}\right),\lambda\in\sigma\left(\mathfrak{p}^{T}\right)}\frac{1}{nT}\left\Vert \mathbf{x}-\gamma\mathcal{M}\left(\mathbf{c}\right)-\gamma\lambda\right\Vert ^{2}\\
 & =\min_{\mathbf{c}\in\mathcal{C}\left(\mathbf{G}\right)}\frac{1}{nT}\left\Vert \left(\mathbf{x}-\gamma\mathcal{M}\left(\mathbf{c}\right)\right)^{*}\right\Vert ^{2},
\end{align*}
in which   $(\cdot)^{*}\triangleq(\cdot)\thinspace\mathrm{mod}\thinspace\sigma\left(\mathfrak{p}^{T}\right)$.
Clearly, $d\left(\mathbf{x},\gamma\Lambda_{c}^{\mathbb{Z}}\right)\leq\frac{\gamma^{2}r_{p}^{2}}{nT}$,
where $r_{p}$ denotes the covering radius of ideal lattice $\sigma\left(\mathfrak{p}^{T}\right)$.
Note that $\mathcal{M}\left(\mathbf{c}\right)$ is uniformly distributed
over the coset leaders $S^{T}$ of $\left(\mathcal{O}_{\mathbb{K}}/\mathfrak{p}\right)^{T}$
as the elements of $\mathbf{G}$ are uniform over $\mathbb{F}_{p}$,
and $|S^{T}|=p^{T}$. With $0<\rho<\alpha$, for any fixed $\mathbf{x}$,
the probability of a small quantization distance is bounded as 
\begin{align}
\varepsilon & \triangleq\mathrm{Pr}\left(d\left(\mathbf{x},\gamma\Lambda_{c}^{\mathbb{Z}}\right)\leq2^{-\rho}\right)\nonumber \\
 & =\mathrm{Pr}\left(\min_{\mathbf{c}\in\mathcal{C}\left(\mathbf{G}\right)}\frac{1}{nT}\left\Vert \left(\mathbf{x}-\gamma\mathcal{M}\left(\mathbf{c}\right)\right)^{*}\right\Vert ^{2}\leq2^{-\rho}\right)\nonumber \\
 & =p^{-T}\left|\gamma S^{T}\cap\mathcal{B}^{*}\left(\mathbf{x},\sqrt{nT2^{-\rho}}\right)\right|\nonumber \\
 & =p^{-T}\left|\gamma\sigma\left(\mathcal{O}_{\mathbb{K}}^{T}\right)\cap\mathcal{B}\left(\mathbf{x},\sqrt{nT2^{-\rho}}\right)\right|\nonumber \\
 & =p^{-T}\left|{\mathrm{disc}_{\mathbb{K}}}^{-1/\left(2n\right)}\sigma\left(\mathcal{O}_{\mathbb{K}}^{T}\right)\cap\mathcal{B}\left(\mathbf{x},\gamma^{-1}{\mathrm{disc}_{\mathbb{K}}}^{-1/\left(2n\right)}\sqrt{nT2^{-\rho}}\right)\right|\nonumber \\
 & \overset{(a)}{\geq}V_{nT}p^{-T}\left(\gamma^{-1}{\mathrm{disc}_{\mathbb{K}}}^{-1/\left(2n\right)}\sqrt{nT2^{-\rho}}-{\mathrm{disc}_{\mathbb{K}}}^{-1/\left(2n\right)}r_{o}\right)^{nT}\nonumber \\
 & \overset{(b)}{\geq}V_{nT}p^{-k}\left(nTV_{nT}^{2/\left(nT\right)}2^{-\alpha}\right)^{-nT/2}\left(nT2^{-\rho}\right)^{nT/2}\left(1-\gamma r_{o}\right)^{nT}\nonumber \\
 & \overset{(c)}{=}V_{nT}2^{-\rho nT/2}O\left(1\right),\label{eq:lastofepsi}
\end{align}
where (a) is from Lemma \ref{lem_points}, (b) is from using ${\mathrm{disc}_{\mathbb{K}}}^{-T/2} \geq p^{\left(T-k\right)}\gamma^{nT}\mathrm{Vol}\left(\gamma\Lambda_{c}^{\mathbb{Z}}\right)^{-1}$
and Eq. (\ref{eq:eqVolume_APA}), 
and (c) has used Eq. (\ref{eq:defpk}) and $\left(1-\gamma r_{o}\right)^{nT}=O\left(1\right)$.
To see this, notice that $r_{o}=O\left(\sqrt{nT}\right)$ as it is
upper bounded by the length of a corner point of a Gram-Schmidt parallelepiped
\cite[Eq. (44)]{Lyu2017}, and that $\mathrm{disc}_{\mathbb{K}}$
is independent of $T$. If we choose $p$ to grow with $T^{cn}$,
$c>1$, e.g., $p=\xi T^{3n/2}$ and minimize $\xi\in[1,2)$ under
the constraint that $p$ is a prime \cite{Ordentlich2012}, then $\left(1-\gamma r_{o}\right)^{nT}=\left(1-\mathrm{disc}_{\mathbb{K}}^{-1/\left(2n\right)}p^{-1/n}O\left({nT}\right)\right)^{nT}=O\left(1\right)$
w.r.t. $T$.

%we have  for a number field with a power basis $\{1,\theta,\thinspace...,\thinspace\theta^{n-1}\}$ that
%\[
%\mathrm{disc}_{\mathbb{K}}=\prod_{1\leq i<j\leq n}\left(\theta_{i}-\theta_{j}\right)^{2},
%\]
%so $\mathrm{disc}_{\mathbb{K}}$ can be regarded as a parameter determined
%by the number of blocks $n$ and the number field basis, and is
%independent of $T$.

For the $p^{k}-1$ non-zero random $\mathbf{w}_{i}\in\mathbb{F}_{p}^{k}$,
define the indicator function 
\[
\chi_{i}=\begin{cases}
1, & \mathrm{if}\thinspace\left\Vert \left(\mathbf{x}-\gamma\mathcal{M}\left(\mathbf{G}\mathbf{w}_{i}\right)\right)^{*}\right\Vert ^{2}\leq2^{-\rho}\\
0, & \mathrm{if}\thinspace\left\Vert \left(\mathbf{x}-\gamma\mathcal{M}\left(\mathbf{G}\mathbf{w}_{i}\right)\right)^{*}\right\Vert ^{2}>2^{-\rho}
\end{cases},
\]
which satisfies $\mathbb{E}\left(\chi_{i}\right)=\varepsilon$. From
Chebyshev's inequality, 
\begin{align*}
\mathrm{Pr}\left(d\left(\mathbf{x},\gamma\Lambda_{c}^{\mathbb{Z}}\right)>2^{-\rho}\right) & \leq\mathrm{Pr}\left(\sum_{i=1}^{p^{k}-1}\chi_{i}=0\right)\\
 & \leq\frac{\mathrm{Var}\left(\frac{1}{p^{k}-1}\sum_{i=1}^{p^{k}-1}\chi_{i}\right)}{\varepsilon^{2}}\\
 & <\frac{p}{\left(p^{k}-1\right)\varepsilon}.
\end{align*}
Together with Eqs. (\ref{eq:defpk}) and (\ref{eq:lastofepsi}), one
has 
\begin{equation}
\mathrm{Pr}\left(d\left(\mathbf{x},\gamma\Lambda_{c}^{\mathbb{Z}}\right)>2^{-\rho}\right)<2^{-\frac{nT}{2}\left(\alpha-\rho+O(1)\right)}.\label{eq:probapx1}
\end{equation}
It follows from (\ref{eq:probapx1}) that we can use the same arguments
as in \cite{Ordentlich2012} to show the expected second moment is
small. Finally, we complete the proof of \eqref{eq:probsmall} by
using Markov's inequality.

\section{\label{sec:Proofgau}Proof of Lemma \ref{lem:seminorm}}

By the law of total probability, 
\begin{align*}
 & \mathrm{Pr}\left(\mathrm{vec}\left(\mathbf{Z}_{\mathrm{eff}}\right)\notin\mathcal{B}\left(\mathbf{0},\sqrt{\left(1+\delta\right)nT\sigma_{\mathrm{eff}}^{2}}\right)\right)=\\
 & \mathrm{Pr}\left(\mathcal{T}=1\right)\mathrm{Pr}\left(\mathrm{vec}\left(\mathbf{Z}_{\mathrm{eff}}\right)\notin\mathcal{B}\left(\mathbf{0},\sqrt{\left(1+\delta\right)nT\sigma_{\mathrm{eff}}^{2}}\right)\mathrel{\Big|}\mathcal{T}=1\right)\\
 & +\mathrm{Pr}\left(\mathcal{T}=0\right)\mathrm{Pr}\left(\mathrm{vec}\left(\mathbf{Z}_{\mathrm{eff}}\right)\notin\mathcal{B}\left(\mathbf{0},\sqrt{\left(1+\delta\right)nT\sigma_{\mathrm{eff}}^{2}}\right)\mathrel{\Big|}\mathcal{T}=0\right),
\end{align*}
where 
\[
\mathcal{T}=\begin{cases}
0, & \mathrm{if}\thinspace\exists\mathbf{x}\in\left\{ \mathrm{vec}\left(\tilde{\mathbf{X}}_{l}\right)\right\} ,\\
 & \thinspace\mathrm{s.t.},\mathbf{x}\notin\mathcal{B}\left(\mathbf{0},\sqrt{\left(1+\delta\right)\mathbb{E}\left(\left\Vert \mathbf{x}\right\Vert ^{2}\right)}\right),\delta>0\\
1, & \mathrm{otherwise}.
\end{cases}
\]
For any $\epsilon>0$, we can make $\mathrm{Pr}\left(\mathcal{T}=0\right)\leq\epsilon$
by increasing $T$ because $\left\{ \mathrm{vec}\left(\tilde{\mathbf{X}}_{l}\right)\right\} $
are all semi norm-ergodic. Then we can confine our discussion to the
case of $\mathcal{T}=1$. This constraint enables us to show the density
of the effective noise is tightly upper bounded by that of a Gaussian
vector with the techniques in \cite[Lemma 11]{Erez2004}, without
proving the algebraic lattices are good for covering.
\begin{prop}
\label{prop:sermi_prop}Assume $\mathcal{T}=1$. Let 
\[
\mathbf{z}_{\mathrm{eff}}=\left(\mathbf{E}_{\mathbf{a}}^{-1}\otimes\mathbf{I}_{T}\right)\left(\sum_{l=1}^{L}\left(\mathbf{B}\mathbf{H}_{l}-\mathbf{A}_{l}\right)\otimes\mathbf{I}_{T}\mathrm{vec}\left(\tilde{\mathbf{X}}_{l}\right)+\mathbf{B}\otimes\mathbf{I}_{T}\mathbf{z}\right).
\]
Then there exists an i.i.d. Gaussian vector 
\[
\mathbf{z}_{\mathrm{eff}}^{*}=\left(\mathbf{E}_{\mathbf{a}}^{-1}\otimes\mathbf{I}_{T}\right)\left(\sum_{l=1}^{L}\left(\mathbf{B}\mathbf{H}_{l}-\mathbf{A}_{l}\right)\otimes\mathbf{I}_{T}\tilde{\mathbf{z}}_{l}^{*}+\mathbf{B}\otimes\mathbf{I}_{T}\mathbf{z}\right)
\]
with density $f_{\mathbf{z}_{\mathrm{eff}}^{*}}\left(\mathbf{z}\right)=\mathcal{N}\left(\mathbf{0},\sigma_{\mathrm{eff}}^{2}\mathbf{I}_{nT}\right)$,
$\sigma_{\mathrm{eff}}^{2}=\prod_{j=1}^{n}\left(|b_{j}|^{2}+P\left\Vert b_{j}\mathbf{h}_{j}-\sigma_{j}(\mathbf{a})\right\Vert ^{2}\right)^{\frac{1}{n}}$,
$\tilde{\mathbf{z}}_{l}^{*}\sim\mathcal{N}\left(\mathbf{0},P\mathbf{I}_{nT}\right)$,
such that the density of $\mathbf{z}_{\mathrm{eff}}$ is upper bounded
as 
\[
f_{\mathbf{z}_{\mathrm{eff}}}\left(\mathbf{z}\right)\leq\left(1-\delta'\right)^{L}e^{Lc(T)nT}f_{\mathbf{z}_{\mathrm{eff}}^{*}}\left(\mathbf{z}\right),
\]
where $c(T)\triangleq\frac{1}{2}\log\left(2\pi eG\left(\Lambda^{(nT)}\right)\right)+\frac{1}{nT}$,
and $\delta',\thinspace c(T)\rightarrow0$ as $T\rightarrow\infty$. 
\end{prop}
\begin{IEEEproof}
i) First, we show that $\mathbf{z}_{\mathrm{eff}}^{*}$ admits density
$\mathcal{N}\left(\mathbf{0},\sigma_{\mathrm{eff}}^{2}\mathbf{I}_{nT}\right)$.
As a linear combination of independent Gaussian random variables,
$\sigma_{\mathrm{eff}}^{-1}\mathbf{z}_{\mathrm{eff}}^{*}$ has a density
\begin{align}
 & f_{\tilde{\mathbf{z}}_{1}^{*}}\left(\sigma_{\mathrm{eff}}^{-1}\mathbf{E}_{\mathbf{a}}^{-1}\left(\mathbf{B}\mathbf{H}_{1}-\mathbf{A}_{1}\right)\otimes\mathbf{I}_{T}\mathbf{z}\right)\circledast\cdots\circledast\nonumber \\
 & f_{\tilde{\mathbf{z}}_{L}^{*}}\left(\sigma_{\mathrm{eff}}^{-1}\mathbf{E}_{\mathbf{a}}^{-1}\left(\mathbf{B}\mathbf{H}_{L}-\mathbf{A}_{L}\right)\otimes\mathbf{I}_{T}\mathbf{z}\right)\circledast f_{\mathbf{z}}\left(\sigma_{\mathrm{eff}}^{-1}\mathbf{E}_{\mathbf{a}}^{-1}\mathbf{B}\otimes\mathbf{I}_{T}\mathbf{z}\right)\nonumber \\
 & =\mathcal{N}\left(\mathbf{0},\mathbf{I}_{nT}\right),\label{eq:gau2}
\end{align}
where $\circledast$ refers to the convolution of density functions.
Thus, we obtain 
\begin{equation}
f_{\mathbf{z}_{\mathrm{eff}}^{*}}\left(\mathbf{z}\right)=\mathcal{N}\left(\mathbf{0},\sigma_{\mathrm{eff}}^{2}\mathbf{I}_{nT}\right).\label{eq:gau3}
\end{equation}

ii) Second, we upper bound the density of each dithered variable $\tilde{\mathbf{X}}_{l}$
by that of $\tilde{\mathbf{z}}_{l}^{*}$. The constrained Voronoi
region for each $\tilde{\mathbf{X}}_{l}$ is $\mathcal{V}_{l}\triangleq\mathcal{V}\left(\gamma\Lambda_{c}^{\mathbb{Z}}\right)\cap\mathcal{B}\left(\mathbf{0},\sqrt{\left(1+\delta\right)nTP}\right)$,
so the density function of $\tilde{\mathbf{X}}_{l}$ becomes 
\[
f_{\tilde{\mathbf{x}}_{l}}\left(\mathbf{z}\right)=\begin{cases}
1/|\mathcal{V}_{l}|, & \mathrm{if}\thinspace\mathbf{z}\in\mathcal{V}_{l}\\
0, & \mathrm{otherwise}.
\end{cases}
\]
Also, $|\mathcal{V}_{l}|\geq V_{nT}r_{\mathrm{eff}}^{nT}\left(\gamma\Lambda_{c}^{\mathbb{Z}}\right)\left(1-\delta'\right)$
for any small $\delta'>0$ as $\mathrm{vec}\left(\tilde{\mathbf{X}}_{l}\right)$
is semi-norm ergodic. Let $\tilde{\mathbf{b}}$ be a random vector
uniformly distributed over a ball of volume $\mathrm{Vol}\left(\gamma\Lambda_{c}^{\mathbb{Z}}\right)$;
its density function $f_{\tilde{\mathbf{b}}}\left(\mathbf{z}\right)$
upper bounds $f_{\tilde{\mathbf{x}}_{l}}\left(\mathbf{z}\right)$:
\[
\frac{f_{\tilde{\mathbf{x}}_{l}}\left(\mathbf{z}\right)}{f_{\tilde{\mathbf{b}}}\left(\mathbf{z}\right)}=\frac{V_{nT}r_{\mathrm{eff}}^{nT}\left(\gamma\Lambda_{c}^{\mathbb{Z}}\right)}{|\mathcal{V}_{l}|}\leq1-\delta'.
\]
The Gaussian variable $\tilde{\mathbf{z}}_{l}^{*}$ with density $f_{\tilde{\mathbf{z}}_{l}^{*}}\left(\mathbf{z}\right)$
has the same second moment as that of the fundamental Voronoi region
$\mathcal{V}\left(\gamma\Lambda_{c}^{\mathbb{Z}}\right)$. Combining
the above, we arrive at 
\begin{align}
\frac{f_{\tilde{\mathbf{x}}_{l}}\left(\mathbf{z}\right)}{f_{\tilde{\mathbf{z}}_{l}^{*}}\left(\mathbf{z}\right)}=\frac{f_{\tilde{\mathbf{x}}_{l}}\left(\mathbf{z}\right)f_{\tilde{\mathbf{b}}}\left(\mathbf{z}\right)}{f_{\tilde{\mathbf{b}}}\left(\mathbf{z}\right)f_{\tilde{\mathbf{z}}_{l}^{*}}\left(\mathbf{z}\right)} & <\left(1-\delta'\right)e^{nTc(T)},\label{eq:gau1}
\end{align}
where $f_{\tilde{\mathbf{b}}}\left(\mathbf{z}\right)/f_{\tilde{\mathbf{z}}_{l}^{*}}\left(\mathbf{z}\right)<e^{nTc(T)}$
due to \cite[Eq. (199)]{Erez2004}.

iii) Finally, notice that the density of $\mathbf{z}_{\mathrm{eff}}$
is: 
\begin{align*}
 & f_{\tilde{\mathbf{x}}_{1}}\left(\mathbf{E}_{\mathbf{a}}^{-1}\left(\mathbf{B}\mathbf{H}_{1}-\mathbf{A}_{1}\right)\otimes\mathbf{I}_{T}\mathbf{z}\right)\circledast\cdots\circledast\\
 & f_{\tilde{\mathbf{x}}_{L}}\left(\mathbf{E}_{\mathbf{a}}^{-1}\left(\mathbf{B}\mathbf{H}_{L}-\mathbf{A}_{L}\right)\otimes\mathbf{I}_{T}\mathbf{z}\right)\circledast f_{\mathbf{z}}\left(\mathbf{E}_{\mathbf{a}}^{-1}\mathbf{B}\otimes\mathbf{I}_{T}\mathbf{z}\right),
\end{align*}
so combining this with the arguments in steps i) and ii) proves the
proposition. 
\end{IEEEproof}
Since the Gaussian vector $\mathbf{z}_{\mathrm{eff}}^{*}$ is semi
norm-ergodic with effective variance $\sigma_{\mathrm{eff}}^{2}$,
we have 
\[
\mathrm{Pr}\left(\mathbf{z}_{\mathrm{eff}}^{*}\notin\mathcal{B}\left(\mathbf{0},\sqrt{\left(1+\delta\right)nT\sigma_{\mathrm{eff}}^{2}}\right)\right)\rightarrow0.
\]
Together with Proposition \ref{prop:sermi_prop}, we have $\mathrm{Pr}\left(\mathrm{vec}\left(\mathbf{Z}_{\mathrm{eff}}\right)\notin\mathcal{B}\left(\mathbf{0},\sqrt{\left(1+\delta\right)nT\sigma_{\mathrm{eff}}^{2}}\right)\mathrel{\Big|}\mathcal{T}=1\right)\rightarrow0$
and the proof is completed.

\section{\label{sec:Proofratep}Derivation of Eq. (\ref{eq:am-1})}

Note that $\sigma_{\mathrm{AM}}^{2}$ is a convex function of $\mathbf{b}$.
By assuming $\mathbf{a}$ to be fixed, the minimum of $\sigma_{\mathrm{AM}}^{2}$
is reached by setting $\partial\sigma_{\mathrm{AM}}^{2}/\partial\mathbf{b}=\mathbf{0}$.
From this we have, for $j=1,2,\ldots,n$, 
\begin{equation}
b_{j}=\frac{P\sigma_{j}(\mathbf{a})^{\top}\mathbf{h}_{j}}{P\left\Vert \mathbf{h}_{j}\right\Vert ^{2}+1}.\label{eq:b coeff}
\end{equation}
By plugging (\ref{eq:b coeff}) into $\frac{1}{P}\sigma_{\mathrm{AM}}^{2}$,
we have 
\begin{align*}
 & \frac{1}{P}\sigma_{\mathrm{AM}}^{2}\\
= & \frac{1}{nP}\sum_{j=1}^{n}\left(b_{j}^{2}\left(1+P\left\Vert \mathbf{h}_{j}\right\Vert ^{2}\right)-2Pb_{j}\sigma_{j}(\mathbf{a})^{\top}\mathbf{h}_{j}+P\left\Vert \sigma_{j}(\mathbf{a})\right\Vert ^{2}\right)\\
= & \frac{1}{nP}\sum_{j=1}^{n}\left(\left(\frac{P\sigma_{j}(\mathbf{a})^{\top}\mathbf{h}_{j}}{P\left\Vert \mathbf{h}_{j}\right\Vert ^{2}+1}\right)^{2}\left(1+P\left\Vert \mathbf{h}_{j}\right\Vert ^{2}\right)\right)+\\
 & \frac{1}{nP}\sum_{j=1}^{n}\left(-2P\left(\frac{P\sigma_{j}(\mathbf{a})^{\top}\mathbf{h}_{j}}{P\left\Vert \mathbf{h}_{j}\right\Vert ^{2}+1}\right)\sigma_{j}(\mathbf{a})^{\top}\mathbf{h}_{j}+P\left\Vert \sigma_{j}(\mathbf{a})\right\Vert ^{2}\right)\\
= & \frac{1}{n}\sum_{j=1}^{n}\left(\left\Vert \sigma_{j}(\mathbf{a})\right\Vert ^{2}-\sigma_{j}(\mathbf{a})^{\top}\left(\frac{P}{P\left\Vert \mathbf{h}_{j}\right\Vert ^{2}+1}\mathbf{h}_{j}\mathbf{h}_{j}^{\top}\right)\sigma_{j}(\mathbf{a})\right)\\
= & \frac{1}{n}\sum_{j=1}^{n}\sigma_{j}(\mathbf{a})^{\top}\left(\underset{\triangleq\mathbf{M}_{j}}{\underbrace{\mathbf{I}-\frac{P}{P\left\Vert \mathbf{h}_{j}\right\Vert ^{2}+1}\mathbf{h}_{j}\mathbf{h}_{j}^{\top}}}\right)\sigma_{j}(\mathbf{a}).
\end{align*}
Then the computation rate in (\ref{eq:am}) can be written as

\[
R_{\mathrm{comp}}\left(\left\{ \mathbf{H}_{l}\right\} ,\mathbf{a}\right)=\frac{n}{2}\log^{+}\left(\frac{1}{\left(1/n\right)\sum_{j=1}^{n}\sigma_{j}(\mathbf{a})^{\top}\mathbf{M}_{j}\sigma_{j}(\mathbf{a})}\right),
\]
in which the free parameter is $\mathbf{a}\in\mathcal{O}_{\mathbb{K}}^{L}$.

\section{\label{sec:Algebraic-approximation}Diophantine approximation by
algebraic conjugates}

Our proof may be viewed as an extension of Khintchine's theorem for
complex numbers given in \cite[Section 4]{Dodson03}, which dealt
with Diophantine approximation by ratios of Gaussian integers. We
first recall a result from \cite[Theorem 5]{Murty2007}, \cite[p. 132]{Lang1971}
to count the number of principal ideals in $\mathcal{O}_{\mathbb{K}}$.
\begin{lem}
\label{thm:num_ideals}Let $J(k,\mathbb{K})$ be the number of principal
ideals in $\mathcal{O}_{\mathbb{K}}$ with norm no larger than $k$.
Then 
\[
|J(k,\mathbb{K})-\rho_{\mathbb{K}}k|\leq\frac{\rho_{\mathbb{K}}}{w}2^{n}k^{\frac{n-1}{n}}\max\left(1,\Phi_{0}^{n}\right),
\]
where $\Phi_{0}=2^{n-1}n^{2n}\bar{\gamma}^{n}e^{rM(n-1)}$, $\rho_{\mathbb{K}}=\frac{2^{r_{1}}(2\pi)^{r_{2}}R_{\mathbb{K}}}{w\sqrt{|\mathrm{disc}_{\mathbb{K}}|}}$,
$w$ denotes the number of roots of unity in $\mathbb{K}$, $R_{\mathbb{K}}$
denotes the regulator of the log-unit lattice, $(r_{1},r_{2})$ is
the signature of $\mathbb{K}$, $r=r_{1}+r_{2}-1$, and $\bar{\gamma},\thinspace M$
are parameters of the log-unit lattice. 
\end{lem}
%	\centering	
%\flushleft
\begin{figure*}
\centering

\subfloat[$q=(5+\sqrt{5})/2$, $\psi(|\mathrm{Nr}(q)|)=5^{-1.2}$.]{\begin{centering}
\includegraphics[width=0.42\textwidth]{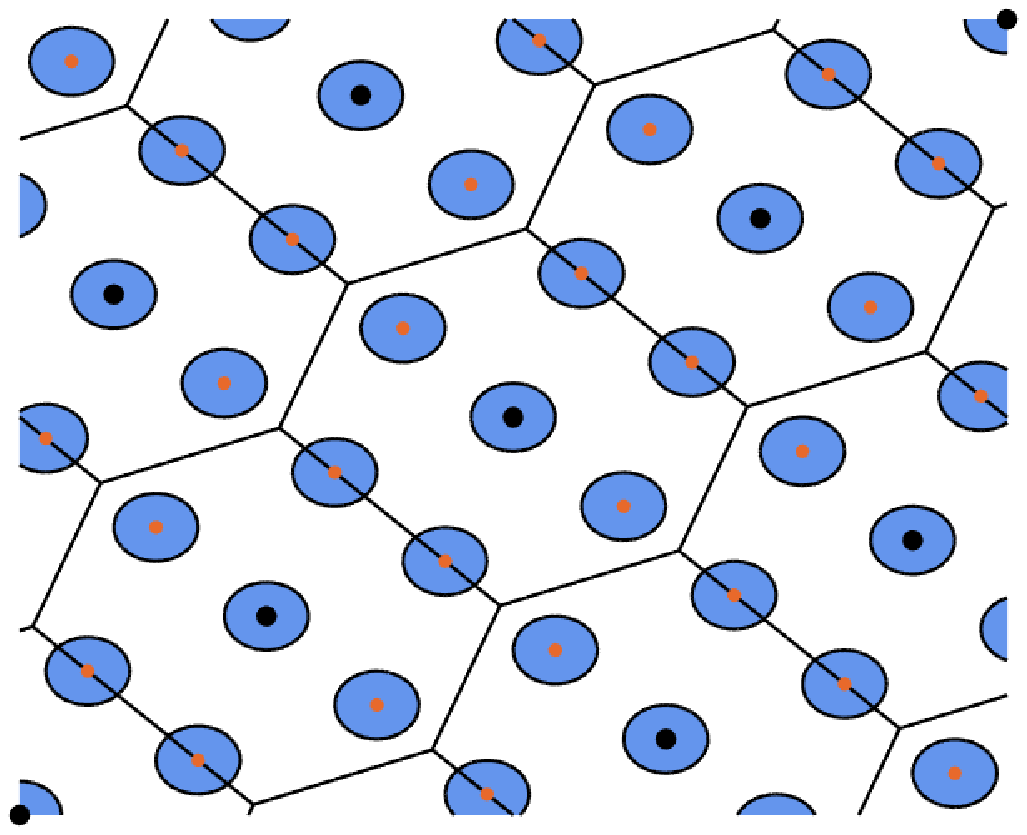} 
\par\end{centering}
}\subfloat[$q=4+\sqrt{5}$, $\psi(|\mathrm{Nr}(q)|)=11^{-1.2}$.]{\begin{centering}
\includegraphics[width=0.42\textwidth]{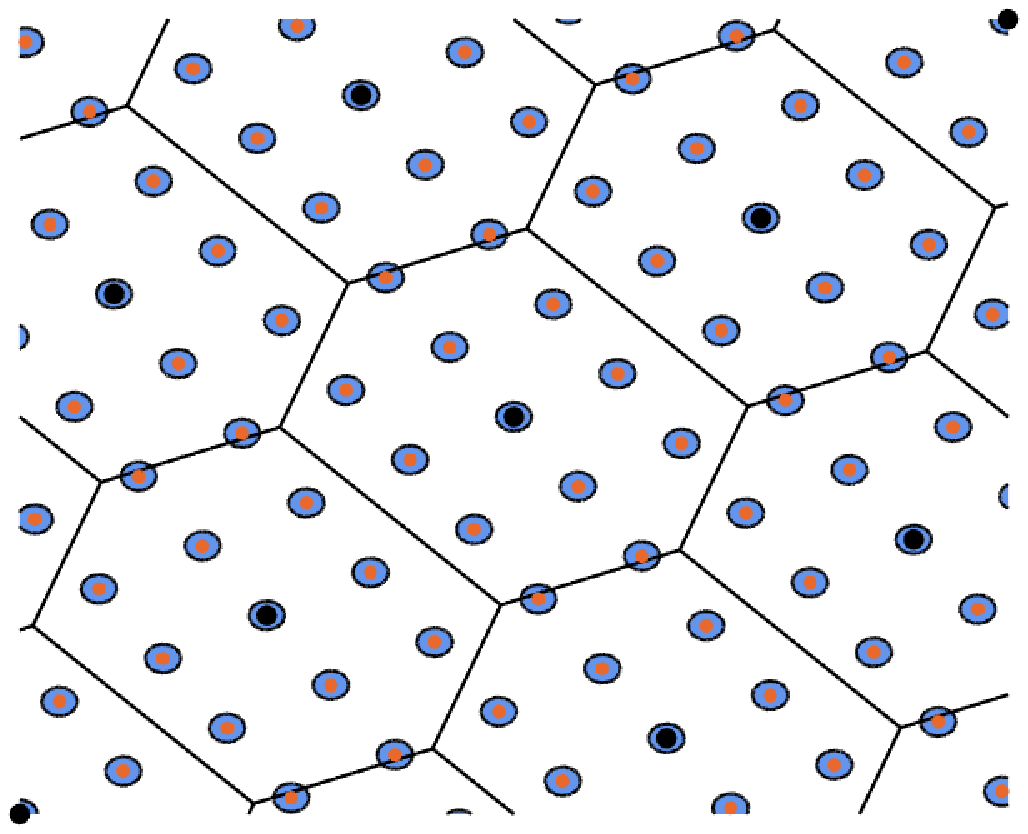} 
\par\end{centering}
}\caption{Approximating $[h_{1},h_{2}]^{\top}\in\mathbb{R}^{2}$ with $\mathcal{O}_{\mathbb{K}}=\mathbb{Z}[\frac{1+\sqrt{5}}{2}]$.
The well approximable set $\left\{ [h_{1},h_{2}]^{\top}\mathrel{\Big|}\min_{a\in\mathcal{O}_{\mathbb{K}}}\left\Vert [h_{1},h_{2}]^{\top}-\sigma(a/q)\right\Vert <\psi(|\mathrm{Nr}(q)|)\right\} $
is shaded in blue. Black dots denote lattice points in $\sigma\left(\mathcal{O}_{\mathbb{K}}\right)$.
Orange dots $\cup_{a\in\mathcal{O}_{\mathbb{K}}}\sigma(a/q)$ denote
the centers of balls in approximating $[h_{1},h_{2}]^{\top}$.}
\label{fig_demo_DA} 
\end{figure*}
\begin{IEEEproof}[Proof of Lemma \ref{lem: DA_conjugate}]
Firstly assume that $\mathbf{H}_{l}$ belongs to $\mathcal{V}_{\mathcal{O}_{\mathbb{K}}}$,
the fundamental Voronoi region of lattice $\sigma\left(\mathcal{O}_{\mathbb{K}}\right)$,
for $1\leq l\leq L$. Let ties on the boundary of $\mathcal{V}_{\mathcal{O}_{\mathbb{K}}}$
be broken in an arbitrary manner. Define 
\begin{align}
& \mathcal{A}_{q,\psi}\triangleq \nonumber \\
& \left\{ \left\{ \mathbf{H}_{l}\right\} \mathrel{\Big|}\max_{l\in\{1,\ldots,L\}}\min_{a\in\mathcal{O}_{\mathbb{K}}}\left\Vert \mathbf{H}_{l}-\mathrm{dg}\left(\sigma(a/q)\right)\right\Vert <\psi(|\mathrm{Nr}(q)|)\right\} \label{eq:vecDA}
\end{align}
for a fixed $q$. Note that $\mathcal{A}_{q,\psi}=\mathcal{A}_{uq,\psi}$
for any unit $u\in\mathcal{U}$, since 
\begin{equation}
\min_{a\in\mathcal{O}_{\mathbb{K}}}\left\Vert \mathbf{H}_{l}-\mathrm{dg}\left(\sigma(a/(uq))\right)\right\Vert =\min_{a\in\mathcal{O}_{\mathbb{K}}}\left\Vert \mathbf{H}_{l}-\mathrm{dg}\left(\sigma(a/q)\right)\right\Vert \label{eq:samelattice}
\end{equation}
and since $|\mathrm{Nr}(q)|=|\mathrm{Nr}(uq)|$. This means that when
investigating a sequence of $\left\{ q\right\} $ with decreasing
approximation error $\psi(|\mathrm{Nr}(q)|)$, we only have to pick
$q$ modulo the unit group. %Fig. \ref{fig_demo_DA} serves as an example to explain
%Eq. (\ref{eq:samelattice}).

%	Since for all $\mathbf{H}_{l}$,
%	\[
%	\min_{a\in\mathcal{O}_{\mathbb{K}}}\left\Vert \mathbf{H}_{l}-\sigma(a/q)\right\Vert =\begin{cases}
%	\min_{a\in\mathcal{O}_{\mathbb{K}}}\left\Vert \mathbf{H}_{l}-\mathcal{Q}_{\mathcal{O}_{\mathbb{K}}}\left(\mathbf{H}_{l}\right)-\sigma(a/q)\right\Vert ,\\
%	\min_{a\in\mathcal{O}_{\mathbb{K}}}\left\Vert \mathbf{H}_{l}-\mathrm{dg}\left(\sigma(uq)^{-1}\sigma(a)\right)\right\Vert ,\thinspace u\in\mathcal{U},
%	\end{cases}
%	\]
%	we can specify some constraints on $\mathbf{H}_{l}$ and $q$. It
%	suffices to i) investigate the well approximable points within the
%	Voronoi region of the $\sigma\left(\mathcal{O}_{\mathbb{K}}\right)$ lattice, i.e., let
%	$\mathrm{dg}\left(\mathbf{H}_{l}\right)\in \mathcal{V}_{\mathcal{O}_{\mathbb{K}}}$.
%	ii) investigate $q$ modulo the unit group,
%	and only the algebraic norm of $q$ dictates the approximation error.
%	We are now interested in a set $\mathcal{A}_{q,\psi}\subset\mathcal{V}_{\mathcal{O}_{\mathbb{K}}}^{L}$
%	w.r.t. $q$ and $\psi$, whose points are all well approximable :
%	\begin{equation}
%	\mathcal{A}_{q,\psi}\triangleq\left\{ \left\{ \mathbf{H}_{l}\cap\mathcal{V}_{\mathcal{O}_{\mathbb{K}}}\right\} \mathrel{\Big|}\max_{l\in \{1,\ldots,L\}}\min_{a\in\mathcal{O}_{\mathbb{K}}}\left\Vert \mathbf{H}_{l}-\sigma(a/q)\right\Vert <\psi(|\mathrm{Nr}(q)|)\right\}. \label{eq:vecDA}
%	\end{equation}

Denote by $\mathcal{B}\left(c,r\right)$ a ball of radius $r$ with
centre at $c$. The Lebesgue measure of a ball of radius $\psi(|\mathrm{Nr}(q)|)$
centred at $\sigma(a/q)$ is given by 
\[
\Upsilon\left(\mathcal{B}\left(\sigma(a/q),\psi(|\mathrm{Nr}(q)|)\right)\right)=\frac{\pi^{n/2}}{\Gamma(\frac{n}{2}+1)}\psi(|\mathrm{Nr}(q)|)^{n}.
\]

A congruence consideration shows that the number of points $\sigma(a/q)\in\mathcal{V}_{\mathcal{O}_{\mathbb{K}}}$
in (\ref{eq:vecDA}) is exactly $|\mathrm{Nr}(q)|$. We further elaborate
counting lattice points inside the fundamental Voronoi region $\mathcal{V}_{\mathcal{O}_{\mathbb{K}}}$
in Fig. \ref{fig_demo_DA}. Then the total measure of $\mathcal{A}_{q,\psi}$
is bounded by 
\begin{align}
 & \Upsilon\left(\mathcal{A}_{q,\psi}\right)\leq\left(\Upsilon(\mathcal{B}\left(\sigma(a/q),\psi(|\mathrm{Nr}(q)|)\right))|\mathrm{Nr}(q)|\right)^{L}\nonumber \\
 & =\left(\frac{\pi^{n/2}}{\Gamma(\frac{n}{2}+1)}\psi(|\mathrm{Nr}(q)|)^{n}|\mathrm{Nr}(q)|\right)^{L}.\label{eq:pr1}
\end{align}

Further define 
\begin{equation}
\mathcal{W}_{\psi}\triangleq\limsup_{|\mathrm{Nr}({q})|\rightarrow\infty}\mathcal{A}_{{q},\psi}=\bigcap_{N=1}^{\infty}\bigcup_{k=N}^{\infty}\bigcup_{{q}:\thinspace|\mathrm{Nr}({q})|=k}\mathcal{A}_{{q},\psi},\label{eq:pr2}
\end{equation}
as the subset of $\left\{ \mathbf{H}_{l}\right\} $ for which (\ref{eq:vecDA})
holds for infinitely many $q$ modulo the unit group.

Let $\mathfrak{q}=(q)$ denote the principal ideal generated by $q$.
Since $(q)=(qu)$ for any unit $u\in\mathcal{U}$, the set of algebraic
integers can be partitioned into different subsets indicated by principle
ideals. Since $\mathrm{Nr}(\mathfrak{q})=|\mathrm{Nr}(q)|$,
the number of subsets $q\mathcal{U}$ whose elements have absolute
norm $k$ is equal to the number of principal ideals with norm $k$.
Consequently, we have, for $N=1,2,\cdots,\infty$, 
\begin{align*}
 & \Upsilon\left(\bigcup_{k=N}^{\infty}\thinspace\bigcup_{{q}:\thinspace|\mathrm{Nr}({q})|=k}\mathcal{A}_{{q},\psi}\right)\\
 & \leq\sum_{k=N}^{\infty}\thinspace\sum_{\mathfrak{q}:\thinspace\mathrm{Nr}(\mathfrak{q})=k}\left(\frac{\pi^{n/2}}{\Gamma(\frac{n}{2}+1)}\psi(\mathrm{Nr}(\mathfrak{q}))\mathrm{Nr}(\mathfrak{q})\right)^{L}\\
 & =\left(\frac{\pi^{n/2}}{\Gamma(\frac{n}{2}+1)}\right)^{L}\sum_{k=N}^{\infty}\psi(k)^{nL}k^{L}\sum_{\mathfrak{q}:\thinspace\mathrm{Nr}(\mathfrak{q})=k}{1}
\end{align*}
where $\mathfrak{q}:\thinspace\mathrm{Nr}(\mathfrak{q})=k$ denotes
a principal ideal with norm $k$. By Lemma \ref{thm:num_ideals},
we have $J(k,\mathbb{K})=\rho_{\mathbb{K}}k+O\left(k^{\frac{n-1}{n}}\right)$.
As $O\left(k^{\frac{n-1}{n}}\right)$ grows no faster than $k$, we
have $\sum_{\mathfrak{q}:\thinspace\mathrm{Nr}(\mathfrak{q})=k}{1}=O\left(1\right)$,
which is bounded by a constant in the limit of $k$.

By the Borel--Cantelli lemma \cite{Cassels1957}, the Lebesgue measure
$\Upsilon\left(\mathcal{W}_{\psi}\right)=0$ if $\Upsilon\left(\bigcup_{k=N}^{\infty}\thinspace\bigcup_{{q}:\thinspace|\mathrm{Nr}({q})|=k}\mathcal{A}_{{q},\psi}\right)<\infty$.
Obviously, the convergence of the series $\sum_{k=1}^{\infty}\psi(k)^{nL}k^{L}$
implies that $\Upsilon\left(\mathcal{W}_{\psi}\right)=0$. Since a
countably infinite number of the Voronoi regions cover the whole space,
our result holds for all $\left\{ \mathbf{H}_{l}\right\} $. In fact,
it is readily verified that the set $\mathcal{W}_{\psi}$ is periodic
with respect to lattice $\sigma(\mathcal{O}_{\mathbb{K}})$. This
establishes an algebraic version of Khintchin's theorem \cite{Cassels1957}
in the convergent part.

Since for almost all $\left\{ \mathbf{H}_{l}\right\} $, (\ref{eq:vecDA})
holds for finitely many $q$ modulo the unit group, there exists a
finite constant $c_{\left\{ \mathbf{H}_{l}\right\} }$ such that 
\[
\max_{l\in\{1,\ldots,L\}}\min_{a\in\mathcal{O}_{\mathbb{K}}}\left\Vert \mathbf{H}_{l}-\sigma(a/q)\right\Vert \geq\psi(|\mathrm{Nr}(q)|)
\]
for all $|\mathrm{Nr}(q)|\geq c_{\left\{ \mathbf{H}_{l}\right\} }$.
So one can claim that 
\[
\max_{l\in\{1,\ldots,L\}}\min_{a\in\mathcal{O}_{\mathbb{K}}}\left\Vert \mathbf{H}_{l}-\sigma(a/q)\right\Vert \geq c_{\left\{ \mathbf{H}_{l}\right\} }'\psi(|\mathrm{Nr}(q)|)
\]
for all algebraic integer $q$ with 
\begin{align*}
& c_{\left\{ \mathbf{H}_{l}\right\} }'=
\\
& \min\left\{ 1,\min_{{q}:\thinspace|\mathrm{Nr}(q)|<c_{\left\{ \mathbf{H}_{l}\right\} }}\frac{\max_{l\in\{1,\ldots,L\}}\min_{a\in\mathcal{O}_{\mathbb{K}}}\left\Vert \mathbf{H}_{l}-\sigma(a/q)\right\Vert }{\psi(|\mathrm{Nr}(q)|)}\right\} .
\end{align*}
\end{IEEEproof}

\section{\label{sec:AlgebraicFields}Real number fields with small discriminants}

\begin{table*}[th!]
\caption{Real quadratic fields with small discriminants.}
\label{tab:title1} \centering %
\begin{tabular}{|c||c|c|c|c|c|}
\hline 
$\mathfrak{m}_{\theta}$  & $\theta^{2}-\theta-1$  & $\theta^{2}-2$  & $\theta^{2}-3$  & $\theta^{2}-\theta-3$  & $\theta^{2}-\theta-4$\tabularnewline
\hline 
$\mathrm{disc}_{\mathbb{K}}$  & $5$  & $8$  & $12$  & $13$  & $17$\tabularnewline
\hline 
basis $\phi$  & $\left\{ 1,\thinspace\theta\right\} $  & $\left\{ 1,\thinspace\theta\right\} $  & $\left\{ 1,\thinspace\theta\right\} $  & $\left\{ 1,\thinspace\theta\right\} $  & $\left\{ 1,\thinspace\theta\right\} $\tabularnewline
\hline 
\end{tabular}
\end{table*}
\begin{table*}[th!]
\caption{Real cubic fields with small discriminants.}
\label{tab:title2} \centering %
\begin{tabular}{|c||c|c|c|c|}
\hline 
$\mathfrak{m}_{\theta}$  & $\theta^{3}+\theta^{2}-2\theta-1$  & $\theta^{3}-3\theta-1$  & $\theta^{3}+\theta^{2}-3\theta-1$  & $\theta^{3}-\theta^{2}-4\theta-1$\tabularnewline
\hline 
$\mathrm{disc}_{\mathbb{K}}$  & $49$  & $81$  & $148$  & $169$ \tabularnewline
\hline 
basis $\phi$  & $\left\{ 1,\thinspace\theta,\thinspace\theta^{2}\right\} $  & $\left\{ 1,\thinspace\theta,\thinspace\theta^{2}\right\} $  & $\left\{ 1,\thinspace\theta,\thinspace\theta^{2}\right\} $  & $\left\{ 1,\thinspace\theta,\thinspace\theta^{2}\right\} $\tabularnewline
\hline 
\end{tabular}
\end{table*}
\begin{table*}[th!]
\caption{Real quartic fields with small discriminants.}
\centering \label{tab:title2-1} %
\begin{tabular}{|c||c|c|}
\hline 
$\mathfrak{m}_{\theta}$  & $\theta^{4}+\theta^{3}-3\theta^{2}-\theta+1$  & $\theta^{4}+\theta^{3}-4\theta^{2}-4\theta+1$\tabularnewline
\hline 
$\mathrm{disc}_{\mathbb{K}}$  & $725$  & $1125$\tabularnewline
\hline 
basis $\phi$  & $\left\{ 1,\thinspace\theta,\thinspace-1+\theta+\theta^{2},\thinspace-1-2\theta+\theta^{2}+\theta^{3}\right\} $  & $\left\{ 1,\thinspace\theta,\thinspace-2+\theta^{2},-1-3\theta+\theta^{3}\right\} $\tabularnewline
\hline 
\hline 
$\mathfrak{m}_{\theta}$  & $\theta^{4}-4\theta^{3}+8\theta-1$  & $\theta^{4}-4\theta^{2}+\theta+1$\tabularnewline
\hline 
$\mathrm{disc}_{\mathbb{K}}$  & $1600$  & $1957$ \tabularnewline
\hline 
basis $\phi$  & $\left\{ 1,\thinspace\theta,\thinspace\frac{1}{2}\left(-1-2\theta+\theta^{2}\right),\thinspace\frac{1}{2}\left(3-\theta-3\theta^{2}+\theta^{3}\right)\right\} $  & $\left\{ 1,\thinspace\theta,\thinspace-2+\theta^{2},1-3\theta+\theta^{3}\right\} $\tabularnewline
\hline 
\end{tabular}
\end{table*}
\begin{table*}[th!]
\caption{Real quintic fields with small discriminants.}
\label{tab:title2-2} \centering

\begin{tabular}{|c||c|c|}
\hline 
$\mathfrak{m}_{\theta}$  & $\theta^{5}+\theta^{4}-4\theta^{3}-3\theta^{2}+3\theta+1$  & $\theta^{5}-5\theta^{3}-\theta^{2}+3\theta+1$\tabularnewline
\hline 
$\mathrm{disc}_{\mathbb{K}}$  & $14641$  & $24217$\tabularnewline
\hline 
basis $\phi$  & $\left\{ 1,\thinspace\theta,\thinspace-2+\theta^{2},\thinspace-3\theta+\theta^{3},\thinspace1-2\theta-3\theta^{2}+\theta^{3}+\theta^{4}\right\} $  & $\left\{ 1,\thinspace\theta,\thinspace-2+\theta^{2},\thinspace-1-4\theta+\theta^{3},\thinspace2-5\theta^{2}+\theta^{4}\right\} $\tabularnewline
\hline 
\hline 
$\mathfrak{m}_{\theta}$  & $\theta^{5}+\theta^{4}-5\theta^{3}-3\theta^{2}+2\theta+1$  & $\theta^{5}+\theta^{4}-5\theta^{3}-\theta^{2}+4\theta-1$\tabularnewline
\hline 
$\mathrm{disc}_{\mathbb{K}}$  & $36497$  & $38569$ \tabularnewline
\hline 
basis $\phi$  & $\left\{ 1,\thinspace\theta,\thinspace-2+\theta^{2},\thinspace-2-4\theta+\theta^{2}+\theta^{3},\thinspace1+2\theta-5\theta^{2}+\theta^{4}\right\} $  & $\left\{ 1,\thinspace\theta,\thinspace-2+\theta+\theta^{2},\thinspace-3\theta+\theta^{2}+\theta^{3},\thinspace3-2\theta-5\theta^{2}+\theta^{3}+\theta^{4}\right\} $\tabularnewline
\hline 
\end{tabular}
\end{table*}
\begin{table*}[th!]
\caption{Maximal real sub-fields of cyclotomic fields $\mathbb{Q}\left(\zeta_{k}\right)$.}
\label{tab:cyclo} \centering

\begin{tabular}{|c||c|c|}
\hline 
$\mathbb{Q}\left(\zeta_{k}+\zeta_{k}^{-1}\right)$  & $\mathbb{Q}\left(\zeta_{23}+\zeta_{23}^{-1}\right)$  & $\mathbb{Q}\left(\zeta_{29}+\zeta_{29}^{-1}\right)$\tabularnewline
\hline 
$\phi\left(k\right)/2$  & $11$  & $14$\tabularnewline
\hline 
$\mathfrak{m}_{\zeta_{k}+\zeta_{k}^{-1}}$  & $\zeta^{11}+\zeta^{10}-10\zeta^{9}-9\zeta^{8}+36\zeta^{7}+28\zeta^{6}$  & $\zeta^{14}+\zeta^{13}-13\zeta^{12}-12\zeta^{11}+66\zeta^{10}+55\zeta^{9}-165\zeta^{8}$ \tabularnewline
 & $-56\zeta^{5}-35\zeta^{4}+35\zeta^{3}+15\zeta^{2}-6\zeta-1$  & $-120\zeta^{7}+210\zeta^{6}+126\zeta^{5}-126\zeta^{4}-56\zeta^{3}+28\zeta^{2}+7\zeta-1$\tabularnewline
\hline 
$\mathrm{disc}_{\mathbb{Q}\left(\zeta_{k}+\zeta_{k}^{-1}\right)}$  & $41426511213649$  & $10260628712958602189$\tabularnewline
\hline 
\end{tabular}
\end{table*}

\section*{Acknowledgment}

The authors acknowledge Dr. Yu-Chih (Jerry) Huang, Prof. Joseph J.
Boutros and Prof. Jean-Claude Belfiore for fruitful discussions, as well as the anonymous reviewers
whose comments improved the presentation of this work.

\bibliographystyle{IEEEtranMine}
\bibliography{lib}

\begin{thebibliography}{10}
\providecommand{\url}[1]{#1}
\csname url@samestyle\endcsname
\providecommand{\newblock}{\relax}
\providecommand{\bibinfo}[2]{#2}
\providecommand{\BIBentrySTDinterwordspacing}{\spaceskip=0pt\relax}
\providecommand{\BIBentryALTinterwordstretchfactor}{4}
\providecommand{\BIBentryALTinterwordspacing}{\spaceskip=\fontdimen2\font plus
\BIBentryALTinterwordstretchfactor\fontdimen3\font minus
  \fontdimen4\font\relax}
\providecommand{\BIBforeignlanguage}[2]{{%
\expandafter\ifx\csname l@#1\endcsname\relax
\typeout{** WARNING: IEEEtran.bst: No hyphenation pattern has been}%
\typeout{** loaded for the language `#1'. Using the pattern for}%
\typeout{** the default language instead.}%
\else
\language=\csname l@#1\endcsname
\fi
#2}}
\providecommand{\BIBdecl}{\relax}
\BIBdecl

\bibitem{Cover1979}
T.~M. Cover and A.~E. Gamal, ``Capacity theorems for the relay channel,''
  \emph{{IEEE} Trans. Inf. Theory}, vol.~25, no.~5, pp. 572--584, Sep. 1979.

\bibitem{Wang2007}
T.~Wang, A.~Cano, G.~B. Giannakis, and J.~N. Laneman, ``High-performance
  cooperative demodulation with decode-and-forward relays,'' \emph{{IEEE}
  Trans. Commun.}, vol.~55, no.~7, pp. 1427--1438, Jul. 2007.

\bibitem{Borade2007}
S.~Borade, L.~Zheng, and R.~G. Gallager, ``Amplify-and-forward in wireless
  relay networks: Rate, diversity, and network size,'' \emph{{IEEE} Trans. Inf.
  Theory}, vol.~53, no.~10, pp. 3302--3318, Oct. 2007.

\bibitem{Patel2007}
C.~S. Patel and G.~L. St{\"{u}}ber, ``Channel estimation for amplify and
  forward relay based cooperation diversity systems,'' \emph{{IEEE} Trans.
  Wirel. Commun.}, vol.~6, no.~6, pp. 2348--2356, Jun. 2007.

\bibitem{Lim2011}
S.~H. Lim, Y.~Kim, A.~E. Gamal, and S.~Chung, ``Noisy network coding,''
  \emph{{IEEE} Trans. Inf. Theory}, vol.~57, no.~5, pp. 3132--3152, May 2011.

\bibitem{Song2013}
Y.~Song and N.~Devroye, ``Lattice codes for the {G}aussian relay channel:
  Decode-and-forward and compress-and-forward,'' \emph{{IEEE} Trans. Inf.
  Theory}, vol.~59, no.~8, pp. 4927--4948, Aug. 2013.

\bibitem{Nazer2011}
B.~Nazer and M.~Gastpar, ``Compute-and-forward: Harnessing interference through
  structured codes,'' \emph{{IEEE} Trans. Inf. Theory}, vol.~57, no.~10, pp.
  6463--6486, Oct. 2011.

\bibitem{Zhan2009}
J.~Zhan, U.~Erez, M.~Gastpar, and B.~Nazer, ``{MIMO} compute-and-forward,'' in
  \emph{Proc. {IEEE} Int. Symp. Inf. Theory, {ISIT} 2009, Seoul, Korea}.\hskip
  1em plus 0.5em minus 0.4em\relax {IEEE}, 2009, pp. 2848--2852.

\bibitem{ElBakoury2015}
I.~E. Bakoury and B.~Nazer, ``The impact of channel variation on
  integer-forcing receivers,'' in \emph{Proc. {IEEE} Int. Symp. Inf. Theory,
  {ISIT} 2015, Hong Kong, China}.\hskip 1em plus 0.5em minus 0.4em\relax
  {IEEE}, 2015, pp. 576--580.

\bibitem{Wang2016}
P.~Wang, Y.~Huang, K.~R. Narayanan, and J.~J. Boutros, ``Physical-layer
  network-coding over block fading channels with root-{LDA} lattice codes,'' in
  \emph{Proc. {IEEE} Int. Conf. Commun., {ICC} 2016, Kuala Lumpur,
  Malaysia}.\hskip 1em plus 0.5em minus 0.4em\relax {IEEE}, 2016, pp. 1--6.

\bibitem{Tse2012}
\BIBentryALTinterwordspacing
D.~Tse and P.~Viswanath, \emph{Fundamentals of Wireless Communication}.\hskip
  1em plus 0.5em minus 0.4em\relax Cambridge University Press, 2012.
\BIBentrySTDinterwordspacing

\bibitem{Kositwattanarerk2015}
W.~Kositwattanarerk, S.~S. Ong, and F.~E. Oggier, ``Construction {A} of
  lattices over number fields and block fading (wiretap) coding,'' \emph{{IEEE}
  Trans. Inf. Theory}, vol.~61, no.~5, pp. 2273--2282, May 2015.

\bibitem{CampelloLingBelfiore2016}
A.~Campello, C.~Ling, and J.~Belfiore, ``Algebraic lattice codes achieve the
  capacity of the compound block-fading channel,'' in \emph{Proc. {IEEE} Int.
  Symp. Inf. Theory, {ISIT} 2016, Barcelona, Spain}.\hskip 1em plus 0.5em minus
  0.4em\relax {IEEE}, 2016, pp. 910--914.

\bibitem{Campello2016}
\BIBentryALTinterwordspacing
------, ``Universal lattice codes for {MIMO} channels,'' \emph{{IEEE} Trans.
  Information Theory}, vol.~64, no.~12, pp. 7847--7865, 2018.
\BIBentrySTDinterwordspacing

\bibitem{FSK13}
C.~Feng, D.~Silva, and F.~R. Kschischang, ``An algebraic approach to
  physical-layer network coding,'' \emph{IEEE Trans. Inf. Theory}, vol.~59,
  no.~11, pp. 7576--7596, Nov 2013.

\bibitem{Sun2013}
Q.~T. Sun, J.~Yuan, T.~Huang, and K.~W. Shum, ``Lattice network codes based on
  {E}isenstein integers,'' \emph{{IEEE} Trans. Commun.}, vol.~61, no.~7, pp.
  2713--2725, Jul. 2013.

\bibitem{Tunali2015}
N.~E. Tunali, Y.~Huang, J.~J. Boutros, and K.~R. Narayanan, ``Lattices over
  {E}isenstein integers for compute-and-forward,'' \emph{{IEEE} Trans. Inf.
  Theory}, vol.~61, no.~10, pp. 5306--5321, 10 2015.

\bibitem{Huang2015b}
\BIBentryALTinterwordspacing
Y.~Huang, K.~R. Narayanan, and P.~Wang, ``Lattices over algebraic integers with
  an application to compute-and-forward,'' \emph{{IEEE} Trans. Information
  Theory}, vol.~64, no.~10, pp. 6863--6877, 2018.
\BIBentrySTDinterwordspacing

\bibitem{Erez2004}
U.~Erez and R.~Zamir, ``Achieving 1/2 log {(1+SNR)} on the {AWGN} channel with
  lattice encoding and decoding,'' \emph{{IEEE} Trans. Inf. Theory}, vol.~50,
  no.~10, pp. 2293--2314, Oct. 2004.

\bibitem{Oggier2013}
F.~E. Oggier and J.~Belfiore, ``Enabling multiplication in lattice codes via
  construction {A},'' in \emph{{IEEE} Information Theory Workshop, {ITW} 2013,
  Sevilla, Spain}.\hskip 1em plus 0.5em minus 0.4em\relax {IEEE}, 2013, pp.
  1--5.

\bibitem{Campello2016b}
A.~Campello, C.~Ling, and J.~Belfiore, ``Algebraic lattices achieving the
  capacity of the ergodic fading channel,'' in \emph{{IEEE} Information Theory
  Workshop, {ITW} 2016, Cambridge, United Kingdom}.\hskip 1em plus 0.5em minus
  0.4em\relax {IEEE}, 2016, pp. 459--463.

\bibitem{Huang2016a}
\BIBentryALTinterwordspacing
Y.~Huang, (2016). ``Construction $\pi_a$ lattices : A review and recent
  results.'' [Online]. Available:
  {https://www.york.ac.uk/media/mathematics/documents/
  Jerry{\_}Huang{\_}York2016.pdf}
\BIBentrySTDinterwordspacing

\bibitem{Lyu2017a}
S.~Lyu, A.~Campello, C.~Ling, and J.~Belfiore, ``Compute-and-forward over
  block-fading channels using algebraic lattices,'' in \emph{Proc. {IEEE} Int.
  Symp. Inf. Theory, {ISIT} 2017, Aachen, Germany}.\hskip 1em plus 0.5em minus
  0.4em\relax {IEEE}, 2017, pp. 1848--1852.

\bibitem{Ordentlich2012}
O.~Ordentlich and U.~Erez, ``A simple proof for the existence of "good" pairs
  of nested lattices,'' \emph{{IEEE} Trans. Inf. Theory}, vol.~62, no.~8, pp.
  4439--4453, Aug. 2016.

\bibitem{Niesen2012}
U.~Niesen and P.~Whiting, ``The degrees of freedom of compute-and-forward,''
  \emph{{IEEE} Trans. Inf. Theory}, vol.~58, no.~8, pp. 5214--5232, Aug. 2012.

\bibitem{Ordentlich2014}
O.~Ordentlich, U.~Erez, and B.~Nazer, ``The approximate sum capacity of the
  symmetric {G}aussian {K}-user interference channel,'' \emph{{IEEE} Trans.
  Inf. Theory}, vol.~60, no.~6, pp. 3450--3482, Jun. 2014.

\bibitem{Nazer2016}
B.~Nazer and O.~Ordentlich, ``Diophantine approximation for network information
  theory: {A} survey of old and new results,'' in \emph{Proc. 54th Annu.
  Allert. Conf. Commun. Control. Comput., Allerton 2016, Monticello, IL,
  USA}.\hskip 1em plus 0.5em minus 0.4em\relax {IEEE}, 2016, pp. 990--996.

\bibitem{Cassels1957}
\BIBentryALTinterwordspacing
J.~W.~S. Cassels, \emph{An {I}ntroduction to {D}iophantine
  {A}pproximation}.\hskip 1em plus 0.5em minus 0.4em\relax Cambridge University
  Press, 1957.
\BIBentrySTDinterwordspacing

\bibitem{Roy04}
D.~Roy and M.~Waldschmidt, ``Diophantine approximation by conjugate algebraic
  integers,'' \emph{Compositio Math.}, vol. 140, no.~3, pp. 593--612, May 2004.

\bibitem{Roy05}
D.~Roy, ``Simultaneous approximation by conjugate algebraic numbers in fields
  of transcendence degree one,'' \emph{Int. J. Number Theory}, vol.~1, no.~3,
  pp. 357--382, 2005.

\bibitem{BK:Mollin-ANT}
R.~A. Mollin, \emph{Algebraic Number Theory}, 2nd~ed.\hskip 1em plus 0.5em
  minus 0.4em\relax Chapman and Hall/CRC, 2011.

\bibitem{ViterboOggier}
F.~Oggier and E.~Viterbo, ``Algebraic number theory and code design for
  {Rayleigh} fading channels,'' \emph{Foundations and Trends on Communications
  and Information Theory}, vol.~1, pp. 336--415, 2004.

\bibitem{BK:Zamir}
R.~Zamir, \emph{Lattice Coding for Signals and Networks}.\hskip 1em plus 0.5em
  minus 0.4em\relax Cambridge University Press, 2014.

\bibitem{Rogers-SD}
\BIBentryALTinterwordspacing
K.~Rogers and H.~P.~F. Swinnerton-Dyer, ``The geometry of numbers over
  algebraic number fields,'' \emph{Transactions of the American Mathematical
  Society}, vol.~88, no.~1, pp. 227--242, 1958.
\BIBentrySTDinterwordspacing

\bibitem{Lekkerkerker1987}
\BIBentryALTinterwordspacing
C.~G. Lekkerkerker and P.~Gruber, \emph{Geometry of Numbers}.\hskip 1em plus
  0.5em minus 0.4em\relax Elsevier Science, 1987.
\BIBentrySTDinterwordspacing

\bibitem{Fieker10}
C.~Fieker and D.~Stehl{\'{e}}, ``Short bases of lattices over number fields,''
  in \emph{Algorithmic Number Theory Symposium (ANTS)}, vol. 6197.\hskip 1em
  plus 0.5em minus 0.4em\relax Springer, 2010, pp. 157--173.

\bibitem{Dummit2003}
\BIBentryALTinterwordspacing
D.~S. Dummit and R.~M. Foote, \emph{Abstract Algebra}.\hskip 1em plus 0.5em
  minus 0.4em\relax John Wiley \& Sons, 2003.
\BIBentrySTDinterwordspacing

\bibitem{Zamir1996}
R.~Zamir and M.~Feder, ``On lattice quantization noise,'' \emph{{IEEE} Trans.
  Inf. Theory}, vol.~42, no.~4, pp. 1152--1159, 1996.

\bibitem{Leibak2005}
A.~Leibak, ``On additive generalization of {V}oronoi's theory to algebraic
  number fields,'' \emph{Proceedings of the Estonian Academy of Science
  Physics/Mathematics}, vol.~54, no.~4, pp. 195--212, 2005.

\bibitem{Baeza1997}
R.~Baeza and M.~Icaza, ``On {H}umbert-{M}inkowski's constant for a number
  field,'' \emph{Proceedings of the American Mathematical Society}, vol. 125,
  no.~11, pp. 3195--3202, 1997.

\bibitem{Micciancio2002}
D.~Micciancio and S.~Goldwasser, \emph{Complexity of Lattice Problems}.\hskip
  1em plus 0.5em minus 0.4em\relax Springer US, 2002.

\bibitem{Sah-TIT}
\BIBentryALTinterwordspacing
S.~Sahraei and M.~Gastpar, ``Polynomially solvable instances of the shortest
  and closest vector problems with applications to compute-and-forward,''
  \emph{{IEEE} Trans. Information Theory}, vol.~63, no.~12, pp. 7780--7792,
  2017.
\BIBentrySTDinterwordspacing

\bibitem{Lyu2017}
S.~Lyu and C.~Ling, ``Boosted {KZ} and {LLL} algorithms,'' \emph{{IEEE} Trans.
  Signal Process.}, vol.~65, no.~18, pp. 4784--4796, Sep. 2017.

\bibitem{Pohst1997}
\BIBentryALTinterwordspacing
M.~Pohst and H.~Zassenhaus, \emph{Algorithmic Algebraic Number Theory}.\hskip
  1em plus 0.5em minus 0.4em\relax Cambridge University Press, 1997.
\BIBentrySTDinterwordspacing

\bibitem{Washington1996}
\BIBentryALTinterwordspacing
L.~C. Washington, \emph{Introduction to Cyclotomic Fields}.\hskip 1em plus
  0.5em minus 0.4em\relax Springer-Verlag New York, 1996.
\BIBentrySTDinterwordspacing

\bibitem{Dodson03}
M.~M. {Dodson} and S.~{Kristensen}, ``{Hausdorff dimension and Diophantine
  approximation},'' \emph{ArXiv Mathematics e-prints}, May 2003.

\bibitem{Murty2007}
M.~R. Murty and J.~V. Order, ``Counting integral ideals in a number field,''
  \emph{Expositiones Mathematicae}, vol.~25, no.~1, pp. 53--66, Feb. 2007.

\bibitem{Lang1971}
\BIBentryALTinterwordspacing
S.~Lang, \emph{Algebraic Number Theory}.\hskip 1em plus 0.5em minus 0.4em\relax
  Springer-Verlag New York, 1994.
\BIBentrySTDinterwordspacing

\end{thebibliography}

\begin{IEEEbiographynophoto}{Shanxiang Lyu}
	received the B.Eng. and M.Eng. degrees in electronic and information engineering from
	South China University of Technology, Guangzhou, China, in 2011
	and 2014, respectively, and the Ph.D. degree from the
	Electrical and Electronic Engineering Department, Imperial College London,
	in 2018. 
	He is currently a lecturer   
	with the College of Cyber Security, Jinan University.
	 His
	research interests are in lattice theory, algebraic number theory, and their applications.
\end{IEEEbiographynophoto}

\begin{IEEEbiographynophoto}{Antonio Campello}
	received the Bachelor and PhD degrees in Applied
	Mathematics from the University of Campinas, Brazil, in 2009 and 2014,
	respectively. He was a visiting researcher at the Complutense University of
	Madrid in 2009, at the École Polytechnique fédérale de Lausanne (EPFL)
	1025 in 2011, and at AT\&T Research Labs - Shannon Labs, New Jersey
	in 2013. He was as a postdoctoral researcher at Télécom ParisTech, France,
	in and at Imperial College London, UK. His research interests are in the
	interplay between discrete geometry, number theory, communications and
	machine learning.
\end{IEEEbiographynophoto}

\begin{IEEEbiographynophoto}{Cong Ling} (S'99-A'01-M'04) 
	received the B.S. and M.S. degrees in electrical engineering from
	the Nanjing Institute of Communications Engineering, Nanjing, China, in 1995
	and 1997, respectively, and the Ph.D. degree in electrical engineering from
	the Nanyang Technological University, Singapore, in 2005.
	He had been on the faculties of the Nanjing Institute of Communications
	Engineering and King's College. He is currently a Reader (Associate Professor) with the Electrical and Electronic Engineering Department, Imperial
	College London. His research interests are coding, information theory, and
	security, with a focus on lattices.
	Dr. Ling has served as an Associate Editor for the IEEE TRANSACTIONS
	ON COMMUNICATIONS and the IEEE TRANSACTIONS ON VEHICULAR
	TECHNOLOGY.\end{IEEEbiographynophoto}

\end{document}